\documentclass[12pt]{article}

\usepackage{jheppub, bm, wrapfig}
\usepackage[utf8]{inputenc}
\numberwithin{equation}{section}
\renewcommand\afterTocRuleSpace{\bigskip}

\usepackage{amsfonts}
\usepackage{amsmath}
\usepackage{color}
\usepackage{tensor}
\usepackage{graphicx}

\usepackage{tikz}
\usetikzlibrary{arrows}
\usetikzlibrary{shapes.geometric,calc,arrows, positioning,shapes.misc,decorations.markings}
\tikzset{
  big arrow/.style={
    decoration={markings,mark=at position 1 with {\arrow[scale=2,#1]{>}}},
    postaction={decorate},
    shorten >=0.4pt},
  big arrow/.default=black}

\newcommand{\bea}{\begin{eqnarray}}
\newcommand{\eea}{\end{eqnarray}}
\newcommand{\be}{\begin{equation}}
\newcommand{\ee}{\end{equation}}
\newcommand{\bit}{\begin{itemize}}
\newcommand{\eit}{\end{itemize}}
\newcommand{\ben}{\begin{enumerate}}
\newcommand{\een}{\end{enumerate}}

\newcommand{\F}{{\mathbb F}}
\renewcommand{\P}{{\mathbb P}}

\newcommand{\cC}{\mathcal{C}}

\newcommand{\cF}{\mathcal{F}}

\newcommand{\cM}{\mathcal{M}}
\newcommand{\cN}{\mathcal{N}}

\newcommand{\cP}{\mathcal{P}}

\newcommand{\cT}{\mathcal{T}}

\newcommand{\fT}{\mathfrak{T}}

\newcommand{\fg}{\mathfrak{g}}

\newcommand{\ubf}[1]{\underline{\bf #1}}

\title{Classifying $5d$ SCFTs via $6d$ SCFTs: Rank one}

\author{Lakshya Bhardwaj\footnote{lbhardwaj@fas.harvard.edu}, Patrick Jefferson\footnote{patrickjefferson@fas.harvard.edu}}

\affiliation{Department of Physics, Harvard University, Cambridge, MA 02138, USA}

\abstract{Following a recent proposal, we delineate a general procedure to classify $5d$ SCFTs via compactifications of $6d$ SCFTs on a circle (possibly with a twist by a discrete global symmetry). The path from $6d$ SCFTs to $5d$ SCFTs can be divided into two steps. The first step involves computing the Coulomb branch data of the $5d$ KK theory obtained by compactifying a $6d$ SCFT on a circle of finite radius. The second step involves computing the limit of the KK theory when the inverse radius along with some other mass parameters is sent to infinity. Under this RG flow, the KK theory reduces to a $5d$ SCFT. We illustrate these ideas in the case of untwisted compactifications of rank one $6d$ SCFTs that can be constructed in F-theory without frozen singularities. The data of the corresponding KK theory can be packaged in the geometry of a Calabi-Yau threefold that we explicitly compute for every case. The RG flows correspond to flopping a collection of curves in the threefold and we formulate a concrete set of criteria which can be used to determine which collection of curves can induce the relevant RG flows, and, in principle, to determine the Calabi-Yau geometries describing the endpoints of these flows. We also comment on how to generalize our results to arbitrary rank.}

\begin{document}

\maketitle

\section{Introduction}
There is a long standing dream that it will be possible to obtain all lower dimensional quantum field theories via compactifications of higher dimensional quantum field theories. The supersymmetric version of this dream states that it will be possible to obtain all supersymmetric quantum field theories in spacetime dimension $d\le5$ by compactifying $6d$ SCFTs since it is believed that all UV complete QFTs can be obtained by deforming CFTs and $d=6$ is the maximum dimension permitting the existence of an SCFT\footnote{It is known that $6d$ SCFTs do not admit relevant deformations \cite{Cordova:2016xhm}.} \cite{Cordova:2016emh}. If we assume that the dream is correct, then we can hope to obtain all $5d$ SCFTs by compactifying $6d$ SCFTs on a circle\footnote{Here we should also include twists by discrete global symmetry transformations as one traverses the compactification circle. The twisted compactifications can lead to $5d$ SCFTs which cannot be obtained from untwisted compactifications of $6d$ SCFTs.}.

A way to make this hope concrete was recently proposed in \cite{Jefferson:2018irk} which classified all the possible rank\footnote{We say that a $5d$ theory has rank $n$ if its Coulomb branch is of dimension $n$. The number of BPS strings in the $5d$ spectrum is also $n$.} two $5d$ SCFTs that can be obtained by compactifying M-theory on a smooth Calabi-Yau threefold. It was found that there are a huge number of such $5d$ SCFTs, and the Coulomb branch of each of these is described by the data of the corresponding Calabi-Yau threefold used for compactifying M-theory. However, it was also noticed that all of these Calabi-Yau threefolds can be obtained by blowing down some curves inside a small number of ``parent'' Calabi-Yau threefolds. The parent Calabi-Yau threefolds also give rise to rank two $5d$ theories but these $5d$ theories are not $5d$ SCFTs. Rather they can be thought of as circle compactified $6d$ SCFTs viewed as $5d$ theories with KK modes. Thus these parent theories were dubbed as $5d$ KK theories in \cite{Jefferson:2018irk}. Now, the process of blowing down curves physically corresponds to performing RG flows where some BPS particles are integrated out of the theory. Based on this observation, the following conjecture was proposed in \cite{Jefferson:2018irk}, which we reiterate:
\begin{center}
\fbox{
\begin{minipage}[c][1cm][c]{15cm}
Every rank $n$ $5d$ SCFT can be obtained by a rank preserving RG flow starting from a $6d$ SCFT compactified on a circle with/without a discrete twist.
\end{minipage}
}
\end{center}

The conjecture can also be justified by the studies of $5d$ gauge theories where it has been found that for low number of flavors, the gauge theory has a $5d$ UV completion; but, if we keep adding flavors then we reach a gauge theory which has a $6d$ UV completion rather than a $5d$ one \cite{Douglas:1996xp, Morrison:1996xf}. Similarly, we expect that if we keep adding BPS particles consistently to a $5d$ SCFT (while keeping the number of BPS strings constant), then at some point we reach a $5d$ KK theory of the same rank.

In this paper, we start a systematic study of circle compactifications of $6d$ SCFTs, which according to the above conjecture can be used to classify $5d$ SCFTs. We will focus our attention on the untwisted compactifications of rank\footnote{Here we are referring to the $6d$ rank which is different from the $5d$ rank used above. The $6d$ rank counts the number of tensor multiplets on the tensor branch of the $6d$ SCFT.} one $6d$ SCFTs which can be constructed in the unfrozen phase\footnote{Frozen singularities in F-theory can construct $6d$ SCFTs which do not admit an F-theory construction without frozen singularities \cite{Bhardwaj:2019hhd,Bhardwaj:2018jgp}. See also \cite{Tachikawa:2015wka}. The incompleteness of the classification of \cite{Heckman:2015bfa, Heckman:2013pva} can also be noticed by comparing their classification with the gauge-theoretic classification of \cite{Bhardwaj:2015xxa}.} \cite{Heckman:2015bfa, Heckman:2013pva} of F-theory. In particular, we will associate a smooth Calabi-Yau threefold $X_\fT$ to each such rank one $6d$ SCFT $\fT$. We will determine $X_\fT$ by resolving the elliptically fibered singular Calabi-Yau threefold $Y_\fT$ appearing in F-theory construction for $\fT$.

Compactifying M-theory on $X_\fT$ leads to the $5d$ KK theory $\fT_{\text{KK}}$ obtained by compactifying $\fT$ on $S^1$ without any twist. The Coulomb branch prepotential for $\fT_{\text{KK}}$ can be recovered from the data of intersection numbers of holomorphic cycles in $X_\fT$. The spectrum of BPS particles relevant for rank preserving RG flows to $5d$ SCFTs can be identified with rational curves of self-intersection $-1$ in $X_\fT$. The KK mode of $\fT_{\text{KK}}$ can be identified with the elliptic fiber in $X_\fT$ whose volume is identified with the inverse radius of compactification $R^{-1}$. We note that some special cases of our results were already obtained by \cite{DelZotto:2017pti} who studied the Calabi-Yau threefolds corresponding to very special rank one $6d$ SCFTs that are completely Higgsed in the sense that they cannot be Higgsed to obtain some other $6d$ theory.

Let us close this introduction with a justification for capturing the data of $5d$ KK theories in terms of Calabi-Yau geometries rather than proceeding field theoretically. As emphasized in \cite{Jefferson:2017ahm}, various important physical processes (e.g. integrating out BPS particles or phase transitions) require us to know the precise functional dependence of the masses of BPS particles in terms of Coulomb branch moduli and mass parameters of the $5d$ theory. This dependence is only known field theoretically for particles that can be seen perturbatively in the $6d$ SCFT on its tensor branch. But the reduction on a circle generates new non-perturbative particles whose masses cannot be determined using present field theoretic methods. The Calabi-Yau geometry makes all these particles manifest and the calculation of their masses a straightforward task.

This paper is organized as follows. In Section \ref{review}, we review aspects of Coulomb branches in $5d$ supersymmetric theories and review how a Calabi-Yau threefold can describe a $5d$ Coulomb branch. In Section \ref{condense}, we review how a $6d$ SCFT compactified on a circle can be viewed as a $5d$ KK theory, and describe the general structure expected of Calabi-Yau threefolds describing the Coulomb branch of $5d$ KK theories. We also introduce our graphical notation which packages the relevant data of a Calabi-Yau in terms of a graph. In Section \ref{results}, we compile the main results of this paper, which is the association of a Calabi-Yau threefold to every\footnote{Actually, due to technical reasons, our methods do not allow us to associate a Calabi Yau threefold to a particular rank one $6d$ SCFT. The tensor branch of this theory is described by $SO(13)$ with a half-hyper in spinor representation and 7 hypers in vector. The F-theory construction of this SCFT involves a non-split I$^{*}_3$ fiber over a $-2$ curve.} rank one $6d$ SCFT. In Section \ref{RG}, we formulate criteria which, in principle, allow one to determine the Calabi-Yaus describing the $5d$ SCFTs arising the end points of RG flows that start from KK theories. In Section \ref{conclusion}, we comment on how to generalize our results to higher rank SCFTs. In Appendix \ref{background}, we collect some mathematical facts and notions that we use throughout the paper. In Appendix \ref{sample}, we provide sample computations of the Calabi-Yau threefolds for some hand-picked KK theories that illustrate some key features of our results.

\section{Coulomb branches in five dimensions: A review} \label{review}

\subsection{Field theoretic aspects}

The minimal supersymmetry algebra in $5d$ has eight supercharges, denoted as $\cN=1$. The $\cN=1$ vector multiplet contains a real scalar, which parametrizes a Coulomb branch of vacua $\cC$ on which the IR physics is described by an $\cN=1$ abelian gauge theory. The kinetic term for the scalars $\phi_{5d,i}$ in the low energy theory provides a natural metric on $\cC$.

A $5d$ $\cN=1$ SCFT has a space of relevant deformations $\cM$ parametrized by mass parameters $m_\alpha$. Each point $p$ in $\cM$ corresponds to a $5d$ $\cN=1$ QFT and we can associate its Coulomb branch to $p$ leading to a fibration $\cP$ of $\cC$ over $\cM$. The spectrum over each point in $\cP$ contains massive BPS particles and strings. The central charge for BPS particles can be written as
\be
Z=n_e^ia_i+f^\alpha m_\alpha
\ee
where $n_e^i$ denotes the electric charge under a low energy gauge group $U(1)_i$, $a_i:=\phi_{5d,i}$, and $f_\alpha$ denotes the charge under a flavor $U(1)_\alpha$ associated to $m_\alpha$. As a gauge $U(1)$ can be redefined by a linear combination flavor $U(1)$s, we have the freedom of shifting $a_i$ by a linear combination of $m_\alpha$. The central charge for BPS strings can be written as
\be
\hat Z=n_{m,i}a_D^i
\ee
where $n_{m,i}$ are the magnetic charges under $U(1)_i$ and $a_D^i=\frac{\partial\cF}{\partial a_i}$ where $\cF$ is the prepotential for the low energy abelian gauge theory.

$\cF$ is in general a cubic polynomial\footnote{$\cF$ can contain some absolute values which means that it is not smooth. More precisely, its third derivatives are not continuous.} \cite{Seiberg:1996bd, Intriligator:1997pq} in $a_i$ and $m_\alpha$. As we have discussed above, its first derivatives with respect to $a_i$ control the tensions of the BPS strings. Its second derivatives with respect to $a_i$ determine the kinetic terms for $\phi_{5d,i}$ and hence control the metric on $\cC$. Its third derivatives with respect to $a_i$ and $m_\alpha$ determine Chern-Simons terms.

The Chern-Simons levels of the low-energy abelian gauge theory can jump across some walls in $\cP$ leading to phase transitions \cite{Witten:1996qb, Seiberg:1996bd}. The locations of these walls are parametrized by some particles becoming massless.


\subsection{Geometric aspects} 

A five dimensional $\mathcal N=1$ field theory can be realized as the low energy effective description of M-theory compactified on a local Calabi-Yau threefold $X'$ \cite{Witten:1996qb}. The hallmark of this correspondence is the identification between the parameter space $\cP$ of the 5d theory and the K\"ahler moduli space of $X'$. The BPS particles arise from M2 branes wrapping compact holomorphic curves and the BPS strings arise from M5 branes wrapping compact holomorphic surfaces in $X'$. Their masses and tensions are proportional to the volumes of the corresponding curves and surfaces. To reach the conformal point, we want to be able to shrink all compact curves and surfaces inside $X'$ at a finite distance in the moduli space. Such a threefold is called shrinkable and the K\''ahler moduli space takes the form of a cone in such a case.

We can determine the geometry and intersection structure of the holomorphic 4-cycles in terms of a basis of (compact and non-compact) divisors $S_i$. To do this, we begin by expanding a K\"ahler class $J$ in terms of this basis:
	\begin{align}
		J = \phi_i S_i,~~ \phi_i \in \mathbb R_{\geq 0}.
	\end{align}
 In order to ensure positive volumes of $\text{vol}(C_p) = \frac{1}{p!} (J^p \cdot C_p)_{X'}$ of holomorphic $p$-cycles $C_p$, $J$ is required to be effective: 
	\begin{align}
		 (J \cdot C)_{X'}  = \text{vol}(C) \geq 0	
	\end{align}
for all holomorphic curves $C$ in $X'$; by a theorem when the above partial inequality is strict, this property holds for all $C_p$. Assuming transverse intersections $S_i \cap S_j$ which locally satisfy the Calabi-Yau condition \ref{CY}, the intersection structure of the divisors $S_i$ is determined in its entirety by the triple intersection numbers\footnote{Mathematically, the triple intersection of three non-compact divisors is not well-defined. Field theoretically, such triple intersections appear as monomials in $\cF$ depending only on mass parameters, and hence can be eliminated.}
	\begin{align}
		k_{ijk} =  (S_i \cdot S_j \cdot S_k)_{X'} 
	\end{align} 
which appear naturally in the expression for the relative volume of $X'$, or equivalently the geometric realization of the prepotential $\mathcal F$:
	\begin{align}
		\text{vol}(X') =  \frac{1}{3!} (J^3)_{X'} =\frac{1}{3!} k_{ijk} \phi_i \phi_j \phi_k =\mathcal F.
	\end{align} 
The following formula is useful to keep in mind when computing triple intersections:
	\begin{align}
			(S_i \cdot S_j \cdot S_k)_{X'} =  (  S_i |_{S_k} \cdot S_j |_{S_k})_{S_k},	
	\end{align}
where $S_{i}|_{S_j}$ indicates the restriction of the complex surface $S_i$ to the surface $S_j$, and the above formula holds for any permutation of the indices $i,j,k$. 

\section{\texorpdfstring{$5d$}{} KK theories and associated Calabi-Yau geometries}\label{condense}
We start out in Section \ref{sketch} by sketching what one expects the general structure of a smooth Calabi-Yau associated to a $5d$ KK theory to look like. We then introduce in Section \ref{package} a notation which packages all the relevant information about the Calabi-Yau into a graph. Section \ref{w_models} describes the tools to compute the resolved Calabi-Yau $\tilde X$ starting from a Weierstrass model for an elliptically fibered singular Calabi-Yau $X$ defining a $6d$ SCFT.
	
\subsection{ \texorpdfstring{$6d\to 5d$}{}} \label{sketch}
Every $6d$ SCFT $\fT$ is believed to admit a tensor branch of vacua $\cT$ on which the low energy effective theory is a non-abelian gauge theory interacting with some tensor multiplets. The scalars $\phi_{6d,a}$ in tensor multiplets parametrize $\cT$. When we compactify $\fT$ on a circle of finite radius $R$, we can view it as a $5d$ KK theory. The KK theory admits a Coulomb branch of vacua $\cC$ parametrized by scalars $\phi_{5d,i}$ which descend from $\phi_{6d,a}$ and from holonomies of the $6d$ gauge fields around the circle. Similarly, the mass parameters $m_\alpha$ for the KK theory descend from the holonomies of the continuous flavor symmetry groups of the $6d$ theory. Since our setup is on a circle, we have an additional mass parameter $m_B=R^{-1}$. The spectrum of BPS particles for a $5d$ KK theory arranges itself into towers such that the masses of two consecutive particles in a tower differ by $m_B$. This mass difference can be attributed to the central charge for the $U(1)_{KK}$ symmetry corresponding to translations along the compactification circle.

As we send $R$ to zero, the KK towers disappear and we land on a $5d$ SCFT. There are multiple ways to send $R$ to zero depending on how we tune $\langle\phi_{5d,i}\rangle$ and $m_\alpha$ in the process. Different ways of taking the $R\to0$ limit give rise to different $5d$ SCFTs starting from the same $5d$ KK theory.

A $6d$ SCFT can be constructed by compactifying F-theory on a (generically singular) non-compact elliptically fibered Calabi-Yau threefold with a smooth base $B$ that is required to satisfy some extra conditions. In particular, the compact holomorphic curves in $B$ must be rational and their intersection pairing matrix must be negative definite. The latter condition follows from the fact that D3 branes wrapping curves in $B$ give rise to BPS strings whose tensions are controlled by the volumes of the curves, and so the conformal point corresponding to tensionless strings exists only if all the curves in $B$ can be shrunk to zero volume simultaneously; this physical requirement translates into a mathematical condition on the intersection pairing of the curves \cite{Grauert}.

The Coulomb branch of KK theories corresponding to such $6d$ SCFTs on a circle of radius $R$ can be described in terms of M-theory compactified on a smooth, resolved version of the same elliptically fibered Calabi-Yau threefold such that the size of the elliptic fiber is\footnote{Here the subscript $B$ stands for the base of the elliptic fibration because $m_B$ is the mass parameter associated to the base.} $m_B$ \cite{Vafa:1996xn, Morrison:1996na, Morrison:1996pp}. The towers of KK particles descend from M2 branes wrapping a holomorphic curve $C$ along with a multiple of the elliptic curve class.

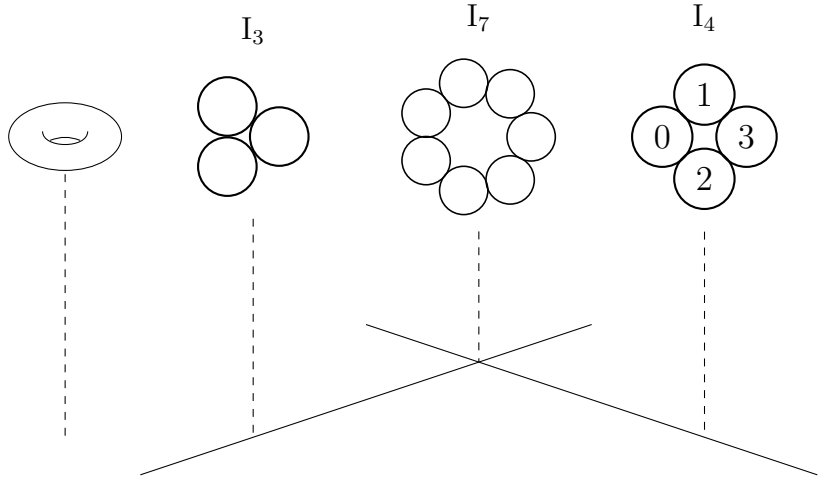
\begin{figure}
\begin{center}
	\begin{tikzpicture}
		\draw[scale=1] (-1,-1) -- (5,1);
		\draw[scale=1] (2,1) -- (8,-1);
		\draw (-2,3.5) ellipse (1.5*.5cm and 1.5*.3cm);
		\draw (-2.3,3.57) to [out=270,in=270] (-1.7,3.57);
		\draw (-2.2,3.45) to [out=20,in=160](-1.8,3.45);
		\node(I4L)[scale=2,label={above:$\text{I}_3$}] at (.5,3.5) {$ 
						\begin{tikzpicture}
							\node[draw,circle](0) at (.23*1,0) {};
							\node[draw,circle](1) at (-.23*.5,.23*.87) {};
							\node[draw,circle](2) at (-.23*.5,-.23*.87) {};
						\end{tikzpicture}
					$};
			\node(I7)[scale=1.5,label={above:$\text{I}_7$}] at (3.5,3.5) {$ 
						\begin{tikzpicture}
							\node[draw,circle,scale=1.1](0) at (.49*1,0) {};
							\node[draw,circle,scale=1.1](1) at (.49*.62,.49*.78) {};
							\node[draw,circle,scale=1.1](2) at (-.49*.22,.49*.97) {};
							\node[draw,circle,scale=1.1](3) at (-.49*.9,.49*.43) {};
							\node[draw,circle,scale=1.1](4) at (-.49*.9,-.49*.43) {};
							\node[draw,circle,scale=1.1](5) at (-.49*.22,-.49*.97) {};
							\node[draw,circle,scale=1.1](6) at (.49*.62,-.49*.78) {};
						\end{tikzpicture}
					$};
					\node[label={above:$\text{I}_4$}](I4R)[scale=2] at (6.5,3.5) {$ 
						\begin{tikzpicture}
							\node[draw,circle,scale=.55](u) at (0,.28*1) {1};
							\node[draw,circle,scale=.55](d) at (0,-.28*1) {2};
							\node[draw,circle,scale=.55](l) at (-.28*1,0) {0};
							\node[draw,circle,scale=.55](r) at (.28*1,0) {3};
						\end{tikzpicture}
					$};
					\draw[dashed](I4R) -- (6.5,-.5);
					\draw[dashed](I7) -- (3.5,.5);
					\draw[dashed](I4L) -- (.5,-.5);
					\draw[dashed](-2,3) -- (-2,-.5);
	\end{tikzpicture}	
\end{center}
\caption{From left to right: A generic point in the base $B$ carries an elliptic fiber. It degenerates to an $\text{I}_3$ fiber at the location of a holomorphic curve in $B$. The fiber degenerates further to an $\text{I}_7$ fiber at a point of intersection with another curve in $B$ carrying an $\text{I}_4$ fiber. The split $\text{I}_4$ fiber is comprised of four rational curves $F_i$ with index $i$ taking values from 0 to 3. $F_i$ moving over the base $\P^1$ gives rise to a Hirzebruch surface $S_i$ with some blow-ups, three out of which come from collision with the $\text{I}_3$ fiber on the adjacent curve adjoining three extra rational curves to $S_i$.}
\label{degenerate}
\end{figure}

Over a generic point on a compact rational curve $C$ in the base $B$, the elliptic fiber degenerates into a collection of rational curves $F_i$ intersecting with each other in some pattern. Each $F_i$ in the collection is then fibered over $C$ giving rise to a ruled surface\footnote{We provide basic mathematical background about Hirzebruch surfaces and more general ruled surfaces in Appendix \ref{ruled}.} $S_i \rightarrow C$ over a smooth base curve $C$. The ruled surfaces $S_i$ then intersect according to the intersection pattern\footnote{We actually find that the intersection pattern of $F_i$ is only part of the full intersection structure of $S_i$. There can be other intersections of $S_i$ which do not change the intersection pattern of $F_i$. For instance, see discussion between equations (\ref{pattern}) and (\ref{patternn}).} of $F_i$ forming the degenerate elliptic fiber, leading to a collection of surfaces $S_C=\cup_i S_i$ in the threefold that are associated to $C$. At the points of intersection of $C$ with other curves, there is a collision of singular elliptic fibers leading to the presence of extra rational curves over the points of intersections. These extra rational curves show up as exceptional curves inside $S_C$. See Figure \ref{degenerate}. If two compact curves $C$ and $D$ in $B$ intersect each other, then the collections $S_C$ and $S_D$ of surfaces over $C$ and $D$ are glued to each other such that components of elliptic fiber in $S_C$ are glued to components of elliptic fiber in $S_D$. 

\subsection{Condensing the data of a Calabi-Yau threefold} \label{package}
We condense the data of a local Calabi-Yau threefold into a graph whose nodes represent compact surfaces $S_i$ which are $p_i$-point blowups of ruled surfaces of degree $n_i$ over a curve of genus $g_i$, denoted 
	\begin{align}
		S_i=\F_{n_i,g_i}^{p_i}.
	\end{align} 
An edge between $S_i$ and $S_j$ represents a transverse intersection between $S_i$ and $S_j$. The locus of such an intersection corresponds to a curve $C_i$ in $S_i$ and some curve $C_j$ in $S_j$. Thus, the intersection can also be viewed as an identification of a curve $C^{ij}_i \subset S_i$ with a curve $C^{ij}_j \subset S_j$. When a pair of curves in two different surfaces are identified with one another in this manner, we say that the surfaces $S_i$ and $S_j$ are \emph{glued} to each other along the curve $C_{ij}$. 

There are some conditions that two curves $C_i$ and $C_j$ have to satisfy for the gluing to be allowed in this setting. Clearly, the genus of $C_i$ must be the same as the genus of $C_j$, i.e.
\be
g(C^{ij}_i)=g(C^{ij}_j)=g
\ee
Moreover, for such a gluing to be consistent with Calabi-Yau condition, it must be the case that
\be
(C^{ij}_i)^2_{S_i}+(C^{ij}_j)^2_{S_j}=2g-2  \label{CY}
\ee

When $g=0$, the nodes represent Hirzebruch surfaces $\mathbb F_{n_i}^{p_i}$. We graphically represent an intersection between two such surfaces as

\begin{align}
\begin{array}{c}
\begin{tikzpicture}[scale=2]
\draw  (-3.9479,4.9946) ellipse (0.8 and 0.5);
\node (v1) at (-3.953,4.9906) {$i^{p_i}_{n_i}$};
\node at (-3.4411,4.795) {\tiny{$C^{ij}_i$}};
\begin{scope}[shift={(-0.1781,0.7187)}]
\draw  (-3.5314,4.4049) rectangle (-3.0269,4.1803);
\node at (-3.256,4.2854) {\tiny{$\left(C^{ij}_i\right)^2$}};
\end{scope}
\begin{scope}[shift={(1.9447,-0.0153)}]
\draw  (-3.9479,4.9946) ellipse (0.8 and 0.5);
\node (v1) at (-3.953,4.9906) {$j^{p_j}_{n_j}$};
\node at (-4.3739,5.2517) {\tiny{$C^{ij}_j$}};
\begin{scope}[shift={(-1.1454,0.7339)}]
\draw  (-3.5314,4.4049) rectangle (-3.0269,4.1803);
\node at (-3.256,4.2854) {\tiny{$\left(C^{ij}_j\right)^2$}};
\end{scope}
\end{scope}
\draw (-3.1553,4.9989) -- (-2.8107,4.9989);
\end{tikzpicture}
\end{array}
\end{align}

which shows the label $i$ of each surface $S_i$, the degree $n_i$ of each surface, the number of blow-ups $p_i$ in each surface, and the self-intersections of the gluing curves for each edge with their names $C^{ij}_i$ and $C^{ij}_j$ adjacent to the corresponding self-intersection numbers.

Sometimes, $2g_i$ blow-ups can be paired up for self-gluing of the surface $S_i$ to itself. The resulting self-glued surface is then a degenerate limit of a ruled surface over a smooth curve of genus $g$. More precisely, if before the self gluing $S_i=\F_{n_i}^{p_i}$, then after the self-gluing it has transformed to $S'_i=\F_{n_i+2g_i,g_i}^{p_i-2g_i}$ which represents a ruled surface of genus $g_i$, degree $n_i+2g_i$ with $p_i-2g_i$ blowups. Keeping this in mind, we denote a general node as 

\begin{align}
\begin{array}{c}
\begin{tikzpicture}[scale=2]
\draw  (-3.9479,4.9946) ellipse (0.5 and 0.5);
\node (v1) at (-3.9528,4.9958) {$i^{p_i}_{n_i,g_i}$};
\end{tikzpicture}
\end{array}
\end{align}

Now, say $S_i$ and $S_j$ are glued along a number $n_{ij}$ of curves which can be represented in $S_i$ as $C^{ij}_{i,\alpha}$ and in $S_j$ as $C^{ij}_{j,\beta}$ which are glued to each other if $\alpha=\beta$, but not otherwise. Also, say that $(C^{ij}_{i,\alpha})^2=(C^{ij}_i)^2$ and $(C^{ij}_{j,\beta})^2=(C^{ij}_j)^2$, i.e. the self-intersections of $C^{ij}_{i,\alpha}$ and $C^{ij}_{j,\beta}$ are independent of $\alpha$ and $\beta$. Then, we denote this configuration by

\begin{align}
\begin{array}{c}
\begin{tikzpicture}[scale=2]
\draw  (-3.9479,4.9946) ellipse (0.8 and 0.5);
\node (v1) at (-4.1698,4.9864) {$i^{p_i}_{n_i,g_i}$};
\node at (-3.4411,4.795) {\tiny{$C^{ij}_{i,\alpha}$}};
\begin{scope}[shift={(-0.1781,0.7187)}]
\draw  (-3.7227,4.4048) rectangle (-3.0269,4.1803);
\node at (-3.3701,4.292) {\tiny{$n_{ij}^2\left(C^{ij}_i\right)^2$}};
\end{scope}
\begin{scope}[shift={(2.5734,-0.0066)}]
\draw  (-3.9479,4.9946) ellipse (0.8 and 0.5);
\node (v1) at (-3.6082,4.996) {$j^{p_j}_{n_j,g_j}$};
\node at (-4.3739,5.2517) {\tiny{$C^{ij}_{j,\alpha}$}};
\begin{scope}[shift={(-1.1454,0.7339)}]
\draw  (-3.5314,4.4049) rectangle (-2.8452,4.1419);
\node at (-3.1687,4.2791) {\tiny{$n_{ij}^2\left(C^{ij}_j\right)^2$}};
\end{scope}
\end{scope}
\draw (-3.1553,4.9989) -- (-2.8107,4.9989);
\begin{scope}[shift={(0.599,0.712)}]
\draw  (-3.4084,4.4041) rectangle (-3.1281,4.1821);
\node at (-3.256,4.2854) {\tiny{$n_{ij}$}};
\end{scope}
\draw (-2.5293,4.9996) -- (-2.1782,4.9979);
\end{tikzpicture}
\end{array}
\end{align}
where we display the self-intersection of the sum of all the gluing curve between the two surfaces. This is the number that enters into the prepotential and not the individual self-intersections. In general, it can happen that the self-intersections of $C^{ij}_{i,\alpha}$ are not independent of $\alpha$. But, this situation will not arise in this paper, so we do not make a notation for this more general case. 

Finally, three distinct surfaces $S_i$, $S_j$ and $S_k$ can intersect in $n_{ijk}$ number of points inside the threefold. We denote this by putting a number in the corresponding face as shown below

\begin{align}
\begin{array}{c}
\begin{tikzpicture}[scale=2]
\draw  (-3.9479,4.9946) ellipse (0.8 and 0.5);
\node (v1) at (-3.953,4.9906) {$i^{p_i}_{n_i,g_i}$};
\node at (-3.455,5.2323) {\tiny{$C^{ij}_{i,\alpha}$}};
\begin{scope}[shift={(-0.1781,0.7187)}]
\draw  (-3.5314,4.4049) rectangle (-3.0269,4.1803);
\node at (-3.256,4.2854) {\tiny{$\left(C^{ij}_i\right)^2$}};
\end{scope}
\begin{scope}[shift={(2.5734,-0.0066)}]
\draw  (-3.9479,4.9946) ellipse (0.8 and 0.5);
\node (v1) at (-3.8633,5.0045) {$j^{p_j}_{n_j,g_j}$};
\node at (-4.3739,5.2517) {\tiny{$C^{ij}_{j,\alpha}$}};
\begin{scope}[shift={(-1.1454,0.7339)}]
\draw  (-3.5314,4.4049) rectangle (-3.0269,4.1803);
\node at (-3.256,4.2854) {\tiny{$\left(C^{ij}_j\right)^2$}};
\end{scope}
\begin{scope}[shift={(-0.6711,0.3586)}]
\draw  (-3.5314,4.4049) rectangle (-3.0269,4.1803);
\node at (-3.256,4.2854) {\tiny{$\left(C^{jk}_j\right)^2$}};
\end{scope}
\node at (-3.5522,4.7028) {\tiny{$C^{jk}_{j,\mu}$}};
\end{scope}
\draw (-3.1553,4.9989) -- (-2.8107,4.9989);
\begin{scope}[shift={(0.599,0.712)}]
\draw  (-3.4084,4.4041) rectangle (-3.1281,4.1821);
\node at (-3.256,4.2854) {\tiny{$n_{ij}$}};
\end{scope}
\begin{scope}[shift={(-0.6779,0.3833)}]
\draw  (-3.5314,4.4049) rectangle (-3.0269,4.1803);
\node at (-3.256,4.2854) {\tiny{$\left(C^{ik}_i\right)^2$}};
\end{scope}
\node at (-4.3553,4.7029) {\tiny{$C^{ik}_{i,a}$}};
\draw (-2.5293,4.9996) -- (-2.1782,4.9979);
\begin{scope}[shift={(1.2515,-1.4206)}]
\draw  (-3.9479,4.9946) ellipse (0.8 and 0.5);
\node (v1) at (-3.9197,4.8234) {$k^{p_k}_{n_k,g_k}$};
\node at (-4.3557,4.9898) {\tiny{$C^{ik}_{k,a}$}};
\begin{scope}[shift={(-1.0278,0.9079)}]
\draw  (-3.5314,4.4049) rectangle (-3.0269,4.1803);
\node at (-3.256,4.2854) {\tiny{$\left(C^{ik}_k\right)^2$}};
\end{scope}
\begin{scope}[shift={(-0.3434,0.9173)}]
\draw  (-3.5314,4.4049) rectangle (-3.0269,4.1803);
\node at (-3.256,4.2854) {\tiny{$\left(C^{jk}_k\right)^2$}};
\end{scope}
\node at (-3.5914,4.9875) {\tiny{$C^{jk}_{k,\mu}$}};
\end{scope}
\begin{scope}[shift={(-0.3135,-0.0571)}]
\draw  (-3.4084,4.4041) rectangle (-3.1281,4.1821);
\node at (-3.256,4.2854) {\tiny{$n_{ik}$}};
\end{scope}
\begin{scope}[shift={(1.4316,-0.0923)}]
\draw  (-3.4084,4.4041) rectangle (-3.1281,4.1821);
\node at (-3.256,4.2854) {\tiny{$n_{jk}$}};
\end{scope}
\begin{scope}[shift={(0.5873,0.1946)}]
\draw  (-3.4084,4.4041) rectangle (-3.1043,4.1814);
\node at (-3.256,4.2854) {\tiny{$n_{ijk}$}};
\end{scope}
\draw (-3.7472,4.5081) -- (-3.6343,4.3458);
\draw (-3.4908,4.1247) -- (-3.3379,3.873) (-1.5246,4.4986) -- (-1.7339,4.3152) (-1.9762,4.12) -- (-2.1667,3.9483);
\end{tikzpicture}
\end{array}
\end{align}

The triple intersection $S_i\cdot S_j\cdot S_k=n_{ijk}$ is invariant under permutation of $i$, $j$, $k$, and can be computed inside any one of the surfaces by using the identity
\be
S_i\cdot S_j\cdot S_k=\left[\left(\sum_\alpha C^{ij}_{j,\alpha}\right)\cdot\left(\sum_\mu C^{jk}_{j,\mu}\right)\right]_{S_j}
\ee

We also have a special non-compact surface for the KK theories, namely the base $B$ of the elliptic fibration. For each curve $C$ in $B$, $B$ joins $S_C$ only along a single node, which we will refer to as the affine node and label it as $S_0$. There is a curve $C_0$ in $S_0$ along which $S_0$ is glued to $B$. We display this curve separately in our figures but we do not associate an edge to this curve, as is evident in the following graph:

\begin{align}
\begin{array}{c}
\begin{tikzpicture}[scale=2]
\draw  (-3.9479,4.9946) ellipse (0.8 and 0.5);
\node (v1) at (-3.953,4.9906) {$0^{p_0}_{n_0,g_0}$};
\node at (-3.4411,4.795) {\tiny{$C^{0j}_{0,\alpha}$}};
\begin{scope}[shift={(-0.1781,0.7187)}]
\draw  (-3.5314,4.4049) rectangle (-3.0269,4.1803);
\node at (-3.256,4.2854) {\tiny{$\left(C^{0j}_0\right)^2$}};
\end{scope}
\begin{scope}[shift={(2.5734,-0.0066)}]
\draw  (-3.9479,4.9946) ellipse (0.8 and 0.5);
\node (v1) at (-3.8633,5.0045) {$j^{p_j}_{n_j,g_j}$};
\node at (-4.3739,5.2517) {\tiny{$C^{0j}_{j,\alpha}$}};
\begin{scope}[shift={(-1.1454,0.7339)}]
\draw  (-3.5314,4.4049) rectangle (-3.0269,4.1803);
\node at (-3.256,4.2854) {\tiny{$\left(C^{0j}_j\right)^2$}};
\end{scope}
\end{scope}
\draw (-3.1553,4.9989) -- (-2.8107,4.9989);
\begin{scope}[shift={(0.599,0.712)}]
\draw  (-3.4084,4.4041) rectangle (-3.1281,4.1821);
\node at (-3.256,4.2854) {\tiny{$n_{0j}$}};
\end{scope}
\draw (-2.5293,4.9996) -- (-2.1782,4.9979);
\begin{scope}[shift={(-1.2107,0.7132)}]
\draw  (-3.4334,4.4036) rectangle (-3.091,4.1583);
\node at (-3.256,4.2854) {\tiny{$C_0^2$}};
\end{scope}
\node at (-4.4699,5.2105) {\tiny{$C_{0}$}};
\end{tikzpicture}
\end{array}
\end{align}

\subsection{Singular elliptically fibered Calabi-Yau threefolds} \label{w_models}

We now discuss an explicit construction of a singular elliptically fibered Calabi-Yau threefold in detail, and the methods by which one can extract the data described in the previous subsection from such a construction. Let $X_0 \rightarrow B$ be an elliptically fibered Calabi-Yau threefold over a complex surface $B$. This elliptic fibration has an explicit realization as a hypersurface 
	\begin{align}
		 W_0=y^2z+a_1 x yz+a_3 yz^2-(x^3 +a_2 x^2z+a_4 xz^2+a_6z^3 ) =0
	\end{align} 
of a rank 2 projective bundle $Y_0 \rightarrow B$ whose $\mathbb P^2$ fibers are parametrized by homogeneous coordinates $[x:y:z]$. Here we are using Tate form of the Weierstrass equation. The parameters $a_n$ are sections of $\mathcal K_B^{-n}$, where $\mathcal K_B^{-1} \rightarrow B$ is the anti-canonical bundle of the base $B$.
For a rank one 6d SCFT, $B$ can be described as the total space of a local $\mathbb P^1$ with self intersection $-k$ in $B$; we use the symbol $C$ to denote this rational curve. Given an explicit realization of $C$ as the zero locus $e_0=0$, the type of Kodaira singular fiber is determined by specifying the order of vanishing $q_n$ of the parameters $a_n$ along $C$ which can be read from Table 2 of \cite{Katz:2011qp}; this procedure leads to an equation of the following form:
	\begin{align}
		W_0=y^2z+a_{1,q_1} e_0^{q_1} x yz+a_{3,q_3} e_0^{q_3} yz^2-(x^3 +a_{2,q_2} e_0^{q_2} x^2z+a_{4,q_4} e_0^{q_4} xz^2+a_{6,q_6} e_0^{q_6} z^3 ) =0.
	\end{align}
The singular locus of $X_0$ is 
	\begin{align}
	\label{eqn:smooth}
		W_0 = \partial_i W_0 =0,
	\end{align}
where $\partial_i$ denote partial derivatives with respect to the complex coordinates $x,y,e_0$. In order to obtain a smooth elliptic fibration, we identify a sequence of blowups which do not change the canonical class of the threefold, 
	\begin{align}
		X_r \overset{f_r}{\rightarrow} X_{r-1} \overset{f_{r-1}}{\rightarrow} \cdots \overset{f_2}{\rightarrow} X_{1} \overset{f_1}{\rightarrow} X_{0},
	\end{align}
such that for some choice of positive $r <\infty $ the elliptic fibration $X_r \rightarrow B$ is smooth. 

We now describe some general facts about blowups in more detail. Suppose the equation $W(y_i) =0$ describes a singular projective variety $X \subset Y$ realized as a hypersurface of the ambient projective space $Y$ with homogeneous coordinates $y_i$. We describe blowups in more detail. Let $W(y_i) =0$ denote a singular projective variety $X \subset Y$ realized as a hypersurface of the ambient projective space $Y$  with homogeneous coordinates $y_i$. A blowup can be described in terms of its center $(g_1,g_2,\dots)$, where $g_i(y_j)$ is a homogeneous polynomial in $y_j$, and a local parameter $e$ whose zero locus $e=0$ is the exceptional divisor $E$ of the blowup. In practice, we use adopt the following succinct notation 
	\begin{align}
		(g_1,g_2,\dots | e),
	\end{align}
which means we make the substitution
	\begin{align}
		(g_1,g_2,\dots) \rightarrow (e g_1, e g_2, \dots)	
	\end{align}
and introduce a new ambient projective space $Y'\rightarrow Y$ in which the locus $g_1=g_2=\cdots =0$ of the original projective bundle $Y$ has been replaced by a projective space $[g_1:g_2:\cdots]$ located at $e=0$ in $Y'$. This procedure defines a map of hypersurfaces $X' \rightarrow X$ where $X' \subset Y'$ is said to be the blowup of $X$ along the center $g_1=g_2=\cdots =W=0$ (note that center must intersect $X$.) Checking smoothness of $X'$ is equivalent to checking that the equations (\ref{eqn:smooth}) have no solutions.

Note that the singular elliptic fibers $F$ are the elliptic fibers located over the rational curve $C \subset B$ described by $e_0 =0$. If one restricts to blowing up sub-loci of the hyperplane $e_0=0$ in $Y_0$, the sequence of blowups $f: X_r \rightarrow X_0$ resolving $X_0$ can be viewed as blowups of singular points on $F$. The elliptic fibers $F$ of the resolved threefold can therefore be viewed as a collection of smooth rational curves $F_i$ with multiplicities $m_i$:
	\begin{align}
	\label{eqn:rat}
		F = \sum_{i=0}^r m_i F_i.
	\end{align}
The components $F_i$ are the fibral divisors of the elliptic fiber $F$; when the Kodaira singular fiber type is associated to a Lie algebra $\mathfrak{g}$, the $F_i$ intersect in the pattern of the affine Dynkin diagram associated to $\mathfrak{g}$. \\

As the irreducible components $F_i$ move over $C$, they sweep out complex surfaces in the threefold $X_r$. Thus the fibral divisors $F_i$ define a natural basis of divisors $S_i$ in $X_r$ which have the structure of $\mathbb P^1$ bundles over $C$. Once we have explicitly computed a resolution $X_r \rightarrow X_0$ of the singular threefold $X_0$, we can describe $X_r$ as local neighborhood of a collection of transversely intersecting surfaces $\cup S_i$ by computing the triple intersection numbers $k_{ijk} = (S_i \cdot S_j \cdot S_k)_{X_r}$ and the degrees $n_i$ of $S_i$. We explain how to perform these computations in the following subsections and illustrate them with detailed examples in Appendix \ref{sample}.

\subsubsection*{Computing degrees}
\label{deg}

First, we explain how to compute the degrees $n_i$ of the divisors $S_i$. For now, let us assume that the Kodaira fiber type is split. Using the fact that the class of the elliptic fiber in $X_r$ can be decomposed into a sum over rational curves as in (\ref{eqn:rat}), we view each divisor $S_i$ as a $\mathbb P^1$ bundle fibered over $C$, namely 
	\begin{align}
		S_i = \mathbb P_{C}[ \mathcal O \oplus \mathcal L_i ],
	\end{align}
 where $c_1(\mathcal L_i) = a_i K_{B} + b_i C$ for some integers $a_i,b_i$. The degree of the line bundle $S_i$ can then be computed as
	\begin{align}
		n_i =  (C \cdot (a_i K_B + b_i C))_B = a_i (k-2) - b_i k.
	\end{align}
In order to determine the number of blowups, we simply use the identity
	\begin{align}
		k_{iii} =(S_i^3)_{X_r} = (K_{S_i}^2)_{S_i} = 8 - p_i
	\end{align}	
to read off $p_i$. 

When the Kodaira fiber type is non-split, a given component $F_i$ of the resolved elliptic fiber may be geometrically reducible, consisting of several components 
	\begin{align}
		F_i = \sum_{j} F_{i,j}
	\end{align}
and moreover there can be non-trivial monodromies permuting the irreducible components $F_{i,j=1,\dots, s_i}$ of the resolved elliptic fiber. For notational clarity, assume in the forthcoming discussion that the elliptic fibration is defined with respect to a curve $C' \subset B$ (as opposed to the usual symbol $C$). In the case of non-split Kodaira fibers over $C'$, the relation between the (identical) line bundles $S'_i = \mathbb P[\mathcal O \oplus \mathcal L_i]_{j=1,\dots,s_i} \rightarrow C'$ and the divisor $S_i$ may be such that $S_i \rightarrow S_i'$ is a ramified $s_i$-cover, ramification locus may consist of a collection of fibers.  

Suppose the ruled surface $\mathbb P[\mathcal O \oplus \mathcal L_i]_j \rightarrow C'$ has degree $n_i'$. Then it is possible to view $S_i \rightarrow C'$ as a projective bundle over a different curve $C$, such that $C \rightarrow C'$ is a ramified $s_i$-cover (see \cite{Esole:2017rgz} for a more precise discussion of this point in the context of the Stein factorization and its relevance to the non-split $\text{IV}^{*}$ model). The genus of the curve $C$ in general depends on the genus of the curve $C'$ along with the ramification locus of the $s_i$-cover $S_i \rightarrow S_i'$. Suppose that the ramification locus of the $s_i$-cover $S_i \rightarrow S_i'$ consists of $2b$ fibers. Then, we have:
	\begin{align}
	\label{eqn:newgen}
		g_i(C) = \frac{s_i}{2} ( 2g_i(C') +b -2 +\frac{2}{s_i}).	
	\end{align}
On the other hand, the degree $n_i$ of the surface $S_i$ depends on the degree $n_i' = -(C')^2_{S_i'}$ of the surface $S'_i$ through the relation
	\begin{align}
		(C^2)_{S_i} = s_i (C'^2)_{S_i'} = -s_in'_i,
	\end{align} 
which is a consequence of the action of the pushforward map associated to the $s_i$ cover $S_i \rightarrow S_i'$ on the intersection product $(C^2)_{S_i}$. See the non-split $\text{I}_0^{*}$ model described in Appendix \ref{2} for an example of this structure. Once we know $g_i(C)$, we can determine the number of blowups $p_i$ using
\be
k_{iii} = K_{S_i}^2 = 8-8g_i(C) -p_i
\ee

\subsubsection*{Computing triple intersection numbers}
\label{trip}

We compute the triple intersection numbers of the divisors of $X_r \rightarrow B$ following the strategy outlined in \cite{Esole:2017kyr}. Suppose that $W_r=0$ describes the resolved threefold $X_r$ as a hypersurface of the ambient projective bundle $Y_r$, and let $\pi \circ f : X_r \rightarrow B$ denote the projection of $X_r$ to the base $B$. The triple intersection numbers of $X_r$ can be considerably simplified by expressing them in terms of geometric data associated to $B$. This simplification can be accomplished by computing the pushforward of the intersection product $(S_i \cdot S_j \cdot S_k)_{X_r}$ to the base of the elliptic fibration:
	\begin{align}
		(S_i \cdot S_j \cdot S_k)_{X_r} = \pi_* \circ f_* (S_i \cdot S_j \cdot S_k)_{X_r} = \pi_* \circ f_* (S_i \cdot S_j \cdot S_k \cdot [ W_r] )_{Y_r}. 
	\end{align}  
 To perform this computation, we need to expand the divisor classes $S_i, [W_r]$, $[g_{i,1}]$, $[g_{i,2}]$, $\dots$ (where $[g_{i,j}]$ are the classes of the divisors $g_{i,j}=0$, associated to the generators of the centers of the blowups $f_i : X_i \rightarrow X_{i-1}$) in the basis $f^* H, f^* \circ \pi^* K_B, E_{i=0,\cdots r}$, where $H= c_1(\mathcal O_{Y_0} (1))$ is the divisor class of a hyperplane in the ambient space $\mathbb P^2$ of the fibers,  $ E_{i=1,\dots,r}$ are the classes of the exceptional divisors of the blowups, and $E_0 = f^* \circ \pi^* C$ is the pullback of the class $C \subset B$ to $X_r$. Once we have expressed $S_i \cdot S_j \cdot S_k \cdot [W_r]$ in terms of this basis, we use the pushforward formula of \cite{Esole:2017kyr} to express $\pi_* \circ f_* (S_i \cdot S_j \cdot S_k \cdot [ W_r] )_{Y_r}$ as an intersection product of the classes $K_B, C$ in $B$. 

\section{KK theories for rank one  \texorpdfstring{$6d$}{} SCFTs}\label{results}
In this section, we collect the Calabi-Yau threefolds associated to rank one $6d$ SCFTs that we obtain via computations using the above mentioned tools. Some sample computations are illustrated in Appendix \ref{sample}. Unless otherwise stated, we will only work with total transforms and avoid proper transforms. This is the reason why some curves can appear with negative sign in what follows. See Appendix \ref{transforms} for a review of this terminology.

Before we start, let us provide a brief review of rank one $6d$ SCFTs constructed in the unfrozen phase of F-theory. The F-theory base for such theories involves a single compact rational curve $C$ in the base of a negative self-intersection $-k$ where $k$ can take values $1\le k\le 8$ or $k=12$. The elliptic fiber degenerates over $C$ to some fiber of Kodaira type which can potentially have a monodromy as one encircles loops in $C$. As is well-known, each Kodaira fiber type along with the specification of monodromy leads to a particular simple gauge algebra $\fg$ in the resulting $6d$ theory. The self-intersection $-k$ of the curve almost always uniquely fixes the associated matter content \cite{Grassi:2011hq}. Sometimes we can tune the corresponding Weierstrass model to change the matter content. In the context of this paper, this is only possible in the situation when $k=1$ and the Kodaira fiber type is $\text{I}_6$ without a monodromy. Generically this model gives rise to $SU(6)$ with a hyper in two-index antisymmetric plus 14 hypers in fundamental. A tuned version of the model gives rise to $SU(6)$ with a half-hyper in three-index antisymmetric plus 15 hypers in fundamental.

Below, each subsection is devoted to a particular class of rank one $6d$ SCFTs. In the starting of each subsection, we specify both the F-theory construction, and the gauge algebra and matter content on the tensor branch of the resulting rank one $6d$ SCFT. We then proceed to provide our choice of resolution and a graphical description of the resulting geometry.

\subsection{ \texorpdfstring{$SU(n)$}{} with  \texorpdfstring{$2n$}{} fundamental hypers} \label{A}
The F-theory construction involves a split $\text{I}_n$ fiber over a $-2$ curve in the base. Note that the zero mass parameter\footnote{`Zero mass parameter resolutions' are resolutions that coincide with the Coulomb branch phases of the corresponding 5d $\mathcal N=1$ gauge theory with the same gauge symmetry and hypermultiplet content as in 6d, in the special case that all mass parameters are set to zero.} resolutions and corresponding triple intersection numbers for the cases $n \leq 5$ were studied in great detail in \cite{Esole:2014bka,Esole:2014hya,Esole:2011sm}; furthermore, a particular set of resolutions and accompanying triple intersection numbers were computed for all $n$ in \cite{Esole:2015xfa}. For even $n=2m > 0$, the Weierstrass model is defined by the following orders of vanishing:
	\begin{align}
		a_1 = a_1,~~ a_2 = a_{2,1} s,~~ a_{3} = 0,~~ a_4 = a_{4,m} e_0^m,~~ a_6= a_{6,2m} e_0^{2m}.
	\end{align}
We consider the resolution \cite{Esole:2015xfa} defined by the following sequence of blowup centers:
	\begin{align}
		(x,y,e_0|e_1)	,~~ (y,e_1|e_2),~~ (x,e_2|e_3) ,~~ \cdots ,~~ (x,e_{2m-2} |e_{2m-1})
	\end{align}
and find that the associated Calabi-Yau threefold is

\begin{align}
\begin{array}{c}
\begin{tikzpicture}[scale=1.5]
\draw  (-3.5,4) ellipse (0.5 and 0.5);
\node (v1) at (-3.5,4) {$0_0$};
\draw  (-2.5,5) ellipse (0.5 and 0.5);
\node (v2) at (-2.5,5) {$1_2$};
\draw  (-1,5) ellipse (0.5 and 0.5);
\node (v4) at (-1,5) {$3_4$};
\draw  (-2.5,3) ellipse (0.5 and 0.5);
\node (v3) at (-2.4683,3.0111) {$2_2$};
\draw  (-1,3) ellipse (0.5 and 0.5);
\node (v5) at (-1,3) {$4_4$};
\draw  (3.4586,4.9793) ellipse (1.4 and 0.5);
\node at (3.4875,4.9958) {$(n-3)_{n-2}$};
\draw  (5.2535,4.018) ellipse (0.8 and 0.6);
\node at (5.2838,4.0004) {$(n-1)^{2n}_{n}$};
\draw (-3.2,4.4) -- (-2.9,4.7);
\draw (-3.2,3.6) -- (-2.9,3.3);
\draw (-2,5) -- (-1.5,5);
\draw (-2,3) -- (-1.5,3);
\draw (4.3555,4.6095) -- (4.6555,4.4095);
\draw (4.8148,3.5151) -- (4.5472,3.2832);
\draw (-0.1,5) -- (0.1,5) (0.4,5) -- (0.6,5);
\draw (0.9,5) -- (1.1,5) (1.4,5) -- (1.6,5);
\begin{scope}[shift={(-0.0558,-1.9909)}]
\draw (0,5) -- (0.2,5) (0.5,5) -- (0.7,5);
\draw (1,5) -- (1.2,5) (1.5,5) -- (1.7,5);
\end{scope}
\draw  (-3.4053,4.3921) rectangle (-3.2204,4.2073);
\node at (-3.3057,4.3031) {\tiny{0}};
\begin{scope}[shift={(0.5477,0.5203)}]
\draw  (-3.4113,4.3642) rectangle (-3.1700,4.1794);
\node at (-3.2917,4.2752) {\tiny{-2}};
\end{scope}
\begin{scope}[shift={(1.1159,0.7257)}]
\draw  (-3.4113,4.3642) rectangle (-3.17,4.1794);
\node at (-3.2917,4.2752) {\tiny{2}};
\end{scope}
\begin{scope}[shift={(1.9717,0.7394)}]
\draw  (-3.4113,4.3642) rectangle (-3.17,4.1794);
\node at (-3.2917,4.2752) {\tiny{-4}};
\end{scope}
\begin{scope}[shift={(2.6153,0.7188)}]
\draw  (-3.4113,4.3642) rectangle (-3.17,4.1794);
\node at (-3.2917,4.2752) {\tiny{4}};
\end{scope}
\begin{scope}[shift={(-0.0137,-0.5476)}]
\draw  (-3.3913,4.3642) rectangle (-3.2064,4.1794);
\node at (-3.2917,4.2752) {\tiny{0}};
\end{scope}
\begin{scope}[shift={(-0.5545,-0.2807)}]
\draw  (-3.3913,4.3642) rectangle (-3.2064,4.1794);
\node at (-3.2917,4.2752) {\tiny{0}};
\end{scope}
\begin{scope}[shift={(0.5067,-1.1502)}]
\draw  (-3.4113,4.3642) rectangle (-3.18,4.1794);
\node at (-3.2917,4.2752) {\tiny{-2}};
\end{scope}
\begin{scope}[shift={(1.1295,-1.2734)}]
\draw  (-3.4113,4.3642) rectangle (-3.17,4.1794);
\node at (-3.2917,4.2752) {\tiny{2}};
\end{scope}
\begin{scope}[shift={(1.958,-1.2597)}]
\draw  (-3.4113,4.3642) rectangle (-3.17,4.1794);
\node at (-3.2917,4.2752) {\tiny{-4}};
\end{scope}
\begin{scope}[shift={(2.6153,-1.2529)}]
\draw  (-3.4113,4.3642) rectangle (-3.17,4.1794);
\node at (-3.2917,4.2752) {\tiny{4}};
\end{scope}
\begin{scope}[shift={(7.2486,0.4223)}]
\draw  (-3.6509,4.3711) rectangle (-2.9441,4.2);
\node at (-3.2917,4.2752) {\tiny{$n-2$}};
\end{scope}
\begin{scope}[shift={(8.3168,-0.0089)}]
\draw  (-3.6509,4.3711) rectangle (-3.2477,4.2136);
\node at (-3.4547,4.3027) {\tiny{$-n$}};
\end{scope}
\begin{scope}[shift={(8.5247,-0.6573)}]
\draw  (-3.6509,4.3711) rectangle (-3.2477,4.2136);
\node at (-3.4547,4.3027) {\tiny{$-n$}};
\end{scope}
\begin{scope}[shift={(5.7562,0.6483)}]
\draw  (-3.6509,4.3711) rectangle (-2.9441,4.2);
\node at (-3.2917,4.2752) {\tiny{$2-n$}};
\end{scope}
\begin{scope}[shift={(0.4434,-2.0286)}]
\draw  (3.0151,5.0088) ellipse (1.4 and 0.5);
\node at (3.1809,4.8747) {$(n-2)_{n-2}$};
\begin{scope}[shift={(7.0448,0.9242)}]
\draw  (-3.6509,4.3711) rectangle (-2.9441,4.2);
\node at (-3.2917,4.2752) {\tiny{$n-2$}};
\end{scope}
\begin{scope}[shift={(5.3469,0.7052)}]
\draw  (-3.6509,4.3711) rectangle (-2.9441,4.2);
\node at (-3.2917,4.2752) {\tiny{$2-n$}};
\end{scope}
\end{scope}
\node at (-3.1608,4.2855) {\tiny{$h$}};
\node at (-2.1625,5.1528) {\tiny{$h$}};
\node at (-3.1558,3.7385) {\tiny{$h$}};
\node at (-2.14,3.1548) {\tiny{$h$}};
\node at (-1.3095,5.1733) {\tiny{$e$}};
\node at (4.399,4.7223) {\tiny{$h$}};
\node at (4.6205,3.1718) {\tiny{$h$}};
\node at (5.1638,4.4149) {\tiny{$h-\sum x_i$}};
\node at (4.7932,3.635) {\tiny{$e$}};
\node at (-1.3312,3.1545) {\tiny{$e$}};
\node at (-2.7714,3.2823) {\tiny{$e$}};
\node at (-2.7148,4.6425) {\tiny{$e$}};
\node at (-0.6595,3.1884) {\tiny{$h$}};
\node at (-0.6684,5.1438) {\tiny{$h$}};
\node at (2.4767,5.0804) {\tiny{$e$}};
\node at (2.4878,3.1019) {\tiny{$e$}};
\node at (-3.8578,4.1551) {\tiny{$e$}};
\end{tikzpicture}
\end{array}
\end{align}
where $\sum x_i$ denotes the sum over (the total transforms of) all the exceptional curves created due to the $2n$ blowups. Since $h-\sum x_i$ is a single curve, all the blow-ups are restricted to happen on the proper transform of the $h$ curve, which means that the blow-ups are not completely generic. We will adopt this notation in what follows unless otherwise stated. 

A degenerate case in this class of models is $n=0$, in which case the F-theory configuration is an $\text{I}_0$ fiber over a $-2$ curve. This configuration preserves $16$ supercharges and hence the surface describing the local threefold splits into a product $S=\P^1\times T^2$. Since $S$ is a product the self-intersections of $\P^1$ and $T^2$ inside $S$ are both zero and their mutual intersection is one.

For odd $n=2m+1 >1$, the Weierstrass model is defined by 
	\begin{align}
		a_1=a_1,~~ a_2=a_{2,1} e_0,~~ a_3=a_{3,m} e_0^r,~~ a_4 = a_{4,m+1} e_0^{m+1},~~ a_6 = a_{6,2m+1} e_0^{2m+1}.
	\end{align}
We consider the resolution \cite{Esole:2015xfa} defined by the following sequence of blowup centers:
	\begin{align}
		(x,y,e_0|e_1) ,~~ (y,e_1|e_2),~~ (x,e_2|e_3),~~ \cdots,~~  (y ,e_{2m-1}|e_{2m} )
	\end{align}
and we find

\begin{align}
\begin{array}{c}
\begin{tikzpicture}[scale=1.5]
\draw  (-3.5,4) ellipse (0.5 and 0.5);
\node (v1) at (-3.5,4) {$0_0$};
\draw  (-2.5,5) ellipse (0.5 and 0.5);
\node (v2) at (-2.5,5) {$1_2$};
\draw  (-1,5) ellipse (0.5 and 0.5);
\node (v4) at (-1,5) {$3_4$};
\draw  (-2.5,3) ellipse (0.5 and 0.5);
\node (v3) at (-2.4683,3.0111) {$2_2$};
\draw  (-1,3) ellipse (0.5 and 0.5);
\node (v5) at (-1,3) {$4_4$};
\draw  (3.4586,4.9793) ellipse (1.4 and 0.5);
\node at (3.5681,5.0448) {$(n-2)^{2n}_{n-1}$};
\draw (-3.2,4.4) -- (-2.9,4.7);
\draw (-3.2,3.6) -- (-2.9,3.3);
\draw (-2,5) -- (-1.5,5);
\draw (-2,3) -- (-1.5,3);
\draw (-0.1,5) -- (0.1,5) (0.4,5) -- (0.6,5);
\draw (0.9,5) -- (1.1,5) (1.4,5) -- (1.6,5);
\begin{scope}[shift={(-0.0558,-1.9909)}]
\draw (0,5) -- (0.2,5) (0.5,5) -- (0.7,5);
\draw (1,5) -- (1.2,5) (1.5,5) -- (1.7,5);
\end{scope}
\draw  (-3.4053,4.3921) rectangle (-3.2204,4.2073);
\node at (-3.3057,4.3031) {\tiny{0}};
\begin{scope}[shift={(0.5477,0.5203)}]
\draw  (-3.4113,4.3642) rectangle (-3.1700,4.1794);
\node at (-3.2917,4.2752) {\tiny{-2}};
\end{scope}
\begin{scope}[shift={(1.1159,0.7257)}]
\draw  (-3.4113,4.3642) rectangle (-3.17,4.1794);
\node at (-3.2917,4.2752) {\tiny{2}};
\end{scope}
\begin{scope}[shift={(1.9717,0.7394)}]
\draw  (-3.4113,4.3642) rectangle (-3.17,4.1794);
\node at (-3.2917,4.2752) {\tiny{-4}};
\end{scope}
\begin{scope}[shift={(2.6153,0.7188)}]
\draw  (-3.4113,4.3642) rectangle (-3.17,4.1794);
\node at (-3.2917,4.2752) {\tiny{4}};
\end{scope}
\begin{scope}[shift={(-0.0137,-0.5476)}]
\draw  (-3.3913,4.3642) rectangle (-3.2064,4.1794);
\node at (-3.2917,4.2752) {\tiny{0}};
\end{scope}
\begin{scope}[shift={(-0.5545,-0.2807)}]
\draw  (-3.3913,4.3642) rectangle (-3.2064,4.1794);
\node at (-3.2917,4.2752) {\tiny{0}};
\end{scope}
\begin{scope}[shift={(0.5067,-1.1502)}]
\draw  (-3.4113,4.3642) rectangle (-3.18,4.1794);
\node at (-3.2917,4.2752) {\tiny{-2}};
\end{scope}
\begin{scope}[shift={(1.1295,-1.2734)}]
\draw  (-3.4113,4.3642) rectangle (-3.17,4.1794);
\node at (-3.2917,4.2752) {\tiny{2}};
\end{scope}
\begin{scope}[shift={(1.958,-1.2597)}]
\draw  (-3.4113,4.3642) rectangle (-3.17,4.1794);
\node at (-3.2917,4.2752) {\tiny{-4}};
\end{scope}
\begin{scope}[shift={(2.6153,-1.2529)}]
\draw  (-3.4113,4.3642) rectangle (-3.17,4.1794);
\node at (-3.2917,4.2752) {\tiny{4}};
\end{scope}
\begin{scope}[shift={(6.7653,0.3348)}]
\draw  (-3.6509,4.3711) rectangle (-2.9441,4.2);
\node at (-3.2917,4.2752) {\tiny{$-n-1$}};
\end{scope}
\begin{scope}[shift={(5.7562,0.6483)}]
\draw  (-3.6509,4.3711) rectangle (-2.9441,4.2);
\node at (-3.2917,4.2752) {\tiny{$1-n$}};
\end{scope}
\begin{scope}[shift={(0.4434,-2.0286)}]
\draw  (3.0151,5.0088) ellipse (1.4 and 0.5);
\node at (3.1809,4.8747) {$(n-1)_{n-1}$};
\begin{scope}[shift={(6.3233,1.0678)}]
\draw  (-3.6509,4.3711) rectangle (-2.9441,4.2);
\node at (-3.2917,4.2752) {\tiny{$n-1$}};
\end{scope}
\begin{scope}[shift={(5.3469,0.7052)}]
\draw  (-3.6509,4.3711) rectangle (-2.9441,4.2);
\node at (-3.2917,4.2752) {\tiny{$1-n$}};
\end{scope}
\end{scope}
\node at (-3.1608,4.2855) {\tiny{$h$}};
\node at (-2.1625,5.1528) {\tiny{$h$}};
\node at (-3.1558,3.7385) {\tiny{$h$}};
\node at (-2.14,3.1548) {\tiny{$h$}};
\node at (-1.3095,5.1733) {\tiny{$e$}};
\node at (3.5059,4.7853) {\tiny{$h-\sum x_i$}};
\node at (3.9039,3.3246) {\tiny{$h$}};
\node at (-1.3312,3.1545) {\tiny{$e$}};
\node at (-2.7714,3.2823) {\tiny{$e$}};
\node at (-2.7003,4.6214) {\tiny{$e$}};
\node at (-0.6595,3.1884) {\tiny{$h$}};
\node at (-0.6684,5.1438) {\tiny{$h$}};
\node at (2.4767,5.0804) {\tiny{$e$}};
\node at (2.4878,3.1019) {\tiny{$e$}};
\draw (3.4681,4.4732) -- (3.4606,3.484);
\node at (-3.8461,4.1604) {\tiny{$e$}};
\end{tikzpicture}
\end{array}
\end{align}

A degenerate case in this class of models is $n=1$, in which case the F-theory configuration is an $\text{I}_1$ fiber over a $-2$ curve. Even though this configuration preserves 8 supercharges, it is known that the corresponding $6d$ theory is $A_1$ $(2,0)$ theory which has 16 supercharges. By taking a limit of geometries described above,  the geometry for this case can be predicted to be

\begin{align}
\begin{array}{c}
\begin{tikzpicture}[scale=2]
\draw  (-3.5,4) ellipse (0.5 and 0.5);
\node (v1) at (-3.5,4) {$0^2_0$};
\draw  (-3.2902,4.2051) rectangle (-3.1053,4.0203);
\node at (-3.1906,4.1161) {\tiny{-2}};
\begin{scope}[shift={(-0.0137,-0.5476)}]
\draw  (-3.3913,4.3642) rectangle (-3.2064,4.1794);
\node at (-3.2917,4.2752) {\tiny{0}};
\end{scope}
\begin{scope}[shift={(-0.5545,-0.2807)}]
\draw  (-3.3913,4.3642) rectangle (-3.2064,4.1794);
\node at (-3.2917,4.2752) {\tiny{0}};
\end{scope}
\node at (-3.4709,4.3287) {\tiny{$h-\sum x_i$}};
\node at (-3.1558,3.7385) {\tiny{$h$}};
\node at (-3.8461,4.1604) {\tiny{$e$}};
\draw (-3.1935,3.5857) .. controls (-1.2581,2.9132) and (-1.2077,4.4972) .. (-3.0281,4.1613);
\end{tikzpicture}
\end{array}
\end{align}
involving a self-gluing. Thus, we have found two geometries for the KK theory corresponding to $A_1$ $(2,0)$ theory. We will see an interesting difference between the two geometries in Section \ref{(2,0)}.

\subsection{ \texorpdfstring{$SO(n)$}{} with  \texorpdfstring{$n-8$}{} hypers in fundamental}\label{D}
For even $n=2r$, the F-theory setup involves a split $\text{I}^*_{r-4}$ fiber, with $r>4$, over a $-4$ curve in the base. The Weierstrass model engineering $SO(n=2r)$ for $r$ odd is defined by
	\begin{align}
		a_1=a_{1,1} e_0 ,~~ a_2 = a_{2,1} e_0,~~ a_{3} = a_{3, \frac{r-1}{2} } e_0^{\frac{r-1}{2}},~~ a_4=a_{4,\frac{r+1}{2}} e_0^{\frac{r+1}{2}},~~ a_6 = a_{6,r} e_0^{r}.
	\end{align}	
We consider the resolution
	\begin{align}
		(x,y,e_0|e_1),~~ (y,e_1|e_2),~~ \cdots ,~~ (x,e_{r-3}|e_{r-2}),~~ (y,e_{r-2}|e_{r-1} ),~~ (e_{r-3},e_{r-2}|e_{r}).
	\end{align}
The Weierstrass model for $r$ even and is defined by 
	\begin{align}
		a_1 = a_{1,1} e_0,~~ a_2=a_{2,1} e_0,~~ a_3=a_{3, \frac{r}{2} } e_0^{\frac{r}{2}},~~ a_4 = a_{4,\frac{r}{2}} e_0^{\frac{r}{2}},~~ a_6 = a_{6,r-1} e_0^{r-1}.
	\end{align}
This model requires an additional split condition, namely
	\begin{align}
		\left. \frac{a_4^2 - 4 a_2 a_6}{e_0^{r} } \right|_{e_0=0}
	\end{align}
must be a perfect square. In practice we satisfy this condition by imposing 
	\begin{align}
		a_{6,r-1} = 0.
	\end{align}	
We consider the resolution
	\begin{align}
		(x,y,e_0|e_1),~~(y,e_1|e_2),~~ \cdots ,~~ (y,e_{r-3}|e_{r-2}),~~(x,e_{r-2}|e_{r-1}),~~ (e_{r-3},e_{r-2}|e_{r}) .
	\end{align}
In combination, we find that the associated collection of surfaces is

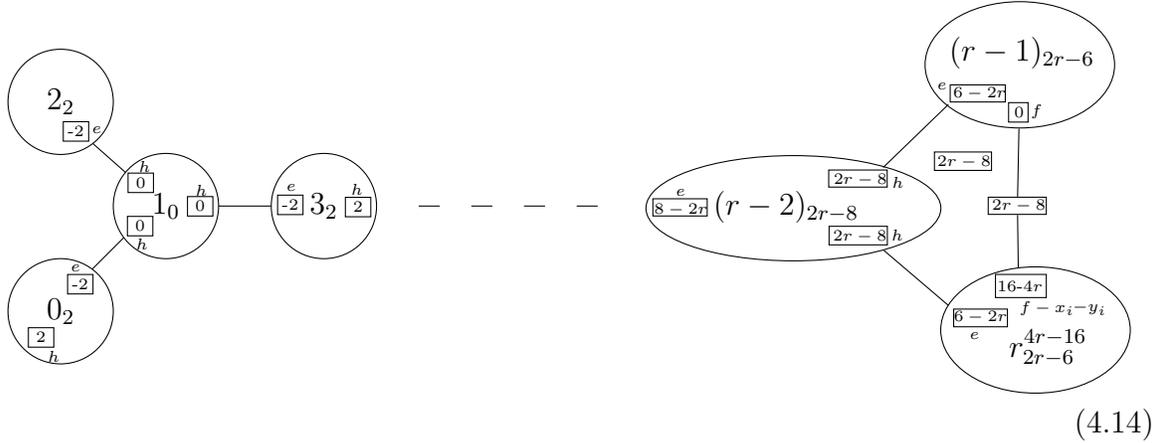
\begin{align}
\begin{array}{c}
\begin{tikzpicture}[scale=1.4]
\draw  (-3.5,4) ellipse (0.5 and 0.5);
\node (v1) at (-3.5,4) {$0_2$};
\draw  (-2.5,5) ellipse (0.5 and 0.5);
\node (v2) at (-2.5,5) {$1_0$};
\draw  (-1,5) ellipse (0.5 and 0.5);
\node (v4) at (-1,5) {$3_2$};
\draw  (3.4586,4.9793) ellipse (1.4 and 0.5);
\node at (3.389,4.9889) {$(r-2)_{2r-8}$};
\draw (-3.2,4.4) -- (-2.9,4.7);
\draw (-2,5) -- (-1.5,5);
\draw (-0.1,5) -- (0.1,5) (0.4,5) -- (0.6,5);
\draw (0.9,5) -- (1.1,5) (1.4,5) -- (1.6,5);
\begin{scope}[shift={(-0.0636,1.4373)}]
\draw  (-3.4113,4.3642) rectangle (-3.17,4.1794);
\node at (-3.2917,4.2752) {\tiny{-2}};
\end{scope}
\begin{scope}[shift={(1.1159,0.7257)}]
\draw  (-3.4113,4.3642) rectangle (-3.17,4.1794);
\node at (-3.2917,4.2752) {\tiny{0}};
\end{scope}
\begin{scope}[shift={(1.9717,0.7394)}]
\draw  (-3.4113,4.3642) rectangle (-3.17,4.1794);
\node at (-3.2917,4.2752) {\tiny{-2}};
\end{scope}
\begin{scope}[shift={(2.6153,0.7188)}]
\draw  (-3.4113,4.3642) rectangle (-3.17,4.1794);
\node at (-3.2917,4.2752) {\tiny{2}};
\end{scope}
\begin{scope}[shift={(7.3822,0.4349)}]
\draw  (-3.5851,4.3727) rectangle (-3.0318,4.196);
\node at (-3.2917,4.2752) {\tiny{$2r-8$}};
\end{scope}
\begin{scope}[shift={(5.6849,0.6908)}]
\draw  (-3.5608,4.3783) rectangle (-3.0453,4.2148);
\node at (-3.2917,4.2752) {\tiny{$8-2r$}};
\end{scope}
\node at (-2.1625,5.1528) {\tiny{$h$}};
\node at (-3.5652,3.5724) {\tiny{$h$}};
\node at (-1.3095,5.1733) {\tiny{$e$}};
\node at (-2.7275,4.6352) {\tiny{$h$}};
\node at (-0.6859,5.1591) {\tiny{$h$}};
\node at (2.3865,5.1337) {\tiny{$e$}};
\node at (-3.3582,4.4131) {\tiny{$e$}};
\begin{scope}[shift={(0.0001,1.9902)}]
\draw  (-3.5,4) ellipse (0.5 and 0.5);
\node (v1) at (-3.5,4) {$2_2$};
\draw (-3.1971,3.6066) -- (-2.8839,3.3294);
\begin{scope}[shift={(-0.0137,-0.5476)}]
\end{scope}
\node at (-3.1558,3.7385) {\tiny{$e$}};
\end{scope}
\begin{scope}[shift={(8.9112,2.3311)}]
\draw  (-3.3027,4.011) ellipse (0.9 and 0.6);
\node (v1) at (-3.2831,4.122) {$(r-1)_{2r-6}$};
\draw (-3.9808,3.6346) -- (-4.6003,3.0354);
\begin{scope}[shift={(-0.0181,-0.7126)}]
\draw  (-3.3913,4.3642) rectangle (-3.2064,4.1794);
\node at (-3.2917,4.2752) {\tiny{0}};
\end{scope}
\node at (-3.1436,3.5684) {\tiny{$f$}};
\node at (-4.0415,3.8069) {\tiny{$e$}};
\end{scope}
\begin{scope}[shift={(8.917,-0.1552)}]
\draw  (-3.1597,3.9773) ellipse (0.9 and 0.6);
\node (v1) at (-3.0415,3.8049) {$r^{4r-16}_{2r-6}$};
\draw (-3.9884,4.2124) -- (-4.6002,4.7342);
\node at (-3.0506,4.1836) {\tiny{$f-x_i$}};
\node at (-3.7387,3.9269) {\tiny{$e$}};
\end{scope}
\begin{scope}[shift={(-0.3948,-0.5205)}]
\draw  (-3.4113,4.3642) rectangle (-3.17,4.1794);
\node at (-3.2917,4.2752) {\tiny{2}};
\end{scope}
\begin{scope}[shift={(-0.0224,-0.0146)}]
\draw  (-3.4113,4.3642) rectangle (-3.17,4.1794);
\node at (-3.2917,4.2752) {\tiny{-2}};
\end{scope}
\begin{scope}[shift={(0.5457,0.5375)}]
\draw  (-3.4113,4.3642) rectangle (-3.17,4.1794);
\node at (-3.2917,4.2752) {\tiny{0}};
\end{scope}
\begin{scope}[shift={(0.5516,0.949)}]
\draw  (-3.4113,4.3642) rectangle (-3.17,4.1794);
\node at (-3.2917,4.2752) {\tiny{0}};
\end{scope}
\node at (-2.7003,5.3738) {\tiny{$h$}};
\begin{scope}[shift={(7.3866,0.9712)}]
\draw  (-3.5928,4.3712) rectangle (-3.0377,4.2089);
\node at (-3.2917,4.2752) {\tiny{$2r-8$}};
\end{scope}
\node at (4.4531,5.2234) {\tiny{$h$}};
\node at (4.4466,4.727) {\tiny{$h$}};
\begin{scope}[shift={(8.5285,-0.3324)}]
\draw  (-3.5483,4.3533) rectangle (-3.0379,4.185);
\node at (-3.2917,4.2752) {\tiny{$6-2r$}};
\end{scope}
\begin{scope}[shift={(8.5214,1.7981)}]
\draw  (-3.5742,4.3532) rectangle (-3.0511,4.1911);
\node at (-3.2917,4.2752) {\tiny{$6-2r$}};
\end{scope}
\begin{scope}[shift={(8.916,-0.0201)}]
\draw  (-3.5388,4.3703) rectangle (-3.0562,4.1503);
\node at (-3.3049,4.2549) {\tiny{16-4$r$}};
\end{scope}
\node at (6.2703,4.0317) {\tiny{$-y_i$}};
\begin{scope}[shift={(8.3714,1.1427)}]
\draw  (-3.5739,4.3711) rectangle (-3.0318,4.196);
\node at (-3.2917,4.2752) {\tiny{$2r-8$}};
\end{scope}
\begin{scope}[shift={(8.8915,0.7249)}]
\draw  (-3.5833,4.3591) rectangle (-3.0318,4.196);
\node at (-3.2917,4.2752) {\tiny{$2r-8$}};
\end{scope}
\draw (5.5873,4.9147) -- (5.5937,4.4173);
\draw (5.5873,5.0892) -- (5.6002,5.7353);
\end{tikzpicture}
\end{array}\label{pattern}
\end{align} 
where the exceptional curves in $S_r$ have been divided into two sets denoted by $x_i$ and $y_i$, $1\le i\le2r-8$. 

Notice that the surfaces intersect in the fashion of an affine $D_r$ Dynkin diagram, but there are extra intersections between the two single valent nodes towards the right end. As we will show now, these extra intersections however do not change the intersection pattern for the components of the degenerate elliptic fiber for $\text{I}^*_{r-4}$. The intersection is computed in a non-compact surface $N$ intersecting (blow-ups of) Hirzebruch surfaces $S_i$ along the total transform of their fibers $f_i$. Then, $\left(f_{r-1}\cdot f_r\right)_N=f\cdot \left( f-x_j-y_j\right)_{S_r}=0$ where we have picked a pair of blow-ups $x_j,y_j$. Similar comments apply to all the following cases where we have extra intersections between the surfaces not accounted for by the corresponding affine Dynkin graphs.

For odd $n=2r+1$, the F-theory setup involves a non-split $\text{I}^*_{r-3}$ fiber over a $-4$ curve, where $r>3$. The Weierstrass model for $r$ even is defined by 
	\begin{align}  \label{patternn}
		a_1 = a_{1,1} e_0,~~ a_2=a_{2,1} e_0 ,~~ a_{3} = a_{3,\frac{r}{2}} e_0^{\frac{r}{2}},~~ a_{4}=a_{4,\frac{r}{2}+1} e_0^{\frac{r}{2}+1} ,~~ a_{6} = a_{6,r} e_0^{r}.
	\end{align}
We consider the following resolution:
	\begin{align}
		(x,y,e_0|e_1) ,~~ (y,e_1|e_2),~~(x,e_2|e_3),~~\cdots ,~~(y,e_{r-3} |e_{r-2}),~~ (x,e_{r-2}|e_{r-1}),~~(e_{r-2},e_{r-1}|e_r).
	\end{align}
The Weierstrass model for $r$ odd is defined by
\begin{align}	
		a_1=a_1,~~a_2 = a_{2,1} e_0,~~ a_3=a_{3,\frac{r+1}{2}} e_0^{\frac{r+1}{2}},~~ a_4= a_{4,\frac{r+1}{2}} e_0^{\frac{r+1}{2}},~~ a_6 = a_{6,r} e_0^r.
	\end{align}	
We consider the following resolution:
	\begin{align}
		(x,y,e_0|e_1),~~ (y,e_1|e_2) ,~~ (x,e_2|e_3) ,~~ \cdots ,~~ (x,e_{r-3}|e_{r-2}),~~ (y,e_{r-2}|e_{r-1}) ,~~ (e_{r-2},e_{r-1}|e_{r}).
	\end{align}
In combination, they give rise to 

\begin{align}
\scalebox{.9}{$
\begin{array}{c}
\noindent\begin{tikzpicture}[scale=1.4]
\draw  (-3.5,4) ellipse (0.5 and 0.5);
\node (v1) at (-3.5,4) {$0_2$};
\draw  (-2.5,5) ellipse (0.5 and 0.5);
\node (v2) at (-2.5,5) {$1_0$};
\draw  (-1,5) ellipse (0.5 and 0.5);
\node (v4) at (-1,5) {$3_2$};
\draw  (3.2324,5.0054) ellipse (1.2 and 0.5);
\node at (3.2306,4.9097) {$r_{2r-6}$};
\draw (-3.2,4.4) -- (-2.9,4.7);
\draw (-2,5) -- (-1.5,5);
\draw (-0.1,5) -- (0.1,5) (0.4,5) -- (0.6,5);
\draw (0.9,5) -- (1.1,5) (1.4,5) -- (1.6,5);
\begin{scope}[shift={(-0.0636,1.4373)}]
\draw  (-3.4113,4.3642) rectangle (-3.17,4.1794);
\node at (-3.2917,4.2752) {\tiny{-2}};
\end{scope}
\begin{scope}[shift={(1.1159,0.7257)}]
\draw  (-3.4113,4.3642) rectangle (-3.17,4.1794);
\node at (-3.2917,4.2752) {\tiny{0}};
\end{scope}
\begin{scope}[shift={(1.9717,0.7394)}]
\draw  (-3.4113,4.3642) rectangle (-3.17,4.1794);
\node at (-3.2917,4.2752) {\tiny{-2}};
\end{scope}
\begin{scope}[shift={(2.6153,0.7188)}]
\draw  (-3.4113,4.3642) rectangle (-3.17,4.1794);
\node at (-3.2917,4.2752) {\tiny{2}};
\end{scope}
\begin{scope}[shift={(7.3594,0.7073)}]
\draw  (-3.6675,4.3771) rectangle (-3.0318,4.196);
\node at (-3.3431,4.2881) {\tiny{$8r-24$}};
\end{scope}
\begin{scope}[shift={(5.6849,0.6908)}]
\draw  (-3.5608,4.3783) rectangle (-3.0178,4.2033);
\node at (-3.2917,4.2752) {\tiny{$6-2r$}};
\end{scope}
\node at (-2.1625,5.1528) {\tiny{$h$}};
\node at (-3.5652,3.5724) {\tiny{$h$}};
\node at (-1.3095,5.1733) {\tiny{$e$}};
\node at (-2.7275,4.6352) {\tiny{$h$}};
\node at (-0.6859,5.1591) {\tiny{$h$}};
\node at (2.3865,5.1337) {\tiny{$e$}};
\node at (-3.3582,4.4131) {\tiny{$e$}};
\begin{scope}[shift={(0.0001,1.9902)}]
\draw  (-3.5,4) ellipse (0.5 and 0.5);
\node (v1) at (-3.5,4) {$2_2$};
\draw (-3.1971,3.6066) -- (-2.8839,3.3294);
\begin{scope}[shift={(-0.0137,-0.5476)}]
\end{scope}
\node at (-3.1558,3.7385) {\tiny{$e$}};
\end{scope}
\begin{scope}[shift={(9.0519,0.9863)}]
\draw  (-2.6505,4.0338) ellipse (1.4 and 0.5);
\node (v1) at (-2.2871,4.0602) {$(r-1)_{4r-8,2r-7}$};
\draw (-4.0584,4.0405) -- (-4.6269,4.0368);
\node at (-3.6487,4.2097) {\tiny{$e$}};
\end{scope}
\begin{scope}[shift={(-0.3948,-0.5205)}]
\draw  (-3.4113,4.3642) rectangle (-3.17,4.1794);
\node at (-3.2917,4.2752) {\tiny{2}};
\end{scope}
\begin{scope}[shift={(-0.0224,-0.0146)}]
\draw  (-3.4113,4.3642) rectangle (-3.17,4.1794);
\node at (-3.2917,4.2752) {\tiny{-2}};
\end{scope}
\begin{scope}[shift={(0.5457,0.5375)}]
\draw  (-3.4113,4.3642) rectangle (-3.17,4.1794);
\node at (-3.2917,4.2752) {\tiny{0}};
\end{scope}
\begin{scope}[shift={(0.5516,0.949)}]
\draw  (-3.4113,4.3642) rectangle (-3.17,4.1794);
\node at (-3.2917,4.2752) {\tiny{0}};
\end{scope}
\node at (-2.7003,5.3738) {\tiny{$h$}};
\node at (3.9908,5.2058) {\tiny{$2h$}};
\begin{scope}[shift={(8.75,0.7231)}]
\draw  (-3.6675,4.3771) rectangle (-3.0318,4.196);
\node at (-3.3431,4.2881) {\tiny{$8-4r$}};
\end{scope}
\end{tikzpicture}
\end{array}
$}
\end{align}

\subsection{ \texorpdfstring{$Sp(n)$}{} with  \texorpdfstring{$2n+8$}{} fundamental hypers}
An F-theory setup for this model involves a non-split $\text{I}_{2n+1}$ fiber, with $n>0$, over a $-1$ curve in the base. The Weierstrass model is defined by 
	\begin{align}
		a_1 = 0,~~ a_2=a_2,~~ a_3 = 0,~~a_4 = a_{4,n+1} e_0^{n+1},~~ a_6 =a_{6,2n+1} e_0^{2n+1}.
	\end{align}
We study the resolution \cite{Esole:2017kyr}
	\begin{align}
		(x,y,e_0|e_1),~~ (x,y,e_1|e_2),~~ \cdots,~~ (x,y,e_{n-1}|e_n)
	\end{align}
and find that the corresponding KK theory is described by (note that the nodes in the following graph ``grow'' from right to left with increasing values of $n$)

\begin{align}
\scalebox{.9}{$
\begin{array}{c}
\noindent\begin{tikzpicture}[scale=1.4]
\draw  (-3.9479,4.9946) ellipse (0.5 and 0.5);
\node (v1) at (-3.9479,4.9946) {$0_1$};
\draw  (-2.5,5) ellipse (0.5 and 0.5);
\node (v2) at (-2.5,5) {$1_6$};
\draw  (-1,5) ellipse (0.5 and 0.5);
\node (v4) at (-1,5) {$2_8$};
\draw  (3.3558,4.9823) ellipse (1.6 and 0.5);
\node at (3.4232,4.9805) {$(n-1)_{2n+2}$};
\draw (-3.4454,5.0233) -- (-2.99,5.0192);
\draw (-2,5) -- (-1.5,5);
\draw (-0.1,5) -- (0.1,5) (0.4,5) -- (0.6,5);
\draw (0.9,5) -- (1.1,5) (1.4,5) -- (1.6,5);
\begin{scope}[shift={(1.1159,0.7257)}]
\draw  (-3.4113,4.3642) rectangle (-3.17,4.1794);
\node at (-3.2917,4.2752) {\tiny{6}};
\end{scope}
\begin{scope}[shift={(1.9717,0.7394)}]
\draw  (-3.4113,4.3642) rectangle (-3.17,4.1794);
\node at (-3.2917,4.2752) {\tiny{-8}};
\end{scope}
\begin{scope}[shift={(2.6153,0.7188)}]
\draw  (-3.4113,4.3642) rectangle (-3.17,4.1794);
\node at (-3.2917,4.2752) {\tiny{8}};
\end{scope}
\begin{scope}[shift={(7.9172,0.6889)}]
\draw  (-3.6209,4.3874) rectangle (-3.0318,4.196);
\node at (-3.3153,4.289) {\tiny{$2n+2$}};
\end{scope}
\begin{scope}[shift={(5.5225,0.7222)}]
\draw  (-3.6329,4.3783) rectangle (-2.9362,4.1728);
\node at (-3.2917,4.2752) {\tiny{$-2n-2$}};
\end{scope}
\node at (-2.1586,5.1627) {\tiny{$h$}};
\node at (-4.2364,4.8087) {\tiny{$e$}};
\node at (-1.3095,5.1733) {\tiny{$e$}};
\node at (-2.7909,4.8388) {\tiny{$e$}};
\node at (-0.6859,5.1591) {\tiny{$h$}};
\node at (2.2489,5.2051) {\tiny{$e$}};
\node at (-3.6501,5.2038) {\tiny{$2h$}};
\begin{scope}[shift={(9.4713,0.9424)}]
\draw  (-2.8576,4.0223) ellipse (1 and 0.5);
\node (v1) at (-2.5786,4.0047) {$n^{2n+8}_{1}$};
\draw (-3.8646,4.0367) -- (-4.5163,4.0368);
\node at (-3.3301,4.2368) {\tiny{$2h$-$\sum x_i$}};
\end{scope}
\begin{scope}[shift={(-0.9773,0.7289)}]
\draw  (-3.4113,4.3642) rectangle (-3.17,4.1794);
\node at (-3.2917,4.2752) {\tiny{-1}};
\end{scope}
\begin{scope}[shift={(-0.3357,0.754)}]
\draw  (-3.4113,4.3642) rectangle (-3.17,4.1794);
\node at (-3.2917,4.2752) {\tiny{4}};
\end{scope}
\begin{scope}[shift={(0.488,0.7442)}]
\draw  (-3.4113,4.3642) rectangle (-3.17,4.1794);
\node at (-3.2917,4.2752) {\tiny{-6}};
\end{scope}
\node at (4.6078,4.7834) {\tiny{$h$}};
\begin{scope}[shift={(9.34,0.6792)}]
\draw  (-3.6675,4.3771) rectangle (-2.9831,4.1983);
\node at (-3.3431,4.2881) {\tiny{$-2n-4$}};
\end{scope}
\end{tikzpicture}
\end{array}
$}
\end{align}
A degenerate case in this class of models is $n=0$, in which case the F-theory configuration is an $\text{I}_0$ fiber over a $-1$ curve which constructs the E-string theory in $6d$. Taking a limit of the above geometries, we can predict the geometry for KK theory corresponding to E-string theory to be

\begin{align}
\begin{array}{c}
\begin{tikzpicture}[scale=2]
\draw  (-3.9479,4.9946) ellipse (0.5 and 0.5);
\node (v1) at (-3.866,5.002) {$0^8_1$};
\node at (-4.2364,4.8087) {\tiny{$e$}};
\begin{scope}[shift={(-0.9773,0.7289)}]
\draw  (-3.4113,4.3642) rectangle (-3.17,4.1794);
\node at (-3.2917,4.2752) {\tiny{-1}};
\end{scope}
\end{tikzpicture}
\end{array}
\end{align}

Reference \cite{Jefferson:2018irk} predicts the geometry as a del Pezzo surface $\text{dP}_9$ which equals $\P^2$ blown up at 9 points. This matches our answer because $\P^2$ blown up at one point equals $\F_1$.

\subsection{ \texorpdfstring{$E_6$}{} with  \texorpdfstring{$(6-k)$}{} fundamental hypers}
The F-theory construction involves a split $\text{IV}^*$ fiber over a $-k$ curve in the base, where $1\le k\le6$. The Weierstrass model is defined by 
	\begin{align}
		a_1 = 0,~~ a_2=0,~~a_3=a_{3,2} e_0^2,~~ a_4 = a_{4,3} e_0^3,~~a_{6}=a_{6,5}e_0^5.	
	\end{align}
We consider the resolution \cite{Esole:2017kyr}
	\begin{align}
		(x,y,e_0|e_1),~~ (y,e_1|e_2),~~(x,e_2|e_3),~~(e_2,e_3|e_4),~~(y,e_3|e_5),~~(y,e_4|e_6)
	\end{align}
and find the KK theory to be

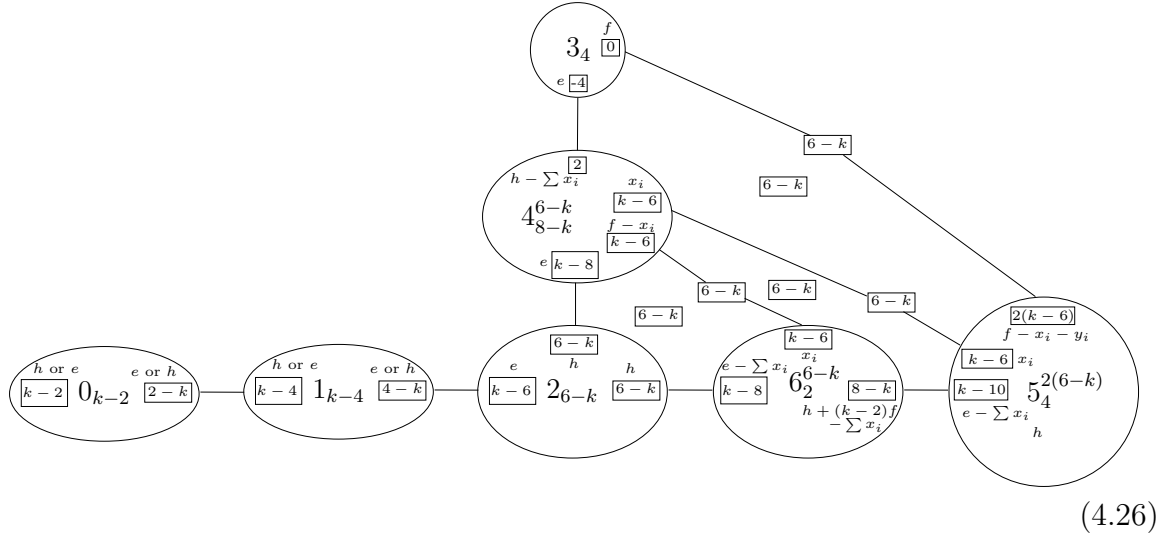
\begin{align}
\scalebox{.9}{$
\begin{array}{c}
\noindent\begin{tikzpicture}[scale=1.4]
\draw  (-4.4603,4.9901) ellipse (1 and 0.5);
\node (v1) at (-4.4603,4.9901) {$0_{k-2}$};
\draw (-3.4464,5.0091) -- (-2.9847,5.0073);
\node at (-3.937,5.2302) {\tiny{$e$ or $h$}};
\node at (-4.9686,5.2627) {\tiny{$h$ or $e$}};
\begin{scope}[shift={(-1.8167,0.7227)}]
\draw  (-3.5199,4.4183) rectangle (-3.0317,4.1342);
\node at (-3.2917,4.2752) {\tiny{$k-2$}};
\end{scope}
\begin{scope}[shift={(-0.4988,0.715)}]
\draw  (-3.5352,4.3891) rectangle (-3.0397,4.1868);
\node at (-3.2901,4.2995) {\tiny{$2-k$}};
\end{scope}
\begin{scope}[shift={(2.4713,0.0223)}]
\draw  (-4.4603,4.9901) ellipse (1 and 0.5);
\node (v1) at (-4.4603,4.9901) {$1_{k-4}$};
\draw (-3.4464,5.0091) -- (-2.9847,5.0073);
\begin{scope}[shift={(-1.8167,0.7227)}]
\draw  (-3.5199,4.4183) rectangle (-3.0317,4.1342);
\node at (-3.2917,4.2752) {\tiny{$k-4$}};
\end{scope}
\begin{scope}[shift={(-0.4988,0.715)}]
\draw  (-3.5352,4.3891) rectangle (-3.0397,4.1868);
\node at (-3.2901,4.2995) {\tiny{$4-k$}};
\end{scope}
\end{scope}
\begin{scope}[shift={(4.9399,0.0223)}]
\draw  (-4.4603,4.9901) ellipse (1 and 0.7);
\node (v1) at (-4.4603,4.9901) {$2_{6-k}$};
\draw (-3.4464,5.0091) -- (-2.9847,5.0073);
\node at (-3.8462,5.2361) {\tiny{$h$}};
\node at (-5.0811,5.2563) {\tiny{$e$}};
\begin{scope}[shift={(-1.8167,0.7227)}]
\draw  (-3.5199,4.4183) rectangle (-3.0317,4.1342);
\node at (-3.2917,4.2752) {\tiny{$k-6$}};
\end{scope}
\begin{scope}[shift={(-0.4988,0.715)}]
\draw  (-3.5352,4.3891) rectangle (-3.0397,4.1868);
\node at (-3.2901,4.2995) {\tiny{$6-k$}};
\end{scope}
\end{scope}
\begin{scope}[shift={(7.426,0.0223)}]
\draw  (-4.4603,4.9901) ellipse (1 and 0.7);
\node (v1) at (-4.3895,5.0952) {$6^{6-k}_2$};
\draw (-3.4464,5.0091) -- (-2.9847,5.0073);
\node at (-3.959,4.6401) {\tiny{$-\sum x_i$}};
\node at (-4.0148,4.7825) {\tiny{$h+(k-2)f$}};
\node at (-5.0165,5.2549) {\tiny{$e-\sum x_i$}};
\begin{scope}[shift={(-1.8625,0.7278)}]
\draw  (-3.5199,4.4183) rectangle (-3.0317,4.1342);
\node at (-3.2917,4.2752) {\tiny{$k-8$}};
\end{scope}
\begin{scope}[shift={(-0.4988,0.715)}]
\draw  (-3.5352,4.3891) rectangle (-3.0397,4.1868);
\node at (-3.2901,4.2995) {\tiny{$8-k$}};
\end{scope}
\begin{scope}[shift={(-1.1701,1.2542)}]
\draw  (-3.5352,4.3891) rectangle (-3.0397,4.1868);
\node at (-3.2901,4.2995) {\tiny{$k-6$}};
\end{scope}
\node at (-4.4422,5.3521) {\tiny{$x_i$}};
\end{scope}
\begin{scope}[shift={(9.9121,0.0223)}]
\draw  (-4.4603,4.9901) ellipse (1 and 1);
\node (v1) at (-4.2276,5.0106) {$5^{2(6-k)}_4$};
\node at (-4.5255,4.5625) {\tiny{$h$}};
\node at (-4.9621,4.7663) {\tiny{$e-\sum x_i$}};
\begin{scope}[shift={(-1.8241,0.7315)}]
\draw  (-3.5803,4.3865) rectangle (-3.0157,4.1716);
\node at (-3.2917,4.2752) {\tiny{$k-10$}};
\end{scope}
\begin{scope}[shift={(-1.7545,1.0524)}]
\draw  (-3.5696,4.3865) rectangle (-3.0341,4.1871);
\node at (-3.2917,4.2752) {\tiny{$k-6$}};
\end{scope}
\begin{scope}[shift={(-1.1716,1.507)}]
\draw  (-3.6354,4.3787) rectangle (-2.9623,4.193);
\node at (-3.2917,4.2752) {\tiny{$2(k-6)$}};
\end{scope}
\node at (-4.6461,5.3094) {\tiny{$x_i$}};
\node at (-4.4334,5.599) {\tiny{$f-x_i-y_i$}};
\end{scope}
\node at (-2.4487,5.3096) {\tiny{$h$ or $e$}};
\node at (-1.4157,5.2653) {\tiny{$e$ or $h$}};
\begin{scope}[shift={(3.7837,1.2228)}]
\draw  (-3.5352,4.3891) rectangle (-3.0397,4.1868);
\node at (-3.2901,4.2995) {\tiny{$6-k$}};
\end{scope}
\node at (0.496,5.3158) {\tiny{$h$}};
\begin{scope}[shift={(4.9906,1.8718)}]
\draw  (-4.4603,4.9901) ellipse (1 and 0.7);
\node (v1) at (-4.7772,4.9799) {$4^{6-k}_{8-k}$};
\node at (-4.7985,5.3804) {\tiny{$h-\sum x_i$}};
\node at (-4.8141,4.5057) {\tiny{$e$}};
\begin{scope}[shift={(-1.2082,0.2019)}]
\draw  (-3.5199,4.4183) rectangle (-3.0317,4.1342);
\node at (-3.2917,4.2752) {\tiny{$k-8$}};
\end{scope}
\begin{scope}[shift={(-0.6096,0.4237)}]
\draw  (-3.5352,4.3891) rectangle (-3.0397,4.1868);
\node at (-3.2901,4.2995) {\tiny{$k-6$}};
\end{scope}
\begin{scope}[shift={(-1.1678,1.2435)}]
\draw  (-3.3836,4.3758) rectangle (-3.1988,4.2014);
\node at (-3.2901,4.2995) {\tiny{2}};
\end{scope}
\begin{scope}[shift={(-0.5382,0.8499)}]
\draw  (-3.5352,4.3891) rectangle (-3.0397,4.1868);
\node at (-3.2901,4.2995) {\tiny{$k-6$}};
\end{scope}
\node at (-3.8901,4.8966) {\tiny{$f-x_i$}};
\node at (-3.8342,5.3391) {\tiny{$x_i$}};
\end{scope}
\begin{scope}[shift={(4.9957,3.6315)}]
\draw  (-4.4603,4.9901) ellipse (0.5 and 0.5);
\node (v1) at (-4.4505,4.9971) {$3_4$};
\node at (-4.1444,5.2221) {\tiny{$f$}};
\node at (-4.6425,4.6524) {\tiny{$e$}};
\begin{scope}[shift={(-1.1627,0.3484)}]
\draw  (-3.3836,4.3758) rectangle (-3.1988,4.2014);
\node at (-3.2901,4.2995) {\tiny{-4}};
\end{scope}
\begin{scope}[shift={(-0.822,0.7248)}]
\draw  (-3.3836,4.3758) rectangle (-3.1988,4.2014);
\node at (-3.2901,4.2995) {\tiny{0}};
\end{scope}
\end{scope}
\begin{scope}[shift={(5.3412,1.7739)}]
\draw  (-3.5352,4.3891) rectangle (-3.0397,4.1868) node (v2) {};
\node at (-3.2901,4.2995) {\tiny{$6-k$}};
\end{scope}
\begin{scope}[shift={(7.1327,1.667)}]
\draw  (-3.5352,4.3891) rectangle (-3.0397,4.1868);
\node at (-3.2901,4.2995) {\tiny{$6-k$}};
\end{scope}
\begin{scope}[shift={(6.4588,3.3302)}]
\draw  (-3.5352,4.3891) rectangle (-3.0397,4.1868);
\node at (-3.2901,4.2995) {\tiny{$6-k$}};
\end{scope}
\draw (0.5114,6.1638) -- (0.5114,5.7178);
\draw (0.5314,7.5621) -- (0.5355,8.1165);
\draw (1.3953,6.5134) -- (1.8855,6.1678) (v2);
\draw (2.2954,5.9991) -- (2.8981,5.7178) (4.5695,5.5048) -- (3.9709,5.8585) (1.5239,6.9272) -- (3.5851,6.0232);
\draw (1.0326,8.6029) -- (2.9927,7.7162) (3.3381,7.5108) -- (5.3636,6.0174);
\begin{scope}[shift={(5.9921,2.8915)}]
\draw  (-3.5352,4.3891) rectangle (-3.0397,4.1868);
\node at (-3.2901,4.2995) {\tiny{$6-k$}};
\end{scope}
\begin{scope}[shift={(6.0855,1.7994)}]
\draw  (-3.5352,4.3891) rectangle (-3.0397,4.1868);
\node at (-3.2901,4.2995) {\tiny{$6-k$}};
\end{scope}
\begin{scope}[shift={(4.6761,1.5194)}]
\draw  (-3.5352,4.3891) rectangle (-3.0397,4.1868);
\node at (-3.2901,4.2995) {\tiny{$6-k$}};
\end{scope}
\end{tikzpicture}
\end{array}
$}
\end{align}
where some of the curves can be either $h$ or $e$ because the degrees of their ambient Hirzebruch surfaces depend on $k$. We notice that our answer matches that of \cite{DelZotto:2017pti} for $k=6$.

\subsection{ \texorpdfstring{$E_7$}{} with \texorpdfstring{$\frac{8-k}{2}$}{} fundamental hypers}
The F-theory construction involves a $\text{III}^*$ fiber over a $-k$ curve in the base, where $1\le k\le8$. The Weierstrass model is defined by 
	\begin{align}
		a_1 = 0,~~ a_2=0,~~a_3=0,~~ a_4 = a_{4,3} e_0^3,~~a_{6}=a_{6,5}e_0^5.	
	\end{align}
We consider the resolution \cite{Esole:2017kyr}
	\begin{align}
		(x,y,e_0|e_1),~~ (y,e_1|e_2),~~(x,e_2|e_3),~~(y,e_3|e_4),~~(e_2,e_3|e_5),~~(e_2,e_4|e_6),~~ (e_4,e_5|e_7)
	\end{align}
and find the KK theory to be described by

\begin{align}
\scalebox{.9}{$
\begin{array}{c}
\noindent\begin{tikzpicture}[scale=1.4]
\draw  (-1.9624,3.1392) ellipse (0.7 and 0.8);
\node (v1) at (-1.9624,3.1392) {$0_{k-2}$};
\node at (-1.9629,3.4946) {\tiny{$e$ or $h$}};
\node at (-1.996,2.8287) {\tiny{$h$ or $e$}};
\begin{scope}[shift={(1.2877,-1.7077)}]
\draw  (-3.5199,4.4183) rectangle (-3.0317,4.1342);
\node at (-3.2917,4.2752) {\tiny{$k-2$}};
\end{scope}
\begin{scope}[shift={(1.3118,-0.5766)}]
\draw  (-3.5352,4.3891) rectangle (-3.0397,4.1868);
\node at (-3.2901,4.2995) {\tiny{$2-k$}};
\end{scope}
\begin{scope}[shift={(2.4713,0.0223)}]
\draw  (-4.4603,4.9901) ellipse (1 and 0.5);
\node (v1) at (-4.4603,4.9901) {$1_{k-4}$};
\draw (-3.4464,5.0091) -- (-2.9847,5.0073);
\begin{scope}[shift={(-1.8167,0.7227)}]
\draw  (-3.5199,4.4183) rectangle (-3.0317,4.1342);
\node at (-3.2917,4.2752) {\tiny{$k-4$}};
\end{scope}
\begin{scope}[shift={(-0.4988,0.715)}]
\draw  (-3.5352,4.3891) rectangle (-3.0397,4.1868);
\node at (-3.2901,4.2995) {\tiny{$4-k$}};
\end{scope}
\end{scope}
\begin{scope}[shift={(4.9399,0.0223)}]
\draw  (-4.4603,4.9901) ellipse (1 and 0.5);
\node (v1) at (-4.4603,4.9901) {$2_{6-k}$};
\draw (-3.4464,5.0091) -- (-2.9847,5.0073);
\node at (-3.8997,5.2056) {\tiny{$h$ or $e$}};
\node at (-5.0091,5.2503) {\tiny{$e$ or $h$}};
\begin{scope}[shift={(-1.8167,0.7227)}]
\draw  (-3.5199,4.4183) rectangle (-3.0317,4.1342);
\node at (-3.2917,4.2752) {\tiny{$k-6$}};
\end{scope}
\begin{scope}[shift={(-0.4988,0.715)}]
\draw  (-3.5352,4.3891) rectangle (-3.0397,4.1868);
\node at (-3.2901,4.2995) {\tiny{$6-k$}};
\end{scope}
\end{scope}
\begin{scope}[shift={(7.426,0.0223)}]
\draw  (-4.4603,4.9901) ellipse (1 and 0.7);
\node (v1) at (-4.4026,4.949) {$5_{8-k}$};
\draw (-3.4464,5.0091) -- (-2.9847,5.0073);
\node at (-3.7958,4.7913) {\tiny{$h$}};
\node at (-5.1627,5.2842) {\tiny{$e$}};
\begin{scope}[shift={(-1.8625,0.7278)}]
\draw  (-3.5199,4.4183) rectangle (-3.0317,4.1342);
\node at (-3.2917,4.2752) {\tiny{$k-8$}};
\end{scope}
\begin{scope}[shift={(-0.4988,0.715)}]
\draw  (-3.5352,4.3891) rectangle (-3.0397,4.1868);
\node at (-3.2901,4.2995) {\tiny{$8-k$}};
\end{scope}
\begin{scope}[shift={(-1.1701,1.2542)}]
\draw  (-3.5352,4.3891) rectangle (-3.0397,4.1868);
\node at (-3.2901,4.2995) {\tiny{$8-k$}};
\end{scope}
\node at (-4.4422,5.3521) {\tiny{$h$}};
\end{scope}
\begin{scope}[shift={(9.9121,0.0223)}]
\draw  (-4.4603,4.9901) ellipse (1 and 1);
\node (v1) at (-4.3411,4.7916) {$7^{8-k}_{k-10}$};
\node at (-4.2701,5.5496) {\tiny{$x_i$}};
\node at (-5.1195,4.8156) {\tiny{$e$}};
\begin{scope}[shift={(-1.8241,0.7315)}]
\draw  (-3.5803,4.3865) rectangle (-3.0157,4.1716);
\node at (-3.2917,4.2752) {\tiny{$k-10$}};
\end{scope}
\begin{scope}[shift={(-1.6086,1.2309)}]
\draw  (-3.5696,4.3865) rectangle (-3.0341,4.1871);
\node at (-3.2917,4.2752) {\tiny{$k-8$}};
\end{scope}
\begin{scope}[shift={(-0.999,1.4473)}]
\draw  (-3.6354,4.3787) rectangle (-2.9623,4.193);
\node at (-3.2917,4.2752) {\tiny{$k-8$}};
\end{scope}
\node at (-4.8758,5.3108) {\tiny{$f-x_i$}};
\node at (-3.8739,5.2177) {\tiny{$h-\sum x_i$}};
\begin{scope}[shift={(-0.3191,0.7086)}]
\draw  (-3.3836,4.3758) rectangle (-3.1988,4.2014);
\node at (-3.2901,4.2995) {\tiny{2}};
\end{scope}
\end{scope}
\node at (-2.4487,5.3096) {\tiny{$h$ or $e$}};
\node at (-1.4157,5.2653) {\tiny{$e$ or $h$}};
\begin{scope}[shift={(7.4323,2.1881)}]
\draw  (-4.4603,4.9901) ellipse (1.1 and 0.7);
\node (v1) at (-4.7772,4.9799) {$6^{8-k}_{10-k}$};
\node at (-4.8666,4.5116) {\tiny{$e$}};
\begin{scope}[shift={(-1.2082,0.2506)}]
\draw  (-3.566,4.4142) rectangle (-3.0317,4.1342);
\node at (-3.2917,4.2752) {\tiny{$k-10$}};
\end{scope}
\begin{scope}[shift={(-0.5934,0.3588)}]
\draw  (-3.5352,4.3891) rectangle (-3.0397,4.1868);
\node at (-3.2901,4.2995) {\tiny{$k-8$}};
\end{scope}
\begin{scope}[shift={(-0.4084,0.8499)}]
\draw  (-3.5352,4.3891) rectangle (-3.0397,4.1868);
\node at (-3.2901,4.2995) {\tiny{$k-8$}};
\end{scope}
\node at (-3.8739,4.8317) {\tiny{$f-x_i$}};
\node at (-3.7044,5.3391) {\tiny{$x_i$}};
\end{scope}
\begin{scope}[shift={(8.6658,2.9811)}]
\draw  (-3.5352,4.3891) rectangle (-3.0397,4.1868);
\node at (-3.2901,4.2995) {\tiny{$8-k$}};
\end{scope}
\begin{scope}[shift={(12.2424,0.0386)}]
\draw  (-4.4603,4.9901) ellipse (0.9 and 0.6);
\node (v1) at (-4.4815,4.8695) {$3_4$};
\begin{scope}[shift={(-1.1677,1.0942)}]
\draw  (-3.6091,4.4021) rectangle (-2.9668,4.1676);
\node at (-3.2917,4.2752) {\tiny{$20-2k$}};
\end{scope}
\begin{scope}[shift={(-1.8978,0.6838)}]
\draw  (-3.3836,4.3758) rectangle (-3.1988,4.2014);
\node at (-3.2901,4.2995) {\tiny{-4}};
\end{scope}
\node at (-4.475,5.1525) {\tiny{$h+(8-k)f$}};
\node at (-5.1889,4.7875) {\tiny{$e$}};
\end{scope}
\begin{scope}[shift={(12.3561,2.1151)}]
\draw  (-4.4603,4.9901) ellipse (1.1 and 0.8);
\node (v1) at (-4.3879,5.034) {$4^{8-k}_{14-k}$};
\node at (-4.2582,4.6008) {\tiny{$e-\sum x_i$}};
\begin{scope}[shift={(-1.1476,0.0987)}]
\draw  (-3.6492,4.4029) rectangle (-2.9749,4.1557);
\node at (-3.2917,4.2752) {\tiny{$2k-22$}};
\end{scope}
\begin{scope}[shift={(-1.91,0.8402)}]
\draw  (-3.5352,4.3891) rectangle (-3.0397,4.1868);
\node at (-3.2901,4.2995) {\tiny{$k-8$}};
\end{scope}
\begin{scope}[shift={(-1.7955,0.3551)}]
\draw  (-3.5352,4.3891) rectangle (-3.0397,4.1868);
\node at (-3.2901,4.2995) {\tiny{$k-8$}};
\end{scope}
\node at (-5.0824,5.3266) {\tiny{$f-x_i$}};
\node at (-5.124,4.8171) {\tiny{$x_i$}};
\end{scope}
\begin{scope}[shift={(7.5626,1.9022)}]
\draw  (-3.5352,4.3891) rectangle (-3.0397,4.1868);
\node at (-3.2901,4.2995) {\tiny{$8-k$}};
\end{scope}
\begin{scope}[shift={(9.8095,1.9184)}]
\draw  (-3.5352,4.3891) rectangle (-3.0397,4.1868);
\node at (-3.2901,4.2995) {\tiny{$8-k$}};
\end{scope}
\draw (2.9474,6.4759) -- (2.9474,5.7158) (6.441,5.0107) -- (6.8668,5.0061);
\draw (7.833,6.3065) -- (7.8421,5.6334);
\draw (5.6168,7.2589) -- (6.8027,7.2451) (4.0554,7.2772) -- (5.1223,7.2634) (3.7761,6.7003) -- (4.1195,6.2927) (4.3027,6.0867) -- (4.7148,5.7021) (7.0317,6.5949) -- (6.615,6.3111) (6.267,6.1233) -- (5.9465,5.8761);
\begin{scope}[shift={(8.6566,2.3126)}]
\draw  (-3.5352,4.3891) rectangle (-3.0397,4.1868);
\node at (-3.2901,4.2995) {\tiny{$8-k$}};
\end{scope}
\begin{scope}[shift={(7.054,1.5388)}]
\draw  (-3.5352,4.3891) rectangle (-3.0397,4.1868);
\node at (-3.2901,4.2995) {\tiny{$8-k$}};
\end{scope}
\begin{scope}[shift={(10.2409,1.5159)}]
\draw  (-3.5352,4.3891) rectangle (-3.0397,4.1868);
\node at (-3.2901,4.2995) {\tiny{$8-k$}};
\end{scope}
\draw (-1.9287,4.5105) -- (-1.937,3.9415);
\end{tikzpicture}
\end{array}
$}
\end{align}
We notice that our answer matches that of \cite{DelZotto:2017pti} for $k=8$ but differs for $k=7$. Noticeably, we have extra edges\footnote{Even though the surfaces do not intersect in the pattern of an affine $E_7$, the components of the degenerate fiber do. This can be seen by a computation similar to the one in between equations (\ref{pattern}) and (\ref{patternn}).} between the nodes of the affine $E_7$ (which implies non-trivial triple intersection numbers between three distinct surfaces) which reference \cite{DelZotto:2017pti} do not have. Moreover, our answer is not flop equivalent to theirs because one cannot get rid of all of the extra edges by doing flops. According to \cite{DelZotto:2017pti}, there is a consistency condition that the correct answer for $k=7$ must satisfy. Namely, one should be able to do flop transitions (along with sending a curve to infinite size) to reach a point where we have two disjoint collection of surfaces. One of them should be $\F_5-\F_3-\P^2$ describing an orbifold CFT and the other one should describe a non-orbifold CFT. In our case, the corresponding flop transition is the flop of $-1$ curve inside $S_2=\F_1$. Expanding the flopped curve to infinite size, we see that we also obtain $\F_5-\F_3-\P^2$ as one of the disjoint pieces. So, the disagreement in the proposal of \cite{DelZotto:2017pti} and our proposal can be phrased as a disagreement in the identification of the non-orbifold piece in the above mentioned limit.

Now, there are two reasons for us to trust that our result is the correct one. First, we found it by an honest computation using the tools described Section \ref{w_models} starting from the Weierstrass model defining this $6d$ SCFT. Second, we give a uniform answer for all $k$, including $k<7$.

\subsection{Pure  \texorpdfstring{$E_8$}{}}
The F-theory construction involves a $\text{II}^*$ fiber over a $-12$ curve in the base. The Weierstrass model is defined by 
	\begin{align}
		a_1 =0,~~a_2=0,~~a_3=0,~~ a_4 =a_{4,4} e_0^4,~~a_6=a_{6,5}e_0^5.
	\end{align}
We consider the resolution \cite{Esole:2017kyr}
	\begin{align}
	\begin{split}
	&	(x,y,e_0|e_1),~~ (y,e_1|e_2),~~(x,e_2|e_3),~~(y,e_3|e_4),~~(e_2,e_3|e_5),~~(e_4,e_5|e_6),~~(e_2,e_4,e_6|e_7),~~\\
		&(e_4,e_7|e_8)
	\end{split}
	\end{align}
and determine the geometry to be

\begin{align}
\scalebox{.95}{$
\begin{array}{c}
\noindent\begin{tikzpicture}[scale=1.5]
\draw  (-3.9479,4.9946) ellipse (0.5 and 0.5);
\node (v1) at (-3.9479,4.9946) {$0_{10}$};
\node at (-4.2364,4.8087) {\tiny{$h$}};
\node at (-3.6501,5.2038) {\tiny{$e$}};
\begin{scope}[shift={(-0.9773,0.7289)}]
\draw  (-3.4113,4.3642) rectangle (-3.17,4.1794);
\node at (-3.2917,4.2752) {\tiny{10}};
\end{scope}
\begin{scope}[shift={(-0.3357,0.754)}]
\draw  (-3.4113,4.3642) rectangle (-3.17,4.1794);
\node at (-3.2917,4.2752) {\tiny{-10}};
\end{scope}
\begin{scope}[shift={(1.3146,-0.0162)}]
\draw  (-3.9479,4.9946) ellipse (0.5 and 0.5);
\node (v1) at (-3.9479,4.9946) {$1_{8}$};
\node at (-4.2364,4.8087) {\tiny{$h$}};
\node at (-3.6501,5.2038) {\tiny{$e$}};
\begin{scope}[shift={(-0.9773,0.7289)}]
\draw  (-3.4113,4.3642) rectangle (-3.17,4.1794);
\node at (-3.2917,4.2752) {\tiny{8}};
\end{scope}
\begin{scope}[shift={(-0.3357,0.754)}]
\draw  (-3.4113,4.3642) rectangle (-3.17,4.1794);
\node at (-3.2917,4.2752) {\tiny{-8}};
\end{scope}
\end{scope}
\begin{scope}[shift={(2.6129,-0.0162)}]
\draw  (-3.9479,4.9946) ellipse (0.5 and 0.5);
\node (v1) at (-3.9479,4.9946) {$2_{6}$};
\node at (-4.2364,4.8087) {\tiny{$h$}};
\node at (-3.6501,5.2038) {\tiny{$e$}};
\begin{scope}[shift={(-0.9773,0.7289)}]
\draw  (-3.4113,4.3642) rectangle (-3.17,4.1794);
\node at (-3.2917,4.2752) {\tiny{6}};
\end{scope}
\begin{scope}[shift={(-0.3357,0.754)}]
\draw  (-3.4113,4.3642) rectangle (-3.17,4.1794);
\node at (-3.2917,4.2752) {\tiny{-6}};
\end{scope}
\end{scope}
\begin{scope}[shift={(3.9112,-0.0162)}]
\draw  (-3.9479,4.9946) ellipse (0.5 and 0.5);
\node (v1) at (-3.9479,4.9946) {$3_{4}$};
\node at (-4.2364,4.8087) {\tiny{$h$}};
\node at (-3.6501,5.2038) {\tiny{$e$}};
\begin{scope}[shift={(-0.9773,0.7289)}]
\draw  (-3.4113,4.3642) rectangle (-3.17,4.1794);
\node at (-3.2917,4.2752) {\tiny{4}};
\end{scope}
\begin{scope}[shift={(-0.3357,0.754)}]
\draw  (-3.4113,4.3642) rectangle (-3.17,4.1794);
\node at (-3.2917,4.2752) {\tiny{-4}};
\end{scope}
\end{scope}
\begin{scope}[shift={(5.2095,-0.0162)}]
\draw  (-3.9479,4.9946) ellipse (0.5 and 0.5);
\node (v1) at (-3.9479,4.9946) {$4_{2}$};
\node at (-4.2364,4.8087) {\tiny{$h$}};
\node at (-3.6501,5.2038) {\tiny{$e$}};
\begin{scope}[shift={(-0.9773,0.7289)}]
\draw  (-3.4113,4.3642) rectangle (-3.17,4.1794);
\node at (-3.2917,4.2752) {\tiny{2}};
\end{scope}
\begin{scope}[shift={(-0.3357,0.754)}]
\draw  (-3.4113,4.3642) rectangle (-3.17,4.1794);
\node at (-3.2917,4.2752) {\tiny{-2}};
\end{scope}
\end{scope}
\begin{scope}[shift={(6.4997,-0.0162)}]
\draw  (-3.9479,4.9946) ellipse (0.5 and 0.5);
\node (v1) at (-3.9479,4.9946) {$5_{0}$};
\node at (-4.2598,4.8234) {\tiny{$h$}};
\node at (-3.6211,5.1939) {\tiny{$e$}};
\begin{scope}[shift={(-0.9773,0.7289)}]
\draw  (-3.4113,4.3642) rectangle (-3.17,4.1794);
\node at (-3.2917,4.2752) {\tiny{0}};
\end{scope}
\begin{scope}[shift={(-0.3357,0.754)}]
\draw  (-3.4113,4.3642) rectangle (-3.17,4.1794);
\node at (-3.2917,4.2752) {\tiny{0}};
\end{scope}
\begin{scope}[shift={(-0.6452,1.0863)}]
\draw  (-3.4113,4.3642) rectangle (-3.17,4.1794);
\node at (-3.2917,4.2752) {\tiny{0}};
\end{scope}
\node at (-4.1293,5.3438) {\tiny{$e$}};
\end{scope}
\begin{scope}[shift={(7.7899,-0.0162)}]
\draw  (-3.9479,4.9946) ellipse (0.5 and 0.5);
\node (v1) at (-3.9479,4.9946) {$6_{2}$};
\node at (-4.2364,4.8087) {\tiny{$e$}};
\node at (-3.6501,5.2038) {\tiny{$h$}};
\begin{scope}[shift={(-0.9773,0.7289)}]
\draw  (-3.4113,4.3642) rectangle (-3.17,4.1794);
\node at (-3.2917,4.2752) {\tiny{-2}};
\end{scope}
\begin{scope}[shift={(-0.3357,0.754)}]
\draw  (-3.4113,4.3642) rectangle (-3.17,4.1794);
\node at (-3.2917,4.2752) {\tiny{2}};
\end{scope}
\end{scope}
\begin{scope}[shift={(9.0801,-0.0162)}]
\draw  (-3.9479,4.9946) ellipse (0.5 and 0.5);
\node (v1) at (-3.9479,4.9946) {$7_{4}$};
\node at (-4.2364,4.8087) {\tiny{$e$}};
\node at (-3.6501,5.2038) {\tiny{$h$}};
\begin{scope}[shift={(-0.9773,0.7289)}]
\draw  (-3.4113,4.3642) rectangle (-3.17,4.1794);
\node at (-3.2917,4.2752) {\tiny{-4}};
\end{scope}
\begin{scope}[shift={(-0.3357,0.754)}]
\draw  (-3.4113,4.3642) rectangle (-3.17,4.1794);
\node at (-3.2917,4.2752) {\tiny{4}};
\end{scope}
\end{scope}
\begin{scope}[shift={(6.5079,1.2497)}]
\draw  (-3.9479,4.9946) ellipse (0.5 and 0.5);
\node (v1) at (-3.9479,4.9946) {$6_{2}$};
\node at (-3.7125,4.6657) {\tiny{$e$}};
\begin{scope}[shift={(-0.669,0.3637)}]
\draw  (-3.4113,4.3642) rectangle (-3.17,4.1794);
\node at (-3.2917,4.2752) {\tiny{-2}};
\end{scope}
\end{scope}
\draw (-3.4455,5.0023) -- (-3.1276,5.0079);
\draw (-0.8335,4.9851) -- (-0.5307,4.985);
\begin{scope}[shift={(1.2939,0)}]
\draw (-0.8335,4.9851) -- (-0.5307,4.985);
\end{scope}
\begin{scope}[shift={(2.5809,-0.0034)}]
\draw (-0.8173,4.9851) -- (-0.5307,4.985);
\end{scope}
\draw (-2.1263,4.9858) -- (-1.8269,4.9858);
\begin{scope}[shift={(3.8744,-0.0116)}]
\draw (-0.8173,4.9851) -- (-0.5307,4.985);
\end{scope}
\begin{scope}[shift={(5.1623,-0.0116)}]
\draw (-0.8173,4.9851) -- (-0.5307,4.985);
\end{scope}
\draw (2.5451,5.7438) -- (2.5423,5.4774);
\end{tikzpicture}
\end{array}
$}
\end{align}
We notice that our answer matches that of \cite{DelZotto:2017pti}.

\subsection{ \texorpdfstring{$F_4$}{} with  \texorpdfstring{$(5-k)$}{} fundamental hypers}

Note that the unique zero mass parameter resolution, triple intersection numbers, and fibral divisor geometry of the $F_4$-model were computed in \cite{Esole:2017rgz}.  The F-theory construction involves a non-split $\text{IV}^*$ fiber over a $-k$ curve in the base, where $1\le k\le5$. The Weierstrass model is defined by 
	\begin{align}
		a_1=0,~~ a_2=0,~~ a_3 =0,~~ a_4 = a_{4,3} e_0^3,~~ a_6= a_{6,4} e_0^4.
	\end{align}
Here, we consider the same resolution, namely
	\begin{align}
		(x,y,e_0|e_1) ,~~ (y,e_1|e_2),~~(x,e_2|e_3),~~(e_2,e_3|e_4).
	\end{align}
In this case, we compute the geometry to be

\begin{align}
\scalebox{.9}{$
\begin{array}{c}
\noindent\begin{tikzpicture}[scale=1.4]
\draw  (-4.4603,4.9901) ellipse (1 and 0.5);
\node (v1) at (-4.4603,4.9901) {$0_{k-2}$};
\draw (-3.4464,5.0091) -- (-2.9847,5.0073);
\node at (-3.937,5.2302) {\tiny{$e$ or $h$}};
\node at (-4.9686,5.2627) {\tiny{$h$ or $e$}};
\begin{scope}[shift={(-1.8167,0.7227)}]
\draw  (-3.5199,4.4183) rectangle (-3.0317,4.1342);
\node at (-3.2917,4.2752) {\tiny{$k-2$}};
\end{scope}
\begin{scope}[shift={(-0.4988,0.715)}]
\draw  (-3.5352,4.3891) rectangle (-3.0397,4.1868);
\node at (-3.2901,4.2995) {\tiny{$2-k$}};
\end{scope}
\begin{scope}[shift={(2.4713,0.0223)}]
\draw  (-4.4603,4.9901) ellipse (1 and 0.5);
\node (v1) at (-4.4603,4.9901) {$1_{k-4}$};
\draw (-3.4464,5.0091) -- (-2.9847,5.0073);
\begin{scope}[shift={(-1.8167,0.7227)}]
\draw  (-3.5199,4.4183) rectangle (-3.0317,4.1342);
\node at (-3.2917,4.2752) {\tiny{$k-4$}};
\end{scope}
\begin{scope}[shift={(-0.4988,0.715)}]
\draw  (-3.5352,4.3891) rectangle (-3.0397,4.1868);
\node at (-3.2901,4.2995) {\tiny{$4-k$}};
\end{scope}
\end{scope}
\begin{scope}[shift={(4.9399,0.0223)}]
\draw  (-4.4603,4.9901) ellipse (1 and 0.7);
\node (v1) at (-4.5471,5.0079) {$2_{6-k}$};
\draw (-3.4464,5.0091) -- (-2.9847,5.0073);
\node at (-3.8462,5.2361) {\tiny{$2h$}};
\node at (-5.0811,5.2563) {\tiny{$e$}};
\begin{scope}[shift={(-1.8447,0.7227)}]
\draw  (-3.5266,4.3959) rectangle (-3.0525,4.1755);
\node at (-3.2917,4.2752) {\tiny{$k-6$}};
\end{scope}
\begin{scope}[shift={(-0.5829,0.7149)}]
\draw  (-3.6193,4.4079) rectangle (-2.9369,4.1961);
\node at (-3.2901,4.2995) {\tiny{$24-4k$}};
\end{scope}
\end{scope}
\begin{scope}[shift={(7.426,0.0223)}]
\draw  (-4.4603,4.9901) ellipse (1 and 0.7);
\node (v1) at (-4.3989,5.3287) {$3_{16-2k,5-k}$};
\draw (-3.4464,5.0091) -- (-2.9847,5.0073);
\node at (-3.8449,4.795) {\tiny{$h$}};
\node at (-5.0763,4.8179) {\tiny{$e$}};
\begin{scope}[shift={(-1.8625,0.7278)}]
\draw  (-3.5199,4.4183) rectangle (-2.9009,4.1903);
\node at (-3.217,4.2845) {\tiny{$2k-16$}};
\end{scope}
\begin{scope}[shift={(-0.4988,0.715)}]
\draw  (-3.6632,4.3751) rectangle (-3.0397,4.1868);
\node at (-3.3477,4.2872) {\tiny{$16-2k$}};
\end{scope}
\end{scope}
\begin{scope}[shift={(9.9121,0.0223)}]
\draw  (-4.4603,4.9901) ellipse (1 and 0.5);
\node (v1) at (-4.3623,4.9594) {$4_{8,5-k}$};
\node at (-5.2162,4.7739) {\tiny{$e$}};
\begin{scope}[shift={(-1.8241,0.7315)}]
\draw  (-3.5803,4.3865) rectangle (-3.2119,4.181);
\node at (-3.4038,4.2752) {\tiny{$-8$}};
\end{scope}
\end{scope}
\node at (-2.4487,5.3096) {\tiny{$h$ or $e$}};
\node at (-1.4157,5.2653) {\tiny{$e$ or $h$}};
\end{tikzpicture}
\end{array}
$}
\end{align}

which matches that of \cite{DelZotto:2017pti} for $k=5$, along with the results of \cite{Esole:2017rgz} for general $k$.

\subsection{ \texorpdfstring{$G_2$}{} with   \texorpdfstring{$(10-3k)$}{} fundamental hypers}
The zero mass parameter resolutions, fibral divisor geometry, and triple intersection numbers of the $G_2$-model were first described in \cite{Esole:2017qeh}. The F-theory construction involves a non-split $\text{I}^*_0$ fiber over a $-k$ curve in the base, where $1\le k\le3$. The Weierstrass model is defined by 
	\begin{align}
		a_1 = 0,~~ a_2 =0 ,~~ a_3 = 0,~~a_4 = a_{4,2} e_0^2,~~ a_6 = a_{6,3} e_0^3.
	\end{align}	
We consider the resolution \cite{Esole:2017kyr}
	\begin{align}
		(x,y,e_0|e_1),~~(y,e_1|e_2),
	\end{align}	
and the geometry for the KK theory turns out to be

\begin{align}
\label{eqn:G2}
\begin{array}{c}
\noindent\begin{tikzpicture}[scale=1.5]
\draw  (-4.4603,4.9901) ellipse (1 and 0.5);
\node (v1) at (-4.4603,4.9901) {$0_{k-2}$};
\draw (-3.4464,5.0091) -- (-2.9847,5.0073);
\node at (-3.937,5.2302) {\tiny{$e$ or $h$}};
\node at (-4.9686,5.2627) {\tiny{$h$ or $e$}};
\begin{scope}[shift={(-1.8167,0.7227)}]
\draw  (-3.5199,4.4183) rectangle (-3.0317,4.1342);
\node at (-3.2917,4.2752) {\tiny{$k-2$}};
\end{scope}
\begin{scope}[shift={(-0.4988,0.715)}]
\draw  (-3.5352,4.3891) rectangle (-3.0397,4.1868);
\node at (-3.2901,4.2995) {\tiny{$2-k$}};
\end{scope}
\begin{scope}[shift={(2.4713,0.0223)}]
\draw  (-4.4603,4.9901) ellipse (1 and 0.5);
\node (v1) at (-4.507,4.9528) {$1_{4-k}$};
\draw (-3.4464,5.0091) -- (-2.9847,5.0073);
\begin{scope}[shift={(-1.8354,0.7134)}]
\draw  (-3.5321,4.3845) rectangle (-3.0504,4.1716);
\node at (-3.2917,4.2752) {\tiny{$k-4$}};
\end{scope}
\begin{scope}[shift={(-0.5642,0.7057)}]
\draw  (-3.6392,4.4011) rectangle (-2.9855,4.1922);
\node at (-3.2901,4.2995) {\tiny{$36-9k$}};
\end{scope}
\end{scope}
\begin{scope}[shift={(5.1361,0.0036)}]
\draw  (-4.4603,4.9901) ellipse (1.2 and 0.5);
\node (v1) at (-4.108,4.9424) {$2_{18-3k,10-3k}$};
\node at (-5.2811,5.2179) {\tiny{$e$}};
\begin{scope}[shift={(-1.9849,0.7134)}]
\draw  (-3.6041,4.3993) rectangle (-2.9476,4.1525);
\node at (-3.2917,4.2752) {\tiny{$3k-18$}};
\end{scope}
\end{scope}
\node at (-2.6169,5.2536) {\tiny{$e$}};
\node at (-1.4157,5.2653) {\tiny{$3h$}};
\end{tikzpicture}
\end{array}
\end{align}

\subsection{ \texorpdfstring{$SU(n)$}{} with one antisymmetric hyper and  \texorpdfstring{$n+8$}{} fundamental hypers}
The F-theory construction involves a split $\text{I}_n$ fiber over a $-1$ curve in the base. The Weierstrass models were already written down in Section \ref{A}. For even $n=2m$, we find the following Calabi-Yau

\begin{align}
\label{eqn:SU2samp1}
\begin{array}{c}
\begin{tikzpicture}[scale=1.5]
\draw  (-3.5,4) ellipse (0.5 and 0.5);
\node (v1) at (-3.5,4) {$0_1$};
\draw  (-2.5,5) ellipse (0.5 and 0.5);
\node (v2) at (-2.5,5) {$1_3$};
\draw  (-1,5) ellipse (0.5 and 0.5);
\node (v4) at (-0.9263,5.1713) {$3_4^1$};
\draw  (-2.5,3) ellipse (0.5 and 0.5);
\node (v3) at (-2.5377,2.7882) {$2_3^2$};
\draw  (-1,3) ellipse (0.5 and 0.5);
\node (v5) at (-1,3) {$4_4^1$};
\draw  (3.4586,4.9793) ellipse (1.4 and 0.5);
\node at (3.6307,5.0449) {$(2m-3)_{m+1}^1$};
\draw  (5.372,3.9893) ellipse (0.9 and 0.6);
\node at (5.2838,4.0004) {$(2m-1)^{2m+8}_{m+3}$};
\draw (-3.2,4.4) -- (-2.9,4.7);
\draw (-3.2,3.6) -- (-2.9,3.3);
\draw (-2,5) -- (-1.5,5);
\draw (-2,3) -- (-1.5,3);
\draw (4.8148,3.5151) -- (4.5472,3.2832);
\draw (-0.1,5) -- (0.1,5) (0.4,5) -- (0.6,5);
\draw (0.9,5) -- (1.1,5) (1.4,5) -- (1.6,5);
\begin{scope}[shift={(-0.0558,-1.9909)}]
\draw (0,5) -- (0.2,5) (0.5,5) -- (0.7,5);
\draw (1,5) -- (1.2,5) (1.5,5) -- (1.7,5);
\end{scope}
\draw  (-3.4053,4.3921) rectangle (-3.2204,4.2073);
\node at (-3.3057,4.3031) {\tiny{1}};
\begin{scope}[shift={(0.5286,0.5394)}]
\draw  (-3.4113,4.3642) rectangle (-3.17,4.1794);
\node at (-3.2917,4.2752) {\tiny{-3}};
\end{scope}
\begin{scope}[shift={(1.1159,0.7257)}]
\draw  (-3.4113,4.3642) rectangle (-3.17,4.1794);
\node at (-3.2917,4.2752) {\tiny{3}};
\end{scope}
\begin{scope}[shift={(1.9717,0.7394)}]
\draw  (-3.4113,4.3642) rectangle (-3.17,4.1794);
\node at (-3.2917,4.2752) {\tiny{-5}};
\end{scope}
\begin{scope}[shift={(2.6153,0.7188)}]
\draw  (-3.4113,4.3642) rectangle (-3.17,4.1794);
\node at (-3.2917,4.2752) {\tiny{4}};
\end{scope}
\begin{scope}[shift={(-0.0137,-0.5476)}]
\draw  (-3.3913,4.3642) rectangle (-3.2064,4.1794);
\node at (-3.2917,4.2752) {\tiny{1}};
\end{scope}
\begin{scope}[shift={(-0.5545,-0.2807)}]
\draw  (-3.3913,4.3642) rectangle (-3.2064,4.1794);
\node at (-3.2917,4.2752) {\tiny{-1}};
\end{scope}
\begin{scope}[shift={(0.5067,-1.1502)}]
\draw  (-3.4113,4.3642) rectangle (-3.18,4.1794);
\node at (-3.2917,4.2752) {\tiny{-3}};
\end{scope}
\begin{scope}[shift={(1.1295,-1.2734)}]
\draw  (-3.4113,4.3642) rectangle (-3.17,4.1794);
\node at (-3.2917,4.2752) {\tiny{2}};
\end{scope}
\begin{scope}[shift={(1.958,-1.2597)}]
\draw  (-3.4113,4.3642) rectangle (-3.17,4.1794);
\node at (-3.2917,4.2752) {\tiny{-4}};
\end{scope}
\begin{scope}[shift={(2.6153,-1.2529)}]
\draw  (-3.4113,4.3642) rectangle (-3.17,4.1794);
\node at (-3.2917,4.2752) {\tiny{3}};
\end{scope}
\begin{scope}[shift={(7.2486,0.4223)}]
\draw  (-3.6509,4.3711) rectangle (-2.9441,4.2);
\node at (-3.2917,4.2752) {\tiny{$m+1$}};
\end{scope}
\begin{scope}[shift={(8.3782,-0.013)}]
\draw  (-3.6509,4.3711) rectangle (-3.0476,4.2282);
\node at (-3.366,4.2929) {\tiny{$-m-3$}};
\end{scope}
\begin{scope}[shift={(8.5247,-0.6573)}]
\draw  (-3.736,4.3764) rectangle (-3.1252,4.2142);
\node at (-3.4526,4.2849) {\tiny{$-m-3$}};
\end{scope}
\begin{scope}[shift={(5.7562,0.6483)}]
\draw  (-3.6509,4.3711) rectangle (-2.9441,4.2);
\node at (-3.2917,4.2752) {\tiny{$-m-2$}};
\end{scope}
\begin{scope}[shift={(0.4434,-2.0286)}]
\draw  (3.0151,5.0088) ellipse (1.4 and 0.5);
\node at (3.1809,4.8747) {$(2m-2)_{m+1}^1$};
\begin{scope}[shift={(7.0448,0.9242)}]
\draw  (-3.6509,4.3711) rectangle (-2.9441,4.2);
\node at (-3.2917,4.2752) {\tiny{$m+1$}};
\end{scope}
\begin{scope}[shift={(5.3469,0.7052)}]
\draw  (-3.6509,4.3711) rectangle (-2.9441,4.2);
\node at (-3.2917,4.2752) {\tiny{$-m-1$}};
\end{scope}
\end{scope}
\node at (-3.1608,4.2855) {\tiny{$h$}};
\node at (-2.1625,5.1528) {\tiny{$h$}};
\node at (-3.1558,3.7385) {\tiny{$h$}};
\node at (-2.1837,2.8313) {\tiny{$h$-$x$}};
\node at (4.399,4.7223) {\tiny{$h$}};
\node at (4.6205,3.1718) {\tiny{$h$}};
\node at (4.7232,3.6563) {\tiny{$e$}};
\node at (-1.3049,2.8215) {\tiny{$e$}};
\node at (-2.7835,2.9631) {\tiny{$e$}};
\node at (-2.7601,4.9723) {\tiny{$e$}};
\node at (-0.6684,5.1438) {\tiny{$h$}};
\node at (2.4878,3.1019) {\tiny{$e$}};
\node at (-3.8578,4.1551) {\tiny{$e$}};
\begin{scope}[shift={(0.7867,0.3557)}]
\draw  (-3.3913,4.3642) rectangle (-3.2064,4.1794);
\node at (-3.2917,4.2752) {\tiny{0}};
\end{scope}
\node at (-2.4975,4.7835) {\tiny{$f$}};
\begin{scope}[shift={(2.3118,0.3634)}]
\draw  (-3.3913,4.3642) rectangle (-3.2064,4.1794);
\node at (-3.2917,4.2752) {\tiny{-1}};
\end{scope}
\begin{scope}[shift={(2.064,0.4663)}]
\draw  (-3.3913,4.3642) rectangle (-3.2064,4.1794);
\node at (-3.2917,4.2752) {\tiny{-1}};
\end{scope}
\node at (-1.385,4.7983) {\tiny{$x$}};
\begin{scope}[shift={(0.7788,-0.9189)}]
\draw  (-3.4113,4.3642) rectangle (-3.17,4.1794);
\node at (-3.2917,4.2752) {\tiny{-2}};
\end{scope}
\begin{scope}[shift={(1.0463,-1.0202)}]
\draw  (-3.3913,4.3642) rectangle (-3.2064,4.1794);
\node at (-3.2917,4.2752) {\tiny{-1}};
\end{scope}
\node at (-2.7835,2.9631) {\tiny{$e$}};
\node at (-2.5237,3.1755) {\tiny{$f$-$x$}};
\node at (-2.4666,3.065) {\tiny{-$y$}};
\node at (-2.093,3.1718) {\tiny{$x$}};
\begin{scope}[shift={(2.2855,-0.91)}]
\draw  (-3.3913,4.3642) rectangle (-3.2064,4.1794);
\node at (-3.2917,4.2752) {\tiny{-1}};
\end{scope}
\begin{scope}[shift={(2.515,-0.9927)}]
\draw  (-3.3913,4.3642) rectangle (-3.2064,4.1794);
\node at (-3.2917,4.2752) {\tiny{-1}};
\end{scope}
\node at (-1.2527,3.2706) {\tiny{$f$-$x$}};
\node at (-0.6244,3.2268) {\tiny{$x$}};
\node at (-0.6814,2.8418) {\tiny{$h$-$x$}};
\begin{scope}[shift={(6.6421,0.3617)}]
\draw  (-3.3913,4.3642) rectangle (-3.2064,4.1794);
\node at (-3.2917,4.2752) {\tiny{-1}};
\end{scope}
\begin{scope}[shift={(6.6649,-0.9309)}]
\draw  (-3.3913,4.3642) rectangle (-3.2064,4.1794);
\node at (-3.2917,4.2752) {\tiny{-1}};
\end{scope}
\begin{scope}[shift={(6.1397,0.4119)}]
\draw  (-3.3913,4.3642) rectangle (-3.2064,4.1794);
\node at (-3.2917,4.2752) {\tiny{-1}};
\end{scope}
\node at (3.3352,4.7989) {\tiny{$f$-$x$}};
\node at (3.3631,3.1684) {\tiny{$f$-$x$}};
\node at (2.6596,4.6838) {\tiny{$x$}};
\node at (-1.2778,5.2008) {\tiny{$e$-$x$}};
\node at (2.4357,5.0912) {\tiny{$e$-$x$}};
\node at (-0.9445,4.8171) {\tiny{$f$-$x$}};
\node at (5.2945,4.4373) {\tiny{$h$+$f$-$\sum x_i$}};
\begin{scope}[shift={(0.5344,-0.1913)}]
\draw  (-3.3913,4.3642) rectangle (-3.2064,4.1794);
\node at (-3.2917,4.2752) {\tiny{1}};
\end{scope}
\begin{scope}[shift={(1.1875,-0.0954)}]
\draw  (-3.3913,4.3642) rectangle (-3.2064,4.1794);
\node at (-3.2917,4.2752) {\tiny{1}};
\end{scope}
\begin{scope}[shift={(1.868,-0.4836)}]
\draw  (-3.3913,4.3642) rectangle (-3.2064,4.1794);
\node at (-3.2917,4.2752) {\tiny{1}};
\end{scope}
\begin{scope}[shift={(2.6947,-0.3009)}]
\draw  (-3.3913,4.3642) rectangle (-3.2064,4.1794);
\node at (-3.2917,4.2752) {\tiny{1}};
\end{scope}
\begin{scope}[shift={(5.1337,0.1924)}]
\draw  (-3.3913,4.3642) rectangle (-3.2064,4.1794);
\node at (-3.2917,4.2752) {\tiny{1}};
\end{scope}
\begin{scope}[shift={(6.0792,-0.2415)}]
\draw  (-3.3913,4.3642) rectangle (-3.2064,4.1794);
\node at (-3.2917,4.2752) {\tiny{1}};
\end{scope}
\begin{scope}[shift={(3.3387,-0.7759)}]
\draw  (-3.3913,4.3642) rectangle (-3.2064,4.1794);
\node at (-3.2917,4.2752) {\tiny{1}};
\end{scope}
\begin{scope}[shift={(7.2896,-0.2826)}]
\draw  (-3.3913,4.3642) rectangle (-3.2064,4.1794);
\node at (-3.2917,4.2752) {\tiny{1}};
\end{scope}
\draw (-2.5045,4.4911) -- (-2.5137,3.4998);
\draw (-2.1694,3.385) -- (-1.3112,4.6105) (-0.9807,4.4911) -- (-0.9991,3.5043) (-0.6411,3.3621) -- (-0.0023,4.0006);
\draw (2.7165,4.5523) -- (1.7208,3.8717) (3.3423,4.4792) -- (3.356,3.4789);
\draw (4.356,4.6016) -- (4.6915,4.3889);
\end{tikzpicture}
\end{array}
\end{align}

and for odd $n=2m+1$ we find

\begin{align}
\begin{array}{c}
\begin{tikzpicture}[scale=1.5]
\draw  (-3.5,4) ellipse (0.5 and 0.5);
\node (v1) at (-3.5,4) {$0_1$};
\draw  (-2.5,5) ellipse (0.5 and 0.5);
\node (v2) at (-2.5,5) {$1_3$};
\draw  (-1,5) ellipse (0.5 and 0.5);
\node (v4) at (-0.9263,5.1713) {$3_4^1$};
\draw  (-2.5,3) ellipse (0.5 and 0.5);
\node (v3) at (-2.5377,2.7882) {$2_3^2$};
\draw  (-1,3) ellipse (0.5 and 0.5);
\node (v5) at (-1,3) {$4_4^1$};
\draw  (3.4586,4.9793) ellipse (1.4 and 0.5);
\node at (3.6307,5.0449) {$(2m-1)_{m+2}^1$};
\draw (-3.2,4.4) -- (-2.9,4.7);
\draw (-3.2,3.6) -- (-2.9,3.3);
\draw (-2,5) -- (-1.5,5);
\draw (-2,3) -- (-1.5,3);
\draw (-0.1,5) -- (0.1,5) (0.4,5) -- (0.6,5);
\draw (0.9,5) -- (1.1,5) (1.4,5) -- (1.6,5);
\begin{scope}[shift={(-0.0558,-1.9909)}]
\draw (0,5) -- (0.2,5) (0.5,5) -- (0.7,5);
\draw (1,5) -- (1.2,5) (1.5,5) -- (1.7,5);
\end{scope}
\draw  (-3.4053,4.3921) rectangle (-3.2204,4.2073);
\node at (-3.3057,4.3031) {\tiny{1}};
\begin{scope}[shift={(0.5286,0.5394)}]
\draw  (-3.4113,4.3642) rectangle (-3.17,4.1794);
\node at (-3.2917,4.2752) {\tiny{-3}};
\end{scope}
\begin{scope}[shift={(1.1159,0.7257)}]
\draw  (-3.4113,4.3642) rectangle (-3.17,4.1794);
\node at (-3.2917,4.2752) {\tiny{3}};
\end{scope}
\begin{scope}[shift={(1.9717,0.7394)}]
\draw  (-3.4113,4.3642) rectangle (-3.17,4.1794);
\node at (-3.2917,4.2752) {\tiny{-5}};
\end{scope}
\begin{scope}[shift={(2.6153,0.7188)}]
\draw  (-3.4113,4.3642) rectangle (-3.17,4.1794);
\node at (-3.2917,4.2752) {\tiny{4}};
\end{scope}
\begin{scope}[shift={(-0.0137,-0.5476)}]
\draw  (-3.3913,4.3642) rectangle (-3.2064,4.1794);
\node at (-3.2917,4.2752) {\tiny{1}};
\end{scope}
\begin{scope}[shift={(-0.5545,-0.2807)}]
\draw  (-3.3913,4.3642) rectangle (-3.2064,4.1794);
\node at (-3.2917,4.2752) {\tiny{-1}};
\end{scope}
\begin{scope}[shift={(0.5067,-1.1502)}]
\draw  (-3.4113,4.3642) rectangle (-3.18,4.1794);
\node at (-3.2917,4.2752) {\tiny{-3}};
\end{scope}
\begin{scope}[shift={(1.1295,-1.2734)}]
\draw  (-3.4113,4.3642) rectangle (-3.17,4.1794);
\node at (-3.2917,4.2752) {\tiny{2}};
\end{scope}
\begin{scope}[shift={(1.958,-1.2597)}]
\draw  (-3.4113,4.3642) rectangle (-3.17,4.1794);
\node at (-3.2917,4.2752) {\tiny{-4}};
\end{scope}
\begin{scope}[shift={(2.6153,-1.2529)}]
\draw  (-3.4113,4.3642) rectangle (-3.17,4.1794);
\node at (-3.2917,4.2752) {\tiny{3}};
\end{scope}
\begin{scope}[shift={(6.7127,0.3364)}]
\draw  (-3.6509,4.3711) rectangle (-2.9441,4.2);
\node at (-3.2917,4.2752) {\tiny{$m+3$}};
\end{scope}
\begin{scope}[shift={(5.7562,0.6483)}]
\draw  (-3.6509,4.3711) rectangle (-2.9441,4.2);
\node at (-3.2917,4.2752) {\tiny{$-m-3$}};
\end{scope}
\begin{scope}[shift={(0.4434,-2.0286)}]
\draw  (3.0151,5.0088) ellipse (1.4 and 0.5);
\node at (3.1103,4.8758) {$2m_{m+2}^{2m+9}$};
\begin{scope}[shift={(6.2021,1.0878)}]
\draw  (-3.6509,4.3711) rectangle (-2.9441,4.2);
\node at (-3.2917,4.2752) {\tiny{$-m-5$}};
\end{scope}
\begin{scope}[shift={(5.3469,0.7052)}]
\draw  (-3.6509,4.3711) rectangle (-2.9441,4.2);
\node at (-3.2917,4.2752) {\tiny{$-m-2$}};
\end{scope}
\end{scope}
\node at (-3.1608,4.2855) {\tiny{$h$}};
\node at (-2.1625,5.1528) {\tiny{$h$}};
\node at (-3.1558,3.7385) {\tiny{$h$}};
\node at (-2.1837,2.8313) {\tiny{$h$-$x$}};
\node at (-1.3049,2.8215) {\tiny{$e$}};
\node at (-2.7835,2.9631) {\tiny{$e$}};
\node at (-2.7601,4.9723) {\tiny{$e$}};
\node at (-0.6684,5.1438) {\tiny{$h$}};
\node at (2.4878,3.1019) {\tiny{$e$}};
\node at (-3.8578,4.1551) {\tiny{$e$}};
\begin{scope}[shift={(0.7867,0.3557)}]
\draw  (-3.3913,4.3642) rectangle (-3.2064,4.1794);
\node at (-3.2917,4.2752) {\tiny{0}};
\end{scope}
\node at (-2.4975,4.7835) {\tiny{$f$}};
\begin{scope}[shift={(2.3118,0.3634)}]
\draw  (-3.3913,4.3642) rectangle (-3.2064,4.1794);
\node at (-3.2917,4.2752) {\tiny{-1}};
\end{scope}
\begin{scope}[shift={(2.064,0.4663)}]
\draw  (-3.3913,4.3642) rectangle (-3.2064,4.1794);
\node at (-3.2917,4.2752) {\tiny{-1}};
\end{scope}
\node at (-1.385,4.7983) {\tiny{$x$}};
\begin{scope}[shift={(0.7788,-0.9189)}]
\draw  (-3.4113,4.3642) rectangle (-3.17,4.1794);
\node at (-3.2917,4.2752) {\tiny{-2}};
\end{scope}
\begin{scope}[shift={(1.0463,-1.0202)}]
\draw  (-3.3913,4.3642) rectangle (-3.2064,4.1794);
\node at (-3.2917,4.2752) {\tiny{-1}};
\end{scope}
\node at (-2.7835,2.9631) {\tiny{$e$}};
\node at (-2.5237,3.1755) {\tiny{$f$-$x$}};
\node at (-2.4666,3.065) {\tiny{-$y$}};
\node at (-2.093,3.1718) {\tiny{$x$}};
\begin{scope}[shift={(2.2855,-0.91)}]
\draw  (-3.3913,4.3642) rectangle (-3.2064,4.1794);
\node at (-3.2917,4.2752) {\tiny{-1}};
\end{scope}
\begin{scope}[shift={(2.515,-0.9927)}]
\draw  (-3.3913,4.3642) rectangle (-3.2064,4.1794);
\node at (-3.2917,4.2752) {\tiny{-1}};
\end{scope}
\node at (-1.2527,3.2706) {\tiny{$f$-$x$}};
\node at (-0.6244,3.2268) {\tiny{$x$}};
\node at (-0.6814,2.8418) {\tiny{$h$-$x$}};
\begin{scope}[shift={(6.1397,0.4119)}]
\draw  (-3.3913,4.3642) rectangle (-3.2064,4.1794);
\node at (-3.2917,4.2752) {\tiny{-1}};
\end{scope}
\node at (3.3846,4.7929) {\tiny{$h$+$f$-$x$}};
\node at (3.3631,3.1684) {\tiny{$h$+$f$-$\sum x_i$}};
\node at (2.6596,4.6838) {\tiny{$x$}};
\node at (-1.2778,5.2008) {\tiny{$e$-$x$}};
\node at (2.4357,5.0912) {\tiny{$e$-$x$}};
\node at (-0.9445,4.8171) {\tiny{$f$-$x$}};
\begin{scope}[shift={(0.5344,-0.1913)}]
\draw  (-3.3913,4.3642) rectangle (-3.2064,4.1794);
\node at (-3.2917,4.2752) {\tiny{1}};
\end{scope}
\begin{scope}[shift={(1.1875,-0.0954)}]
\draw  (-3.3913,4.3642) rectangle (-3.2064,4.1794);
\node at (-3.2917,4.2752) {\tiny{1}};
\end{scope}
\begin{scope}[shift={(1.868,-0.4836)}]
\draw  (-3.3913,4.3642) rectangle (-3.2064,4.1794);
\node at (-3.2917,4.2752) {\tiny{1}};
\end{scope}
\begin{scope}[shift={(2.6947,-0.3009)}]
\draw  (-3.3913,4.3642) rectangle (-3.2064,4.1794);
\node at (-3.2917,4.2752) {\tiny{1}};
\end{scope}
\begin{scope}[shift={(5.1337,0.1924)}]
\draw  (-3.3913,4.3642) rectangle (-3.2064,4.1794);
\node at (-3.2917,4.2752) {\tiny{1}};
\end{scope}
\begin{scope}[shift={(6.0792,-0.2415)}]
\draw  (-3.3913,4.3642) rectangle (-3.2064,4.1794);
\node at (-3.2917,4.2752) {\tiny{1}};
\end{scope}
\begin{scope}[shift={(3.3387,-0.7759)}]
\draw  (-3.3913,4.3642) rectangle (-3.2064,4.1794);
\node at (-3.2917,4.2752) {\tiny{1}};
\end{scope}
\draw (-2.5045,4.4911) -- (-2.5137,3.4998);
\draw (-2.1694,3.385) -- (-1.3112,4.6105) (-0.9807,4.4911) -- (-0.9991,3.5043) (-0.6411,3.3621) -- (-0.0023,4.0006);
\draw (2.7165,4.5523) -- (1.7208,3.8717) (3.3423,4.4792) -- (3.356,3.4789);
\end{tikzpicture}
\end{array}
\end{align}

For the degenerate case of $SU(3)$ we have

\begin{align}
\begin{array}{c}
\begin{tikzpicture}[scale=1.5]
\begin{scope}[shift={(2.4713,0.0223)}]
\draw  (-4.4603,4.9901) ellipse (1 and 0.5);
\node (v1) at (-4.507,4.9528) {$0_{1}$};
\draw (-3.4464,5.0091) -- (-2.9847,5.0073);
\begin{scope}[shift={(-1.8354,0.7134)}]
\draw  (-3.5321,4.3845) rectangle (-3.0504,4.1716);
\node at (-3.2917,4.2752) {\tiny{$-1$}};
\end{scope}
\begin{scope}[shift={(-0.5642,0.7057)}]
\draw  (-3.6392,4.4011) rectangle (-2.9855,4.1922);
\node at (-3.2901,4.2995) {\tiny{$1$}};
\end{scope}
\begin{scope}[shift={(-1.1756,1.0455)}]
\draw  (-3.5321,4.3845) rectangle (-3.0504,4.1716);
\node at (-3.2917,4.2752) {\tiny{$1$}};
\end{scope}
\node at (-4.0918,5.3074) {\tiny{$h$}};
\end{scope}
\begin{scope}[shift={(5.1361,0.0036)}]
\draw  (-4.4603,4.9901) ellipse (1.2 and 0.5);
\node (v1) at (-4.4811,4.879) {$2_{3}^{12}$};
\node at (-5.3519,5.2155) {\tiny{$e$}};
\begin{scope}[shift={(-2.1142,0.7232)}]
\draw  (-3.4342,4.3914) rectangle (-3.1098,4.1794);
\node at (-3.2917,4.2752) {\tiny{$-3$}};
\end{scope}
\begin{scope}[shift={(-1.3924,1.0617)}]
\draw  (-3.4542,4.367) rectangle (-3.1597,4.1724);
\node at (-3.308,4.2726) {\tiny{$-7$}};
\end{scope}
\node at (-4.1605,5.3287) {\tiny{$h$+$f$-$\sum x_i$}};
\end{scope}
\node at (-2.6424,5.2031) {\tiny{$e$}};
\node at (-1.423,4.7873) {\tiny{$h$}};
\begin{scope}[shift={(2.4199,1.5062)}]
\draw  (-4.4603,4.9901) ellipse (1 and 0.5);
\node (v1) at (-4.5578,4.9971) {$1_{3}$};
\node at (-3.8099,5.0963) {\tiny{$h$+$f$}};
\node at (-4.7699,4.6439) {\tiny{$e$}};
\begin{scope}[shift={(-0.5576,0.611)}]
\draw  (-3.5896,4.3807) rectangle (-3.0131,4.1849);
\node at (-3.292,4.2758) {\tiny{$5$}};
\end{scope}
\begin{scope}[shift={(-1.1411,0.3576)}]
\draw  (-3.5352,4.3891) rectangle (-3.0397,4.1868);
\node at (-3.2901,4.2995) {\tiny{$-3$}};
\end{scope}
\end{scope}
\begin{scope}[shift={(2.4325,1.2108)}]
\draw  (-3.6392,4.4011) rectangle (-2.9855,4.1922);
\node at (-3.2901,4.2995) {\tiny{$1$}};
\end{scope}
\draw (-2.019,5.9932) -- (-2.019,5.5182);
\draw (-1.1458,6.2729) -- (0.2707,5.464);
\end{tikzpicture}
\end{array}
\end{align}

\subsection{ \texorpdfstring{$SU(6)$}{} with half-hyper in three-index antisymmetric and \texorpdfstring{$15$}{} fundamental hypers}

The F-theory construction of this theory involves an alternate tuning of a split $\text{I}_{2n}$ fiber over a $-1$ curve \cite{Huang:2018gpl}. The Weierstrass model for general $n$ is defined by 
	\begin{align}
		a_1 = a_1,~~ a_2 = e_0^2 a_{2,2},~~ a_{3} = a_{3,n-1} e_0^{n-1} ,~~ a_{4} = a_{4,n+1} e_0^{n+1},~~ a_6 = a_{6,2n} e_0^{2n}.
	\end{align}
For $n=3$, which is the only case of interest for $6d$ SCFTs, we consider the resolution
	\begin{align}
			(x,y,e_0|e_1),~~ (y,e_1|e_2),~~ (x,e_2|e_3),~~(y,e_3|e_4),~~(y,e_4|e_5).
	\end{align}
and we identify the geometry to be
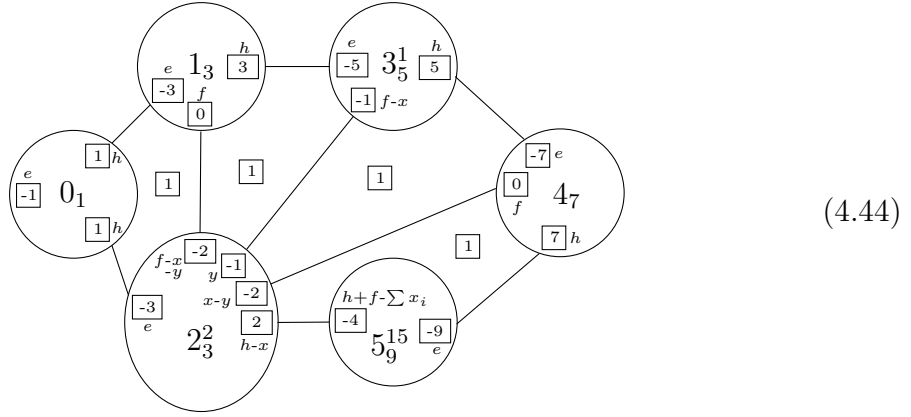
\begin{align}
\begin{array}{c}
\begin{tikzpicture}[scale=1.7]
\draw  (-3.5,4) ellipse (0.5 and 0.5);
\node (v1) at (-3.5,4) {$0_1$};
\draw  (-2.5,5) ellipse (0.5 and 0.5);
\node (v2) at (-2.5,5) {$1_3$};
\draw  (-1,5) ellipse (0.5 and 0.5);
\node (v4) at (-0.9827,5.0196) {$3_5^1$};
\draw  (-2.5,3) ellipse (0.6 and 0.7);
\node (v3) at (-2.5056,2.8479) {$2_3^2$};
\draw  (-1,3) ellipse (0.5 and 0.5);
\node (v5) at (-1.025,2.8104) {$5_9^{15}$};
\draw  (0.3058,4.0117) ellipse (0.5 and 0.5);
\node at (0.3576,3.9889) {$4_7$};
\draw (-3.2,4.4) -- (-2.9,4.7);
\draw (-3.2,3.6) -- (-3.0699,3.2091);
\draw (-2,5) -- (-1.5,5);
\draw (-1.9057,2.997) -- (-1.5,3);
\begin{scope}[shift={(0.7818,-0.7181)}]
\draw  (-3.4113,4.3642) rectangle (-3.17,4.1794);
\node at (-3.2917,4.2752) {\tiny{-2}};
\end{scope}
\draw  (-3.4053,4.3921) rectangle (-3.2204,4.2073);
\node at (-3.3057,4.3031) {\tiny{1}};
\begin{scope}[shift={(0.5286,0.5394)}]
\draw  (-3.4113,4.3642) rectangle (-3.17,4.1794);
\node at (-3.2917,4.2752) {\tiny{-3}};
\end{scope}
\begin{scope}[shift={(1.1159,0.7257)}]
\draw  (-3.4113,4.3642) rectangle (-3.17,4.1794);
\node at (-3.2917,4.2752) {\tiny{3}};
\end{scope}
\begin{scope}[shift={(1.9717,0.7394)}]
\draw  (-3.4113,4.3642) rectangle (-3.17,4.1794);
\node at (-3.2917,4.2752) {\tiny{-5}};
\end{scope}
\begin{scope}[shift={(2.6153,0.7188)}]
\draw  (-3.4113,4.3642) rectangle (-3.17,4.1794);
\node at (-3.2917,4.2752) {\tiny{5}};
\end{scope}
\begin{scope}[shift={(-0.0137,-0.5476)}]
\draw  (-3.3913,4.3642) rectangle (-3.2064,4.1794);
\node at (-3.2917,4.2752) {\tiny{1}};
\end{scope}
\begin{scope}[shift={(-0.5545,-0.2807)}]
\draw  (-3.3913,4.3642) rectangle (-3.2064,4.1794);
\node at (-3.2917,4.2752) {\tiny{-1}};
\end{scope}
\begin{scope}[shift={(0.3718,-1.1552)}]
\draw  (-3.4113,4.3642) rectangle (-3.18,4.1794);
\node at (-3.2917,4.2752) {\tiny{-3}};
\end{scope}
\begin{scope}[shift={(1.2238,-1.2764)}]
\draw  (-3.4113,4.3642) rectangle (-3.17,4.1794);
\node at (-3.2917,4.2752) {\tiny{2}};
\end{scope}
\begin{scope}[shift={(1.958,-1.2597)}]
\draw  (-3.4113,4.3642) rectangle (-3.17,4.1794);
\node at (-3.2917,4.2752) {\tiny{-4}};
\end{scope}
\begin{scope}[shift={(2.6188,-1.342)}]
\draw  (-3.4113,4.3642) rectangle (-3.17,4.1794);
\node at (-3.2917,4.2752) {\tiny{-9}};
\end{scope}
\node at (-3.1608,4.2855) {\tiny{$h$}};
\node at (-2.1625,5.1528) {\tiny{$h$}};
\node at (-3.1558,3.7385) {\tiny{$h$}};
\node at (-2.0843,2.828) {\tiny{$h$-$x$}};
\node at (-0.0276,3.8905) {\tiny{$f$}};
\node at (0.4264,3.6659) {\tiny{$h$}};
\node at (0.2942,4.3073) {\tiny{$e$}};
\node at (-2.7601,4.9723) {\tiny{$e$}};
\node at (-0.6607,5.1664) {\tiny{$h$}};
\node at (-3.8578,4.1551) {\tiny{$e$}};
\begin{scope}[shift={(0.7867,0.3557)}]
\draw  (-3.3913,4.3642) rectangle (-3.2064,4.1794);
\node at (-3.2917,4.2752) {\tiny{0}};
\end{scope}
\node at (-2.4975,4.7835) {\tiny{$f$}};
\begin{scope}[shift={(2.064,0.4663)}]
\draw  (-3.3913,4.3642) rectangle (-3.2064,4.1794);
\node at (-3.2917,4.2752) {\tiny{-1}};
\end{scope}
\begin{scope}[shift={(1.1834,-1.0477)}]
\draw  (-3.4113,4.3642) rectangle (-3.17,4.1794);
\node at (-3.2917,4.2752) {\tiny{-2}};
\end{scope}
\begin{scope}[shift={(1.0492,-0.8312)}]
\draw  (-3.3913,4.3642) rectangle (-3.2064,4.1794);
\node at (-3.2917,4.2752) {\tiny{-1}};
\end{scope}
\node at (-2.9184,2.9581) {\tiny{$e$}};
\node at (-2.7588,3.4811) {\tiny{$f$-$x$}};
\node at (-2.715,3.3757) {\tiny{-$y$}};
\node at (-2.4079,3.3499) {\tiny{$y$}};
\node at (-2.3781,3.1478) {\tiny{$x$-$y$}};
\begin{scope}[shift={(3.43,0.0318)}]
\draw  (-3.3913,4.3642) rectangle (-3.2064,4.1794);
\node at (-3.2917,4.2752) {\tiny{-7}};
\end{scope}
\begin{scope}[shift={(3.5558,-0.6124)}]
\draw  (-3.3913,4.3642) rectangle (-3.2064,4.1794);
\node at (-3.2917,4.2752) {\tiny{7}};
\end{scope}
\node at (-0.6601,2.7787) {\tiny{$e$}};
\node at (-1.0658,3.1864) {\tiny{$h$+$f$-$\sum x_i$}};
\node at (-1.3237,5.1789) {\tiny{$e$}};
\node at (-0.9942,4.7211) {\tiny{$f$-$x$}};
\begin{scope}[shift={(0.5344,-0.1913)}]
\draw  (-3.3913,4.3642) rectangle (-3.2064,4.1794);
\node at (-3.2917,4.2752) {\tiny{1}};
\end{scope}
\begin{scope}[shift={(1.1875,-0.0954)}]
\draw  (-3.3913,4.3642) rectangle (-3.2064,4.1794);
\node at (-3.2917,4.2752) {\tiny{1}};
\end{scope}
\begin{scope}[shift={(2.1945,-0.1439)}]
\draw  (-3.3913,4.3642) rectangle (-3.2064,4.1794);
\node at (-3.2917,4.2752) {\tiny{1}};
\end{scope}
\begin{scope}[shift={(3.2583,-0.1941)}]
\draw  (-3.3913,4.3642) rectangle (-3.2064,4.1794);
\node at (-3.2917,4.2752) {\tiny{0}};
\end{scope}
\begin{scope}[shift={(2.882,-0.68)}]
\draw  (-3.3913,4.3642) rectangle (-3.2064,4.1794);
\node at (-3.2917,4.2752) {\tiny{1}};
\end{scope}
\draw (-2.5045,4.4911) -- (-2.5107,3.7006);
\draw (-2.1452,3.571) -- (-1.3112,4.6105);
\draw (-0.5158,4.9219) -- (0.0237,4.4304) (-0.198,4.0458) -- (-1.9581,3.298) (-0.5042,2.9842) -- (0.1412,3.5357);
\end{tikzpicture}
\end{array}
\end{align}

\subsection{ \texorpdfstring{$SO(7)$}{} with  \texorpdfstring{$8-2k$}{} spinor hypers and \texorpdfstring{$3-k$}{} fundamental hypers}
The zero mass parameter resolutions, geometry of the fibral divisors, and corresponding triple intersection numbers of the $SO(7)$-model were studied in \cite{Esole:2017qeh}. The F-theory construction involves a semi-split $\text{I}^*_0$ fiber over a $-k$ curve in the base, where $1\le k\le3$. The Weierstrass model is defined by
	\begin{align}
		a_1 = 0,~~ a_2 = a_{2,1} e_0,~~ a_3 = 0,~~ a_4 = a_{4,2} e_0^2,~~ a_6 = a_{6,4}e_0^4.
	\end{align}
We consider the resolution
	\begin{align}
		(x,y,e_0|e_1),~~ (y,e_1|e_2),~~(x,e_2|e_3),
	\end{align}
and the corresponding geometry is described by

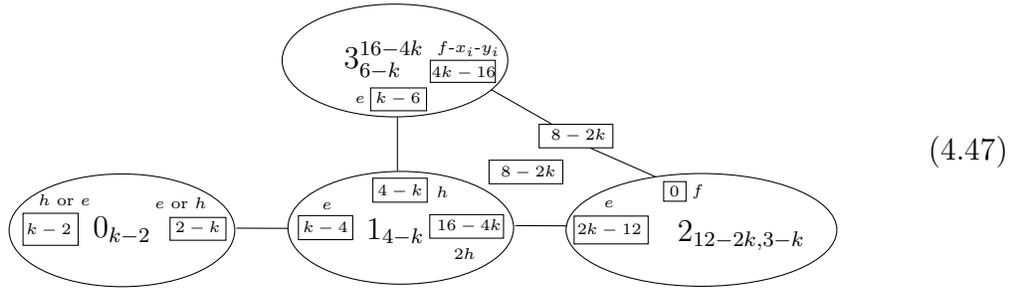
\begin{align}
\begin{array}{c}
\begin{tikzpicture}[scale=1.5]
\draw  (-4.4603,4.9901) ellipse (1 and 0.5);
\node (v1) at (-4.4603,4.9901) {$0_{k-2}$};
\draw (-3.4464,5.0091) -- (-2.9847,5.0073);
\node at (-3.937,5.2302) {\tiny{$e$ or $h$}};
\node at (-4.9686,5.2627) {\tiny{$h$ or $e$}};
\begin{scope}[shift={(-1.8167,0.7227)}]
\draw  (-3.5199,4.4183) rectangle (-3.0317,4.1342);
\node at (-3.2917,4.2752) {\tiny{$k-2$}};
\end{scope}
\begin{scope}[shift={(-0.4988,0.715)}]
\draw  (-3.5352,4.3891) rectangle (-3.0397,4.1868);
\node at (-3.2901,4.2995) {\tiny{$2-k$}};
\end{scope}
\begin{scope}[shift={(2.4713,0.0223)}]
\draw  (-4.4603,4.9901) ellipse (1 and 0.5);
\node (v1) at (-4.507,4.9528) {$1_{4-k}$};
\draw (-3.4464,5.0091) -- (-2.9847,5.0073);
\begin{scope}[shift={(-1.8354,0.7134)}]
\draw  (-3.5321,4.3845) rectangle (-3.0504,4.1716);
\node at (-3.2917,4.2752) {\tiny{$k-4$}};
\end{scope}
\begin{scope}[shift={(-0.5642,0.7057)}]
\draw  (-3.6392,4.4011) rectangle (-2.9855,4.1922);
\node at (-3.2901,4.2995) {\tiny{$16-4k$}};
\end{scope}
\begin{scope}[shift={(-1.1756,1.0455)}]
\draw  (-3.5321,4.3845) rectangle (-3.0504,4.1716);
\node at (-3.2917,4.2752) {\tiny{$4-k$}};
\end{scope}
\node at (-4.0918,5.3074) {\tiny{$h$}};
\end{scope}
\begin{scope}[shift={(5.1361,0.0036)}]
\draw  (-4.4603,4.9901) ellipse (1.2 and 0.5);
\node (v1) at (-4.108,4.9424) {$2_{12-2k,3-k}$};
\node at (-5.2811,5.2179) {\tiny{$e$}};
\begin{scope}[shift={(-1.9849,0.7134)}]
\draw  (-3.6041,4.3993) rectangle (-2.9476,4.1525);
\node at (-3.2917,4.2752) {\tiny{$2k-12$}};
\end{scope}
\begin{scope}[shift={(-1.3924,1.0617)}]
\draw  (-3.4027,4.3566) rectangle (-3.2043,4.1827);
\node at (-3.308,4.2726) {\tiny{$0$}};
\end{scope}
\node at (-4.4923,5.3311) {\tiny{$f$}};
\end{scope}
\node at (-2.6424,5.2031) {\tiny{$e$}};
\node at (-1.423,4.7873) {\tiny{$2h$}};
\begin{scope}[shift={(2.4199,1.5062)}]
\draw  (-4.4603,4.9901) ellipse (1 and 0.5);
\node (v1) at (-4.5578,4.9971) {$3_{6-k}^{16-4k}$};
\node at (-3.8099,5.0963) {\tiny{$f$-$x_i$-$y_i$}};
\node at (-4.7699,4.6439) {\tiny{$e$}};
\begin{scope}[shift={(-0.5576,0.611)}]
\draw  (-3.5896,4.3807) rectangle (-3.0131,4.1849);
\node at (-3.292,4.2758) {\tiny{$4k-16$}};
\end{scope}
\begin{scope}[shift={(-1.1411,0.3576)}]
\draw  (-3.5352,4.3891) rectangle (-3.0397,4.1868);
\node at (-3.2901,4.2995) {\tiny{$k-6$}};
\end{scope}
\end{scope}
\begin{scope}[shift={(2.8763,1.5294)}]
\draw  (-3.6392,4.4011) rectangle (-2.9855,4.1922);
\node at (-3.2901,4.2995) {\tiny{$8-2k$}};
\end{scope}
\begin{scope}[shift={(2.4325,1.2108)}]
\draw  (-3.6392,4.4011) rectangle (-2.9855,4.1922);
\node at (-3.2901,4.2995) {\tiny{$8-2k$}};
\end{scope}
\draw (-2.019,5.9932) -- (-2.019,5.5182);
\draw (-0.3102,5.7204) -- (0.2707,5.464);
\draw (-0.6531,5.9337) -- (-1.1906,6.2395);
\end{tikzpicture}
\end{array}
\end{align}

\subsection{ \texorpdfstring{$SO(8)$}{} with  \texorpdfstring{$4-k$}{} fundamentals,  \texorpdfstring{$4-k$}{} spinors and  \texorpdfstring{$4-k$}{} conjugate spinors}
The zero mass parameter resolutions, geometry of the fibral divisors, and corresponding triple intersection numbers of the $SO(8)$-model were studied in \cite{Esole:2017qeh}. The F-theory construction involves a split $\text{I}^*_0$ fiber over a $-k$ curve in the base, where $1\le k\le4$. We use the Weierstrass model defined by the following orders of vanishing:
	\begin{align}
		a_1 = a_{1,1} e_0,~~ a_2=a_{2,1} e_0,~~ a_3 = a_{3,2} e_0^2,~~ a_4 = a_{4,2} e_0^2 ,~~ a_6 = a_{6,2} e_0^4,
	\end{align}
where we also impose the split condition
	\begin{align}
		4 a_{4,2} - a_{2,1}^2 = \alpha^2	.
	\end{align}
We consider the resolution
	\begin{align}
		(x,y,e_0|e_1),~~(y,e_1|e_2) ,~~(x ,e_2|e_3),~~ (e_0 \alpha_2^{\pm{}} z+ 2 e_3 x,e_2 | e_4)
	\end{align}
where $\alpha_2^{\pm{}} = \alpha \pm{} a_2$. The geometry is 

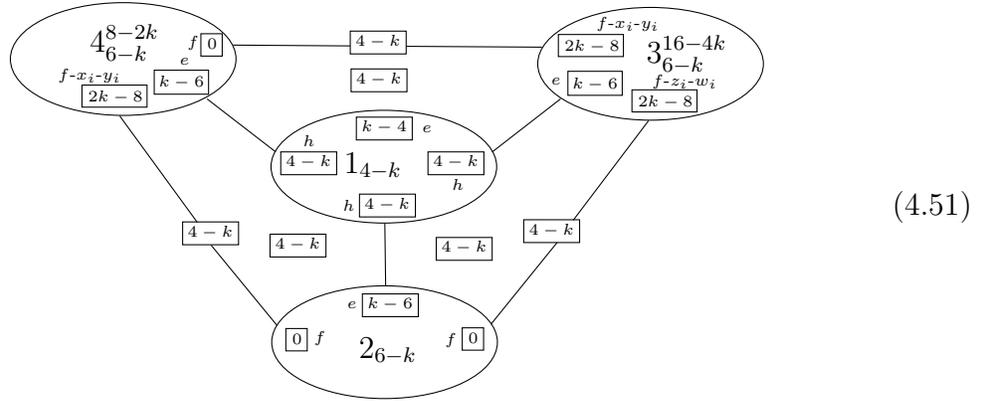
\begin{align}
\begin{array}{c}
\begin{tikzpicture}[scale=1.5]
\draw  (-4.4603,4.9901) ellipse (1 and 0.5);
\node (v1) at (-4.4475,5.1322) {$4_{6-k}^{8-2k}$};
\node at (-3.84,5.1334) {\tiny{$f$}};
\begin{scope}[shift={(-0.3826,0.8248)}]
\draw  (-3.3893,4.3897) rectangle (-3.2001,4.1915);
\node at (-3.2901,4.2995) {\tiny{$0$}};
\end{scope}
\begin{scope}[shift={(4.674,-0.0423)}]
\draw  (-4.4603,4.9901) ellipse (1 and 0.5);
\node (v1) at (-4.1388,5.0769) {$3_{6-k}^{16-4k}$};
\begin{scope}[shift={(-1.0356,0.3784)}]
\draw  (-3.5896,4.3807) rectangle (-3.0131,4.1849);
\node at (-3.292,4.2758) {\tiny{$2k-8$}};
\end{scope}
\begin{scope}[shift={(-1.6879,0.8693)}]
\draw  (-3.5896,4.3807) rectangle (-3.0131,4.1849);
\node at (-3.292,4.2758) {\tiny{$2k-8$}};
\end{scope}
\begin{scope}[shift={(-1.6609,0.539)}]
\draw  (-3.5321,4.3845) rectangle (-3.0504,4.1716);
\node at (-3.2917,4.2752) {\tiny{$k-6$}};
\end{scope}
\node at (-4.6626,5.3455) {\tiny{$f$-$x_i$-$y_i$}};
\node at (-4.1588,4.8352) {\tiny{$f$-$z_i$-$w_i$}};
\node at (-5.2977,4.8371) {\tiny{$e$}};
\end{scope}
\begin{scope}[shift={(2.3165,-2.5115)}]
\draw  (-4.4603,4.9901) ellipse (1 and 0.5);
\node (v1) at (-4.4317,4.9136) {$2_{6-k}$};
\node at (-4.7493,5.3185) {\tiny{$e$}};
\begin{scope}[shift={(-1.1217,1.0294)}]
\draw  (-3.5352,4.3891) rectangle (-3.0397,4.1868);
\node at (-3.2901,4.2995) {\tiny{$k-6$}};
\end{scope}
\begin{scope}[shift={(-0.3826,0.7279)}]
\draw  (-3.3893,4.3897) rectangle (-3.2001,4.1915);
\node at (-3.2901,4.2995) {\tiny{$0$}};
\end{scope}
\begin{scope}[shift={(-1.9458,0.7215)}]
\draw  (-3.3893,4.3897) rectangle (-3.2001,4.1915);
\node at (-3.2901,4.2995) {\tiny{$0$}};
\end{scope}
\node at (-3.8721,5.0056) {\tiny{$f$}};
\node at (-5.0347,5.025) {\tiny{$f$}};
\end{scope}
\begin{scope}[shift={(2.3101,-0.9548)}]
\draw  (-4.4603,4.9901) ellipse (1 and 0.5);
\node (v1) at (-4.5578,4.9971) {$1_{4-k}$};
\node at (-4.7699,4.6439) {\tiny{$h$}};
\begin{scope}[shift={(-1.1411,0.3576)}]
\draw  (-3.5352,4.3891) rectangle (-3.0397,4.1868);
\node at (-3.2901,4.2995) {\tiny{$4-k$}};
\end{scope}
\begin{scope}[shift={(-1.8387,0.7322)}]
\draw  (-3.5352,4.3891) rectangle (-3.0397,4.1868);
\node at (-3.2901,4.2995) {\tiny{$4-k$}};
\end{scope}
\begin{scope}[shift={(-0.5339,0.7322)}]
\draw  (-3.5352,4.3891) rectangle (-3.0397,4.1868);
\node at (-3.2901,4.2995) {\tiny{$4-k$}};
\end{scope}
\node at (-5.1241,5.2216) {\tiny{$h$}};
\node at (-3.8037,4.8291) {\tiny{$h$}};
\begin{scope}[shift={(-1.1688,1.0404)}]
\draw  (-3.5352,4.3891) rectangle (-3.0397,4.1868);
\node at (-3.2901,4.2995) {\tiny{$k-4$}};
\end{scope}
\node at (-4.0789,5.3238) {\tiny{$e$}};
\end{scope}
\begin{scope}[shift={(-1.2359,0.3784)}]
\draw  (-3.5896,4.3807) rectangle (-3.0131,4.1849);
\node at (-3.292,4.2758) {\tiny{$2k-8$}};
\end{scope}
\node at (-4.7594,4.8445) {\tiny{$f$-$x_i$-$y_i$}};
\begin{scope}[shift={(-0.6533,0.5067)}]
\draw  (-3.5321,4.3845) rectangle (-3.0504,4.1716);
\node at (-3.2917,4.2752) {\tiny{$k-6$}};
\end{scope}
\node at (-3.9176,4.9718) {\tiny{$e$}};
\begin{scope}[shift={(-0.3983,-0.8438)}]
\draw  (-3.5352,4.3891) rectangle (-3.0397,4.1868);
\node at (-3.2901,4.2995) {\tiny{$4-k$}};
\end{scope}
\begin{scope}[shift={(2.6246,-0.8244)}]
\draw  (-3.5352,4.3891) rectangle (-3.0397,4.1868);
\node at (-3.2901,4.2995) {\tiny{$4-k$}};
\end{scope}
\begin{scope}[shift={(1.0873,0.8486)}]
\draw  (-3.5352,4.3891) rectangle (-3.0397,4.1868);
\node at (-3.2901,4.2995) {\tiny{$4-k$}};
\end{scope}
\begin{scope}[shift={(1.0946,0.5132)}]
\draw  (-3.5352,4.3891) rectangle (-3.0397,4.1868);
\node at (-3.2901,4.2995) {\tiny{$4-k$}};
\end{scope}
\begin{scope}[shift={(0.3763,-0.9525)}]
\draw  (-3.5352,4.3891) rectangle (-3.0397,4.1868);
\node at (-3.2901,4.2995) {\tiny{$4-k$}};
\end{scope}
\begin{scope}[shift={(1.8494,-0.9817)}]
\draw  (-3.5352,4.3891) rectangle (-3.0397,4.1868);
\node at (-3.2901,4.2995) {\tiny{$4-k$}};
\end{scope}
\draw (-3.4869,5.1137) -- (-2.4478,5.1064) (-1.9519,5.1028) -- (-0.7487,5.0955);
\draw (-0.5846,4.6361) -- (-1.1899,4.1658) (-3.7166,4.6361) -- (-3.1077,4.1658) (-4.4912,4.4856) -- (-3.8042,3.5423) (-3.6802,3.3381) -- (-3.0968,2.6271) (-1.2008,2.638) -- (-0.6138,3.3636) (-0.4716,3.5605) -- (0.1956,4.4429);
\draw (-2.1415,3.5313) -- (-2.1342,2.9771);
\end{tikzpicture}
\end{array}
\end{align}

where we have split the $16-4k$ blow-ups on $S_3$ into four equal pieces denoted by $x_i$, $y_i$, $z_i$ and $w_i$. We have also chosen to hide the affine node $S_0$ so that the diagram would be planar. $S_0$ has degree $2-k$ and no blow-ups, and is attached to $S_1$ along the curve $e$ in $S_1$. This matches the answer of \cite{DelZotto:2017pti} for $k=4$.

\subsection{ \texorpdfstring{$SO(9)$}{} with  \texorpdfstring{$4-k$}{} spinor hypers and  \texorpdfstring{$5-k$}{} fundamental hypers}
The F-theory construction involves a non-split $\text{I}^*_1$ fiber over a $-k$ curve in the base, where $1\le k\le4$. The Weierstrass model appears in Section \ref{D}. The geometry is

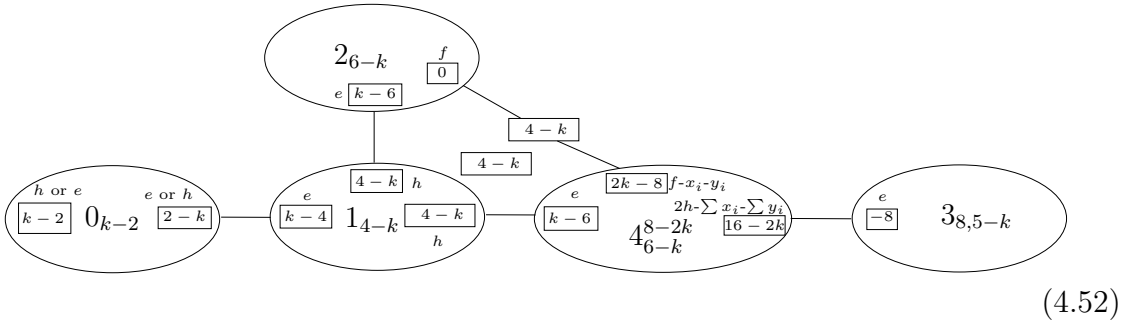
\begin{align}
\scalebox{.95}{$
\begin{array}{c}
\begin{tikzpicture}[scale=1.5]
\draw  (-4.4603,4.9901) ellipse (1 and 0.5);
\node (v1) at (-4.4603,4.9901) {$0_{k-2}$};
\draw (-3.4464,5.0091) -- (-2.9847,5.0073);
\node at (-3.937,5.2302) {\tiny{$e$ or $h$}};
\node at (-4.9686,5.2627) {\tiny{$h$ or $e$}};
\begin{scope}[shift={(-1.8167,0.7227)}]
\draw  (-3.5199,4.4183) rectangle (-3.0317,4.1342);
\node at (-3.2917,4.2752) {\tiny{$k-2$}};
\end{scope}
\begin{scope}[shift={(-0.4988,0.715)}]
\draw  (-3.5352,4.3891) rectangle (-3.0397,4.1868);
\node at (-3.2901,4.2995) {\tiny{$2-k$}};
\end{scope}
\begin{scope}[shift={(2.4713,0.0223)}]
\draw  (-4.4603,4.9901) ellipse (1 and 0.5);
\node (v1) at (-4.507,4.9528) {$1_{4-k}$};
\draw (-3.4464,5.0091) -- (-2.9847,5.0073);
\begin{scope}[shift={(-1.8354,0.7134)}]
\draw  (-3.5321,4.3845) rectangle (-3.0504,4.1716);
\node at (-3.2917,4.2752) {\tiny{$k-4$}};
\end{scope}
\begin{scope}[shift={(-0.5642,0.7057)}]
\draw  (-3.6392,4.4011) rectangle (-2.9855,4.1922);
\node at (-3.2901,4.2995) {\tiny{$4-k$}};
\end{scope}
\begin{scope}[shift={(-1.1756,1.0455)}]
\draw  (-3.5321,4.3845) rectangle (-3.0504,4.1716);
\node at (-3.2917,4.2752) {\tiny{$4-k$}};
\end{scope}
\node at (-4.0918,5.3074) {\tiny{$h$}};
\end{scope}
\begin{scope}[shift={(5.1361,0.0036)}]
\draw  (-4.4603,4.9901) ellipse (1.2 and 0.5);
\node (v1) at (-4.4744,4.8202) {$4_{6-k}^{8-2k}$};
\node at (-5.2811,5.2179) {\tiny{$e$}};
\begin{scope}[shift={(-1.9849,0.7134)}]
\draw  (-3.5835,4.3891) rectangle (-3.0825,4.1694);
\node at (-3.3438,4.2761) {\tiny{$k-6$}};
\end{scope}
\begin{scope}[shift={(-1.4125,1.0405)}]
\draw  (-3.5769,4.3717) rectangle (-3.0077,4.19);
\node at (-3.2917,4.2752) {\tiny{$2k-8$}};
\end{scope}
\node at (-4.1392,5.3182) {\tiny{$f$-$x_i$-$y_i$}};
\begin{scope}[shift={(-0.3134,0.6475)}]
\draw  (-3.5769,4.3717) rectangle (-3.0077,4.19);
\node at (-3.2917,4.2752) {\tiny{$16-2k$}};
\end{scope}
\node at (-3.8406,5.0988) {\tiny{$2h$-$\sum x_i$-$\sum y_i$}};
\end{scope}
\node at (-2.6424,5.2031) {\tiny{$e$}};
\node at (-1.423,4.7873) {\tiny{$h$}};
\begin{scope}[shift={(2.4199,1.5062)}]
\draw  (-4.4603,4.9901) ellipse (1 and 0.5);
\node (v1) at (-4.5578,4.9971) {$2_{6-k}$};
\node at (-3.7953,5.0293) {\tiny{$f$}};
\node at (-4.7699,4.6439) {\tiny{$e$}};
\begin{scope}[shift={(-1.1411,0.3576)}]
\draw  (-3.5352,4.3891) rectangle (-3.0397,4.1868);
\node at (-3.2901,4.2995) {\tiny{$k-6$}};
\end{scope}
\begin{scope}[shift={(-0.5147,0.5858)}]
\draw  (-3.4282,4.3553) rectangle (-3.1432,4.1635);
\node at (-3.2917,4.2752) {\tiny{$0$}};
\end{scope}
\end{scope}
\begin{scope}[shift={(2.8763,1.5294)}]
\draw  (-3.6392,4.4011) rectangle (-2.9855,4.1922);
\node at (-3.2901,4.2995) {\tiny{$4-k$}};
\end{scope}
\begin{scope}[shift={(2.4325,1.2108)}]
\draw  (-3.6392,4.4011) rectangle (-2.9855,4.1922);
\node at (-3.2901,4.2995) {\tiny{$4-k$}};
\end{scope}
\draw (-2.019,5.9932) -- (-2.019,5.5182);
\draw (-0.3102,5.7204) -- (0.2707,5.464);
\draw (-0.6531,5.9337) -- (-1.1906,6.2395);
\begin{scope}[shift={(7.901,0.0069)}]
\draw  (-4.4603,4.9901) ellipse (1 and 0.5);
\node (v1) at (-4.2995,4.9925) {$3_{8,5-k}$};
\node at (-5.1761,5.1861) {\tiny{$e$}};
\begin{scope}[shift={(-1.8953,0.7227)}]
\draw  (-3.4282,4.3553) rectangle (-3.1432,4.1635);
\node at (-3.2917,4.2752) {\tiny{$-8$}};
\end{scope}
\end{scope}
\draw (1.8669,4.9968) -- (2.432,4.9939);
\end{tikzpicture}
\end{array}
$}
\end{align}

\subsection{ \texorpdfstring{$SO(10)$}{} with  \texorpdfstring{$4-k$}{} spinor hypers and  \texorpdfstring{$6-k$}{} fundamental hypers}
The F-theory construction involves a split $\text{I}^*_1$ fiber over a $-k$ curve in the base, where $1\le k\le4$. The Weierstrass model appears in Section \ref{D}. The threefold is

\begin{align}
\label{eqn:SO10samp}
\begin{array}{c}
\begin{tikzpicture}[scale=1.5]
\draw  (-4.4603,4.9901) ellipse (1 and 0.5);
\node (v1) at (-4.4603,4.9901) {$0_{k-2}$};
\draw (-3.4464,5.0091) -- (-2.9847,5.0073);
\node at (-3.937,5.2302) {\tiny{$e$ or $h$}};
\node at (-4.9686,5.2627) {\tiny{$h$ or $e$}};
\begin{scope}[shift={(-1.8167,0.7227)}]
\draw  (-3.5199,4.4183) rectangle (-3.0317,4.1342);
\node at (-3.2917,4.2752) {\tiny{$k-2$}};
\end{scope}
\begin{scope}[shift={(-0.4988,0.715)}]
\draw  (-3.5352,4.3891) rectangle (-3.0397,4.1868);
\node at (-3.2901,4.2995) {\tiny{$2-k$}};
\end{scope}
\begin{scope}[shift={(2.4713,0.0223)}]
\draw  (-4.4603,4.9901) ellipse (1 and 0.5);
\node (v1) at (-4.507,4.9528) {$1_{4-k}$};
\draw (-3.4464,5.0091) -- (-2.9847,5.0073);
\begin{scope}[shift={(-1.8354,0.7134)}]
\draw  (-3.5321,4.3845) rectangle (-3.0504,4.1716);
\node at (-3.2917,4.2752) {\tiny{$k-4$}};
\end{scope}
\begin{scope}[shift={(-0.5642,0.7057)}]
\draw  (-3.6392,4.4011) rectangle (-2.9855,4.1922);
\node at (-3.2901,4.2995) {\tiny{$4-k$}};
\end{scope}
\begin{scope}[shift={(-1.1756,1.0455)}]
\draw  (-3.5321,4.3845) rectangle (-3.0504,4.1716);
\node at (-3.2917,4.2752) {\tiny{$4-k$}};
\end{scope}
\node at (-4.0918,5.3074) {\tiny{$h$}};
\end{scope}
\begin{scope}[shift={(5.1361,0.0036)}]
\draw  (-4.4603,4.9901) ellipse (1.2 and 0.5);
\node (v1) at (-4.3814,4.9083) {$5_{6-k}^{4-k}$};
\node at (-5.3928,5.1471) {\tiny{$e$}};
\begin{scope}[shift={(-2.0408,0.7209)}]
\draw  (-3.5524,4.3484) rectangle (-3.1035,4.1956);
\node at (-3.3438,4.2761) {\tiny{$k-6$}};
\end{scope}
\begin{scope}[shift={(-1.595,1.033)}]
\draw  (-3.5156,4.3674) rectangle (-3.0604,4.1989);
\node at (-3.2917,4.2752) {\tiny{$k-4$}};
\end{scope}
\node at (-4.8506,5.1441) {\tiny{$f$-$x_i$}};
\begin{scope}[shift={(-0.1234,0.722)}]
\draw  (-3.4041,4.3635) rectangle (-3.206,4.2062);
\node at (-3.3103,4.2939) {\tiny{$2$}};
\end{scope}
\node at (-3.651,4.8569) {\tiny{$h$-$\sum x_i$}};
\begin{scope}[shift={(-0.8189,1.0748)}]
\draw  (-3.5524,4.3484) rectangle (-3.1035,4.1956);
\node at (-3.3438,4.2761) {\tiny{$6-k$}};
\end{scope}
\node at (-3.8317,5.3301) {\tiny{$h$}};
\end{scope}
\node at (-2.6424,5.2031) {\tiny{$e$}};
\node at (-1.423,4.7873) {\tiny{$h$}};
\begin{scope}[shift={(2.4199,1.5062)}]
\draw  (-4.4603,4.9901) ellipse (1 and 0.5);
\node (v1) at (-4.7515,5.1201) {$2_{6-k}^{4-k}$};
\node at (-4.1752,5.115) {\tiny{$x_i$}};
\node at (-4.7699,4.6439) {\tiny{$e$}};
\begin{scope}[shift={(-1.1411,0.3576)}]
\draw  (-3.5352,4.3891) rectangle (-3.0397,4.1868);
\node at (-3.2901,4.2995) {\tiny{$k-6$}};
\end{scope}
\begin{scope}[shift={(-0.6121,0.5364)}]
\draw  (-3.5352,4.3891) rectangle (-3.0397,4.1868);
\node at (-3.2901,4.2995) {\tiny{$k-4$}};
\end{scope}
\begin{scope}[shift={(-0.5525,0.8269)}]
\draw  (-3.5352,4.3891) rectangle (-3.0397,4.1868);
\node at (-3.2901,4.2995) {\tiny{$k-4$}};
\end{scope}
\node at (-4.3291,4.8865) {\tiny{$f$-$x_i$}};
\end{scope}
\begin{scope}[shift={(2.8763,1.5294)}]
\draw  (-3.6392,4.4011) rectangle (-2.9855,4.1922);
\node at (-3.2901,4.2995) {\tiny{$4-k$}};
\end{scope}
\begin{scope}[shift={(2.4325,1.2108)}]
\draw  (-3.6392,4.4011) rectangle (-2.9855,4.1922);
\node at (-3.2901,4.2995) {\tiny{$4-k$}};
\end{scope}
\draw (-2.019,5.9932) -- (-2.019,5.5182);
\draw (-0.3102,5.7204) -- (0.1701,5.4491);
\draw (-0.6531,5.9337) -- (-1.1906,6.2395);
\begin{scope}[shift={(7.901,0.0069)}]
\draw  (-4.4603,4.9901) ellipse (1 and 0.5);
\node (v1) at (-4.2995,4.9925) {$3_4$};
\node at (-5.1985,5.1414) {\tiny{$e$}};
\begin{scope}[shift={(-1.8953,0.7227)}]
\draw  (-3.4282,4.3553) rectangle (-3.1432,4.1635);
\node at (-3.2917,4.2752) {\tiny{$-4$}};
\end{scope}
\begin{scope}[shift={(-1.159,1.0796)}]
\draw  (-3.4041,4.3635) rectangle (-3.206,4.2062);
\node at (-3.3103,4.2939) {\tiny{$0$}};
\end{scope}
\node at (-4.2784,5.3538) {\tiny{$f$}};
\end{scope}
\draw (1.8669,4.9968) -- (2.432,4.9939);
\begin{scope}[shift={(7.8974,1.5551)}]
\draw  (-4.4603,4.9901) ellipse (1 and 0.5);
\node (v1) at (-4.2364,5.0884) {$4_4^{16-3k}$};
\node at (-4.9626,5.3376) {\tiny{$z_i$}};
\begin{scope}[shift={(-1.7979,0.8707)}]
\draw  (-3.5112,4.3662) rectangle (-3.0665,4.1777);
\node at (-3.2917,4.2752) {\tiny{$k-4$}};
\end{scope}
\begin{scope}[shift={(-1.0659,0.3588)}]
\draw  (-3.5906,4.3632) rectangle (-3.0156,4.1777);
\node at (-3.3001,4.2753) {\tiny{$2k-12$}};
\end{scope}
\begin{scope}[shift={(-1.654,0.4786)}]
\draw  (-3.5112,4.3662) rectangle (-3.0665,4.1777);
\node at (-3.2917,4.2752) {\tiny{$k-8$}};
\end{scope}
\node at (-4.7957,4.9353) {\tiny{$e$-$\sum z_i$}};
\node at (-4.0809,4.8324) {\tiny{$f$-$x_i$-$y_i$}};
\end{scope}
\begin{scope}[shift={(4.0467,2.3257)}]
\draw  (-3.6392,4.4011) rectangle (-2.9855,4.1922);
\node at (-3.2901,4.2995) {\tiny{$4-k$}};
\end{scope}
\begin{scope}[shift={(6.7072,1.5173)}]
\draw  (-3.6392,4.4011) rectangle (-2.9855,4.1922);
\node at (-3.2901,4.2995) {\tiny{$6-k$}};
\end{scope}
\begin{scope}[shift={(4.0834,1.8194)}]
\draw  (-3.6392,4.4011) rectangle (-2.9855,4.1922);
\node at (-3.2901,4.2995) {\tiny{$4-k$}};
\end{scope}
\begin{scope}[shift={(5.834,1.3412)}]
\draw  (-3.6392,4.4011) rectangle (-2.9855,4.1922);
\node at (-3.2901,4.2995) {\tiny{$6-k$}};
\end{scope}
\draw (-1.0684,6.6169) -- (0.4036,6.6109) (1.0491,6.6048) -- (2.4486,6.641);
\draw (1.2421,5.4285) -- (2.6779,6.2127) (3.4396,6.0414) -- (3.4263,5.9185) (3.4237,5.7112) -- (3.4186,5.5015);
\end{tikzpicture}
\end{array}
\end{align}

where we have split the blow-ups on $S_4$ into $(6-k)+(6-k)+(4-k)$ blowups denoted respectively by $x_i$, $y_i$ and $z_i$.

\subsection{ \texorpdfstring{$SO(11)$}{} with  \texorpdfstring{$\frac{4-k}{2}$}{} spinor hypers and  \texorpdfstring{$7-k$}{} fundamental hypers}
The F-theory construction involves a non-split $\text{I}^*_2$ fiber over a $-k$ curve in the base, where $1\le k\le4$. The Weierstrass model was written down in Section \ref{D}. We find the corresponding threefold to be

\begin{align}
\begin{array}{c}
\begin{tikzpicture}[scale=1.5]
\draw  (-4.4603,4.9901) ellipse (1 and 0.5);
\node (v1) at (-4.4603,4.9901) {$1_{4-k}$};
\draw (-3.4464,5.0091) -- (-2.9847,5.0073);
\node at (-3.786,4.819) {\tiny{$h$}};
\node at (-4.7504,4.6753) {\tiny{$e$}};
\begin{scope}[shift={(-1.1557,0.4151)}]
\draw  (-3.5199,4.4183) rectangle (-3.0317,4.1342);
\node at (-3.2917,4.2752) {\tiny{$k-2$}};
\end{scope}
\begin{scope}[shift={(-0.4988,0.715)}]
\draw  (-3.5352,4.3891) rectangle (-3.0397,4.1868);
\node at (-3.2901,4.2995) {\tiny{$4-k$}};
\end{scope}
\begin{scope}[shift={(2.4713,0.0223)}]
\draw  (-4.4603,4.9901) ellipse (1 and 0.5);
\node (v1) at (-4.507,4.9528) {$3_{6-k}$};
\draw (-3.4464,5.0091) -- (-2.9847,5.0073);
\begin{scope}[shift={(-1.8354,0.7134)}]
\draw  (-3.5321,4.3845) rectangle (-3.0504,4.1716);
\node at (-3.2917,4.2752) {\tiny{$k-6$}};
\end{scope}
\begin{scope}[shift={(-0.5642,0.7057)}]
\draw  (-3.6392,4.4011) rectangle (-2.9855,4.1922);
\node at (-3.2901,4.2995) {\tiny{$6-k$}};
\end{scope}
\begin{scope}[shift={(-1.1756,1.0455)}]
\draw  (-3.5321,4.3845) rectangle (-3.0504,4.1716);
\node at (-3.2917,4.2752) {\tiny{$0$}};
\end{scope}
\node at (-4.0918,5.3074) {\tiny{$f$}};
\end{scope}
\begin{scope}[shift={(5.1361,0.0036)}]
\draw  (-4.4603,4.9901) ellipse (1.2 and 0.5);
\node (v1) at (-4.4306,4.9371) {$5_{8-k}^{4-k}$};
\node at (-5.3353,5.184) {\tiny{$e$}};
\begin{scope}[shift={(-1.9849,0.7134)}]
\draw  (-3.5835,4.3891) rectangle (-3.0825,4.1694);
\node at (-3.3438,4.2761) {\tiny{$k-8$}};
\end{scope}
\begin{scope}[shift={(-1.4125,1.0405)}]
\draw  (-3.5769,4.3717) rectangle (-3.0077,4.19);
\node at (-3.2917,4.2752) {\tiny{$k-4$}};
\end{scope}
\node at (-4.2128,5.3193) {\tiny{$f$-$x_i$}};
\begin{scope}[shift={(-0.3134,0.6475)}]
\draw  (-3.5769,4.3717) rectangle (-3.0077,4.19);
\node at (-3.2917,4.2752) {\tiny{$24-3k$}};
\end{scope}
\node at (-3.6705,5.1134) {\tiny{$2h$-$\sum x_i$}};
\end{scope}
\node at (-2.6424,5.2031) {\tiny{$e$}};
\node at (-1.423,4.7873) {\tiny{$h$}};
\begin{scope}[shift={(2.4199,1.5062)}]
\draw  (-4.4603,4.9901) ellipse (1 and 0.5);
\node (v1) at (-4.5201,5.2268) {$2_{6-k}^{8-2k}$};
\node at (-4.5643,4.833) {\tiny{$f$-$x_i$-$y_i$}};
\node at (-5.1686,4.8012) {\tiny{$e$}};
\begin{scope}[shift={(-1.1411,0.3576)}]
\draw  (-3.5352,4.3891) rectangle (-3.0397,4.1868);
\node at (-3.2901,4.2995) {\tiny{$2k-8$}};
\end{scope}
\begin{scope}[shift={(-0.5781,0.566)}]
\draw  (-3.5769,4.3717) rectangle (-3.0077,4.19);
\node at (-3.2917,4.2752) {\tiny{$k-4$}};
\end{scope}
\begin{scope}[shift={(-0.5259,0.8157)}]
\draw  (-3.5352,4.3891) rectangle (-3.0397,4.1868);
\node at (-3.2901,4.2995) {\tiny{$2k-8$}};
\end{scope}
\begin{scope}[shift={(-1.8715,0.6971)}]
\draw  (-3.5321,4.3845) rectangle (-3.0504,4.1716);
\node at (-3.2917,4.2752) {\tiny{$k-6$}};
\end{scope}
\node at (-3.8761,4.6703) {\tiny{$x_i$}};
\node at (-3.9312,5.2928) {\tiny{$y_i$-$x_i$}};
\end{scope}
\begin{scope}[shift={(2.8763,1.5294)}]
\draw  (-3.6392,4.4011) rectangle (-2.9855,4.1922);
\node at (-3.2901,4.2995) {\tiny{$4-k$}};
\end{scope}
\begin{scope}[shift={(2.4325,1.2108)}]
\draw  (-3.6392,4.4011) rectangle (-2.9855,4.1922);
\node at (-3.2901,4.2995) {\tiny{$4-k$}};
\end{scope}
\draw (-0.3102,5.7204) -- (0.2707,5.464);
\draw (-0.6531,5.9337) -- (-1.1906,6.2395);
\begin{scope}[shift={(7.901,0.0069)}]
\draw  (-4.4603,4.9901) ellipse (1 and 0.5);
\node (v1) at (-4.2995,4.9925) {$4_{16-k,7-k}$};
\node at (-5.1718,5.134) {\tiny{$e$}};
\begin{scope}[shift={(-1.4464,1.0579)}]
\draw  (-3.4282,4.3553) rectangle (-3.1432,4.1635);
\node at (-3.2917,4.2752) {\tiny{$0$}};
\end{scope}
\begin{scope}[shift={(-1.835,0.69)}]
\draw  (-3.5769,4.3717) rectangle (-3.0077,4.19);
\node at (-3.2917,4.2752) {\tiny{$k-16$}};
\end{scope}
\node at (-4.508,5.3152) {\tiny{$f$}};
\end{scope}
\draw (1.8669,4.9968) -- (2.432,4.9939);
\begin{scope}[shift={(0.0621,-1.4787)}]
\draw  (-4.4603,4.9901) ellipse (1 and 0.5);
\node (v1) at (-4.4603,4.9901) {$0_{k-2}$};
\node at (-3.9592,5.2666) {\tiny{$e$ or $h$}};
\node at (-4.9368,4.7662) {\tiny{$h$ or $e$}};
\begin{scope}[shift={(-1.1524,0.3889)}]
\draw  (-3.5199,4.4183) rectangle (-3.0317,4.1342);
\node at (-3.2917,4.2752) {\tiny{$k-2$}};
\end{scope}
\begin{scope}[shift={(-1.1565,1.0455)}]
\draw  (-3.5352,4.3891) rectangle (-3.0397,4.1868);
\node at (-3.2901,4.2995) {\tiny{$2-k$}};
\end{scope}
\end{scope}
\begin{scope}[shift={(-0.8358,0.9735)}]
\draw  (-3.5352,4.3891) rectangle (-3.0397,4.1868);
\node at (-3.2901,4.2995) {\tiny{$4-k$}};
\end{scope}
\node at (-4.4311,5.2727) {\tiny{$h$}};
\begin{scope}[shift={(1.3101,1.4693)}]
\draw  (-3.6392,4.4011) rectangle (-2.9855,4.1922);
\node at (-3.2901,4.2995) {\tiny{$4-k$}};
\end{scope}
\begin{scope}[shift={(4.2198,1.8269)}]
\draw  (-3.6392,4.4011) rectangle (-2.9855,4.1922);
\node at (-3.2901,4.2995) {\tiny{$4-k$}};
\end{scope}
\begin{scope}[shift={(4.1725,1.4333)}]
\draw  (-3.6392,4.4011) rectangle (-2.9855,4.1922);
\node at (-3.2901,4.2995) {\tiny{$4-k$}};
\end{scope}
\begin{scope}[shift={(0.2396,1.3688)}]
\draw  (-3.6392,4.4011) rectangle (-2.9855,4.1922);
\node at (-3.2901,4.2995) {\tiny{$4-k$}};
\end{scope}
\draw (-4.4097,4.4919) -- (-4.4172,4.0136) (-4.0368,5.4449) -- (-3.0273,6.4355) (-2.0104,5.9872) -- (-2.0104,5.8629) (-2.0141,5.6558) -- (-2.0104,5.5051) (-1.0725,6.6163) -- (0.5735,6.1944) (1.2251,6.055) -- (3.1197,5.4675);
\end{tikzpicture}
\end{array}
\end{align}

\subsection{ \texorpdfstring{$SO(12)$}{} with  \texorpdfstring{$\frac{4-k}{2}$}{} spinor hypers and fundamental \texorpdfstring{$8-k$}{}}
The F-theory construction involves a split $\text{I}^*_2$ fiber over a $-k$ curve in the base, where $1\le k\le4$. The Weierstrass model was written down in Section \ref{D}. In this case, our answer is

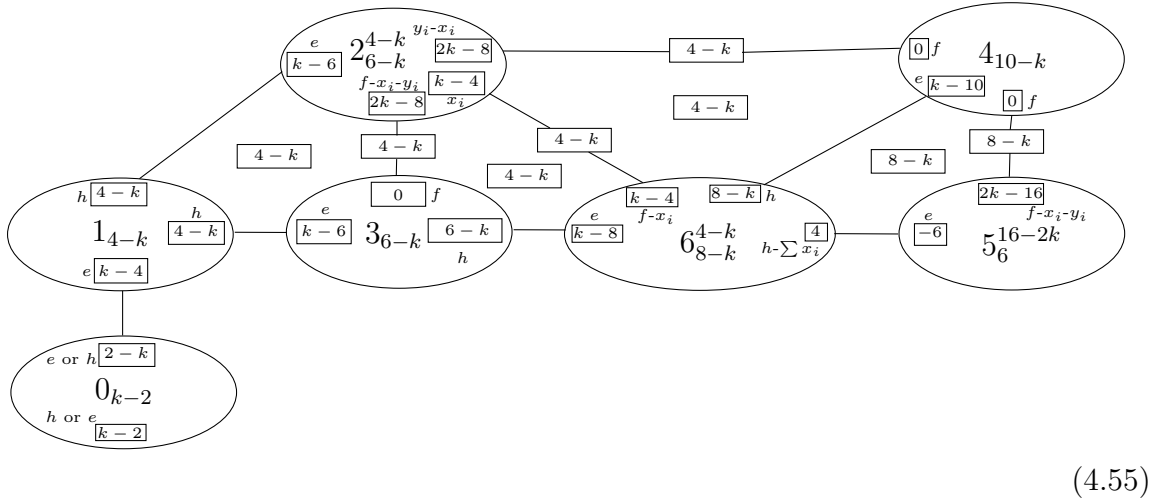
\begin{align}
\begin{array}{c}
\begin{tikzpicture}[scale=1.5]
\draw  (-4.4603,4.9901) ellipse (1 and 0.5);
\node (v1) at (-4.4603,4.9901) {$1_{4-k}$};
\draw (-3.4464,5.0091) -- (-2.9847,5.0073);
\node at (-3.785,5.1805) {\tiny{$h$}};
\node at (-4.7982,5.3284) {\tiny{$h$}};
\node at (-4.7578,4.6477) {\tiny{$e$}};
\begin{scope}[shift={(-1.1644,0.3871)}]
\draw  (-3.5262,4.3886) rectangle (-3.0539,4.1721);
\node at (-3.2917,4.2752) {\tiny{$k-4$}};
\end{scope}
\begin{scope}[shift={(-0.4988,0.715)}]
\draw  (-3.5352,4.3891) rectangle (-3.0397,4.1868);
\node at (-3.2901,4.2995) {\tiny{$4-k$}};
\end{scope}
\begin{scope}[shift={(-1.186,1.057)}]
\draw  (-3.5352,4.3891) rectangle (-3.0397,4.1868);
\node at (-3.2901,4.2995) {\tiny{$4-k$}};
\end{scope}
\begin{scope}[shift={(2.4713,0.0223)}]
\draw  (-4.4603,4.9901) ellipse (1 and 0.5);
\node (v1) at (-4.507,4.9528) {$3_{6-k}$};
\draw (-3.4464,5.0091) -- (-2.9847,5.0073);
\begin{scope}[shift={(-1.8354,0.7134)}]
\draw  (-3.5321,4.3845) rectangle (-3.0504,4.1716);
\node at (-3.2917,4.2752) {\tiny{$k-6$}};
\end{scope}
\begin{scope}[shift={(-0.5642,0.7057)}]
\draw  (-3.6392,4.4011) rectangle (-2.9855,4.1922);
\node at (-3.2901,4.2995) {\tiny{$6-k$}};
\end{scope}
\begin{scope}[shift={(-1.1756,1.0455)}]
\draw  (-3.5321,4.3845) rectangle (-3.0504,4.1716);
\node at (-3.2917,4.2752) {\tiny{$0$}};
\end{scope}
\node at (-4.133,5.3201) {\tiny{$f$}};
\end{scope}
\begin{scope}[shift={(5.1361,0.0036)}]
\draw  (-4.4603,4.9901) ellipse (1.2 and 0.5);
\node (v1) at (-4.3814,4.9083) {$6_{8-k}^{4-k}$};
\node at (-5.3928,5.1471) {\tiny{$e$}};
\begin{scope}[shift={(-2.0408,0.7209)}]
\draw  (-3.5524,4.3484) rectangle (-3.1035,4.1956);
\node at (-3.3438,4.2761) {\tiny{$k-8$}};
\end{scope}
\begin{scope}[shift={(-1.595,1.033)}]
\draw  (-3.5156,4.3674) rectangle (-3.0604,4.1989);
\node at (-3.2917,4.2752) {\tiny{$k-4$}};
\end{scope}
\node at (-4.8506,5.1441) {\tiny{$f$-$x_i$}};
\begin{scope}[shift={(-0.1234,0.722)}]
\draw  (-3.4041,4.3635) rectangle (-3.206,4.2062);
\node at (-3.3103,4.2939) {\tiny{$4$}};
\end{scope}
\node at (-3.651,4.8569) {\tiny{$h$-$\sum x_i$}};
\begin{scope}[shift={(-0.8189,1.0748)}]
\draw  (-3.5524,4.3484) rectangle (-3.1035,4.1956);
\node at (-3.3438,4.2761) {\tiny{$8-k$}};
\end{scope}
\node at (-3.8317,5.3301) {\tiny{$h$}};
\end{scope}
\node at (-2.6424,5.2031) {\tiny{$e$}};
\node at (-1.423,4.7873) {\tiny{$h$}};
\begin{scope}[shift={(2.4199,1.5062)}]
\draw  (-4.4603,4.9901) ellipse (1 and 0.5);
\node (v1) at (-4.593,5.1091) {$2_{6-k}^{4-k}$};
\node at (-3.9028,4.6583) {\tiny{$x_i$}};
\node at (-4.0954,5.2962) {\tiny{$y_i$-$x_i$}};
\begin{scope}[shift={(-1.1411,0.3576)}]
\draw  (-3.5352,4.3891) rectangle (-3.0397,4.1868);
\node at (-3.2901,4.2995) {\tiny{$2k-8$}};
\end{scope}
\begin{scope}[shift={(-0.6121,0.5364)}]
\draw  (-3.5352,4.3891) rectangle (-3.0397,4.1868);
\node at (-3.2901,4.2995) {\tiny{$k-4$}};
\end{scope}
\begin{scope}[shift={(-0.5525,0.8269)}]
\draw  (-3.5352,4.3891) rectangle (-3.0397,4.1868);
\node at (-3.2901,4.2995) {\tiny{$2k-8$}};
\end{scope}
\node at (-5.1598,5.1746) {\tiny{$e$}};
\node at (-4.496,4.823) {\tiny{$f$-$x_i$-$y_i$}};
\begin{scope}[shift={(-1.8702,0.7134)}]
\draw  (-3.5321,4.3845) rectangle (-3.0504,4.1716);
\node at (-3.2917,4.2752) {\tiny{$k-6$}};
\end{scope}
\end{scope}
\begin{scope}[shift={(2.8763,1.5294)}]
\draw  (-3.6392,4.4011) rectangle (-2.9855,4.1922);
\node at (-3.2901,4.2995) {\tiny{$4-k$}};
\end{scope}
\begin{scope}[shift={(2.4325,1.2108)}]
\draw  (-3.6392,4.4011) rectangle (-2.9855,4.1922);
\node at (-3.2901,4.2995) {\tiny{$4-k$}};
\end{scope}
\draw (-0.3102,5.7204) -- (0.1701,5.4491);
\draw (-0.6531,5.9337) -- (-1.1906,6.2395);
\begin{scope}[shift={(7.901,0.0069)}]
\draw  (-4.4603,4.9901) ellipse (1 and 0.5);
\node (v1) at (-4.3723,4.8912) {$5_6^{16-2k}$};
\node at (-5.1985,5.1414) {\tiny{$e$}};
\begin{scope}[shift={(-1.8953,0.7227)}]
\draw  (-3.4282,4.3553) rectangle (-3.1432,4.1635);
\node at (-3.2917,4.2752) {\tiny{$-6$}};
\end{scope}
\begin{scope}[shift={(-1.1641,1.0713)}]
\draw  (-3.5906,4.3632) rectangle (-3.0156,4.1777);
\node at (-3.3001,4.2753) {\tiny{$2k-16$}};
\end{scope}
\node at (-4.0633,5.1687) {\tiny{$f$-$x_i$-$y_i$}};
\end{scope}
\draw (1.8669,4.9968) -- (2.432,4.9939);
\begin{scope}[shift={(7.8974,1.5551)}]
\draw  (-4.4603,4.9901) ellipse (1 and 0.5);
\node (v1) at (-4.4453,5.0113) {$4_{10-k}$};
\begin{scope}[shift={(-1.1387,0.3525)}]
\draw  (-3.3931,4.3603) rectangle (-3.2271,4.1785);
\node at (-3.3131,4.2709) {\tiny{$0$}};
\end{scope}
\begin{scope}[shift={(-1.654,0.4786)}]
\draw  (-3.5428,4.3599) rectangle (-3.045,4.1827);
\node at (-3.2917,4.2752) {\tiny{$k-10$}};
\end{scope}
\node at (-5.2824,4.8066) {\tiny{$e$}};
\node at (-4.2741,4.6108) {\tiny{$f$}};
\begin{scope}[shift={(-1.9683,0.8116)}]
\draw  (-3.3931,4.3603) rectangle (-3.2271,4.1785);
\node at (-3.3131,4.2709) {\tiny{$0$}};
\end{scope}
\node at (-5.1264,5.0805) {\tiny{$f$}};
\end{scope}
\begin{scope}[shift={(4.0467,2.3257)}]
\draw  (-3.6392,4.4011) rectangle (-2.9855,4.1922);
\node at (-3.2901,4.2995) {\tiny{$4-k$}};
\end{scope}
\begin{scope}[shift={(6.7072,1.5173)}]
\draw  (-3.6392,4.4011) rectangle (-2.9855,4.1922);
\node at (-3.2901,4.2995) {\tiny{$8-k$}};
\end{scope}
\begin{scope}[shift={(4.0834,1.8194)}]
\draw  (-3.6392,4.4011) rectangle (-2.9855,4.1922);
\node at (-3.2901,4.2995) {\tiny{$4-k$}};
\end{scope}
\begin{scope}[shift={(5.834,1.3412)}]
\draw  (-3.6392,4.4011) rectangle (-2.9855,4.1922);
\node at (-3.2901,4.2995) {\tiny{$8-k$}};
\end{scope}
\draw (-1.0684,6.6169) -- (0.4036,6.6109) (1.0491,6.6048) -- (2.4486,6.641);
\draw (1.2421,5.4285) -- (2.6779,6.2127) (3.4396,6.0414) -- (3.4263,5.9185) (3.4237,5.7112) -- (3.4186,5.5015);
\begin{scope}[shift={(0.0298,-1.4012)}]
\draw  (-4.4603,4.9901) ellipse (1 and 0.5);
\node (v1) at (-4.4603,4.9901) {$0_{k-2}$};
\node at (-4.9208,5.2898) {\tiny{$e$ or $h$}};
\node at (-4.9231,4.778) {\tiny{$h$ or $e$}};
\begin{scope}[shift={(-1.2024,0.349)}]
\draw  (-3.5067,4.3572) rectangle (-3.0602,4.2134);
\node at (-3.2917,4.2752) {\tiny{$k-2$}};
\end{scope}
\begin{scope}[shift={(-1.1448,1.0332)}]
\draw  (-3.5352,4.3891) rectangle (-3.0397,4.1868);
\node at (-3.2901,4.2995) {\tiny{$2-k$}};
\end{scope}
\end{scope}
\begin{scope}[shift={(1.3147,1.4736)}]
\draw  (-3.6392,4.4011) rectangle (-2.9855,4.1922);
\node at (-3.2901,4.2995) {\tiny{$4-k$}};
\end{scope}
\draw (-2.0086,5.9938) -- (-2.0086,5.8672) (-2.0117,5.6582) -- (-2.0149,5.5125);
\draw (-4.2916,5.484) -- (-3.025,6.434) (-4.4404,4.4897) -- (-4.4404,4.0939);
\begin{scope}[shift={(0.2145,1.3862)}]
\draw  (-3.6392,4.4011) rectangle (-2.9855,4.1922);
\node at (-3.2901,4.2995) {\tiny{$4-k$}};
\end{scope}
\end{tikzpicture}
\end{array}
\end{align}

\subsection{Pure  \texorpdfstring{$SU(3)$}{}}
This is constructed by taking a split $\text{IV}$ fiber over a $-3$ curve in the base. The corresponding collection of surfaces is

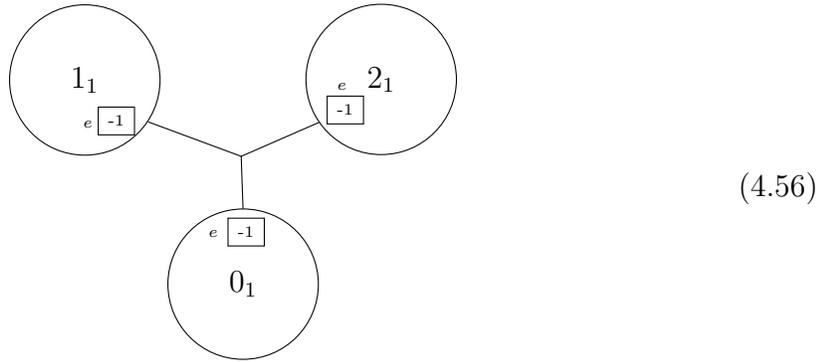
\begin{align}
\begin{array}{c}
\begin{tikzpicture}[scale=2]
\draw  (-3.9479,4.9946) ellipse (0.5 and 0.5);
\node (v1) at (-3.9479,4.9946) {$0_1$};
\node at (-4.1458,5.3303) {\tiny{$e$}};
\begin{scope}[shift={(-0.6371,1.0692)}]
\draw  (-3.4113,4.3642) rectangle (-3.17,4.1794);
\node at (-3.2917,4.2752) {\tiny{-1}};
\end{scope}
\begin{scope}[shift={(-1.0523,1.3581)}]
\draw  (-3.9479,4.9946) ellipse (0.5 and 0.5);
\node (v1) at (-3.9479,4.9946) {$1_1$};
\node at (-3.9282,4.7027) {\tiny{$e$}};
\begin{scope}[shift={(-0.4478,0.4502)}]
\draw  (-3.4113,4.3642) rectangle (-3.17,4.1794);
\node at (-3.2917,4.2752) {\tiny{-1}};
\end{scope}
\end{scope}
\begin{scope}[shift={(0.9176,1.3618)}]
\draw  (-3.9479,4.9946) ellipse (0.5 and 0.5);
\node (v1) at (-3.9479,4.9946) {$2_1$};
\node at (-4.2086,4.9466) {\tiny{$e$}};
\begin{scope}[shift={(-0.8956,0.5194)}]
\draw  (-3.4113,4.3642) rectangle (-3.17,4.1794);
\node at (-3.2917,4.2752) {\tiny{-1}};
\end{scope}
\end{scope}
\draw (-4.5795,6.0742) -- (-3.9605,5.8448) -- (-3.4435,6.0706);
\draw (-3.9485,5.4963) -- (-3.9609,5.8453);
\end{tikzpicture}
\end{array}
\end{align}

which represents three $\F_1$ glued together along their $e$ curve. There are actually four surfaces glued along a single locus because the base is also glued to the $e$ curve of $S_0$. Such a multi-valent gluing still satisfies the Calabi-Yau condition \ref{CY} pairwise as all the curves involved in the gluing are $-1$ curves. The data of triple intersection numbers can be depicted as

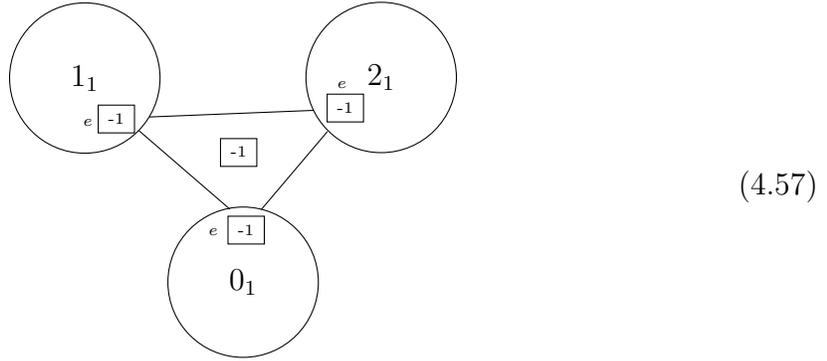
\begin{align}
\begin{array}{c}
\begin{tikzpicture}[scale=2]
\draw  (-3.9479,4.9946) ellipse (0.5 and 0.5);
\node (v1) at (-3.9479,4.9946) {$0_1$};
\node at (-4.1458,5.3303) {\tiny{$e$}};
\begin{scope}[shift={(-0.6371,1.0692)}]
\draw  (-3.4113,4.3642) rectangle (-3.17,4.1794);
\node at (-3.2917,4.2752) {\tiny{-1}};
\end{scope}
\begin{scope}[shift={(-1.0523,1.3581)}]
\draw  (-3.9479,4.9946) ellipse (0.5 and 0.5);
\node (v1) at (-3.9479,4.9946) {$1_1$};
\node at (-3.9282,4.7027) {\tiny{$e$}};
\begin{scope}[shift={(-0.4478,0.4502)}]
\draw  (-3.4113,4.3642) rectangle (-3.17,4.1794);
\node at (-3.2917,4.2752) {\tiny{-1}};
\end{scope}
\end{scope}
\begin{scope}[shift={(0.9176,1.3618)}]
\draw  (-3.9479,4.9946) ellipse (0.5 and 0.5);
\node (v1) at (-3.9479,4.9946) {$2_1$};
\node at (-4.2086,4.9466) {\tiny{$e$}};
\begin{scope}[shift={(-0.8956,0.5194)}]
\draw  (-3.4113,4.3642) rectangle (-3.17,4.1794);
\node at (-3.2917,4.2752) {\tiny{-1}};
\end{scope}
\end{scope}
\begin{scope}[shift={(-0.687,1.5856)}]
\draw  (-3.4113,4.3642) rectangle (-3.17,4.1794);
\node at (-3.2917,4.2752) {\tiny{-1}};
\end{scope}
\draw (-4.6404,6.0026) -- (-4.036,5.4804) (-3.8277,5.4774) -- (-3.3876,5.9967) (-3.4786,6.1376) -- (-4.57,6.0936);
\end{tikzpicture}
\end{array}
\end{align}

but it should be kept in mind that the latter figure does not represent the actual geometry and is only a bookkeeping device for triple intersection numbers.

\section{RG flows}\label{RG}
\subsection{Geometric criteria}
An RG flow is induced by taking the volumes of some compact curves to infinity. We cannot simply expand a compact curve $C$ living inside a compact surface $S$ to infinite volume, because that sends the volume of $S$ to infinity as well, leading to a rank changing RG flow. This would seem to suggest that there are no RG flows possible. However, we should recall that there are two ways in which a mass parameter can be take to infinity, i.e. either to $+\infty$ or to $-\infty$. This suggests that we should try to send the volumes of some curves to negative infinity. This is indeed possible if we have a rational curve $E$ of self-intersection $-1$ (henceforth referred to as a ``$-1$ curve'') inside a compact surface $S$. The volume of $E$ can be sent to negative infinity, which is a formal way of saying that we first flop $E$ to another $-1$ curve $E'$ and then send the volume of $E'$ to infinity. This RG flow is allowed whenever $E'$ lives in a non-compact surface inside the threefold, as then sending the volume of $E'$ to infinity doesn't remove any BPS string from the spectrum.

\begin{figure}
\begin{center}
	\begin{tikzpicture}
		\node(1) at (0,0) {$
			\begin{tikzpicture}[scale=1.2]
				\draw (0,0) -- (0,2) -- (1,2) -- (1,0) -- (0,0);
				\draw (0,1) -- node[above,midway]{\tiny$E$} (1,1);
				\node at (.5,.4) {\tiny $T$};
				\node at (.5,1.6) {\tiny $S$};
				\draw[dashed] (.5,2) to [out=90,in=180] (1.5,2.5) to [out=0,in=90] (2,2) to (2,0) to [out=270,in=0] (1.5,-0.5) to [out=180,in=270] (.5,0);
			\end{tikzpicture}
		$};	
		\node(3) at (5,0) {$
			\begin{tikzpicture}[scale=1.2]
				\draw (0,0) -- (1,0) -- (.5,.4) -- (0,0);
				\draw (0,2) -- (1,2) -- (.5,1.6) -- (0,2);
				\draw (.5,1.6) -- node[right,midway]{\tiny $E'$}(.5,.4);
				\node at (.5,.2) {\tiny $T'$};
				\node at (.5,1.8) {\tiny $S'$};
				\draw[dashed] (.5,2) to [out=90,in=180] (1.5,2.5) to [out=0,in=90] (2,2) to (2,0) to [out=270,in=0] (1.5,-0.5) to [out=180,in=270] (.5,0);
			\end{tikzpicture}
		$};	
		\node(4) at (10,0) {$
			\begin{tikzpicture}[scale=1.2]
				\draw (.5,.2) -- (1,.2) -- (.5,.6) -- (0,.2) -- (.5,.2);
				\draw (.5,-.2) -- (1,-.2) -- (.5,-.6) -- (0,-.2) -- (.5,-.2);
				\draw[dashed] (.5,.2) -- (.5,-.2);
				\draw (.5,.6) --++ (0,1);
				\draw[dotted] (.5,1.4) --++ (0,.5);
				\draw (.5,-.6) --++ (0,-1);
				\draw[dotted] (.5,-1.4) --++ (0,-.5);
				\node at (.5,.4) {\tiny $T'$};
				\node at (.5,-.4) {\tiny $S'$};
			\end{tikzpicture}
		$};	
		\draw[big arrow] ($(1)+(2,0)$) -- ($(3)+(-2,0)$);
		\draw[big arrow] ($(3)+(2,0)$) -- ($(4)+(-2,0)$);
	\end{tikzpicture}
\end{center}
\caption{From left to right: Two surfaces $S$ and $T$ are glued to each other along a $-1$ curve $E$. The dashed line represents a collection of surfaces joining $S$ and $T$ in some other direction. Flopping $E$ leads to the second figure with $S'$ being a blow-down of $S$ and $T'$ being a blow-down of $T$. The volume of $E'$ can then be sent to infinity leading to a geometry (shown in the last figure) in which $S'$ and $T'$ are joined only via the collection of surfaces denoted by dashed line.}
\label{disjoin}
\end{figure}
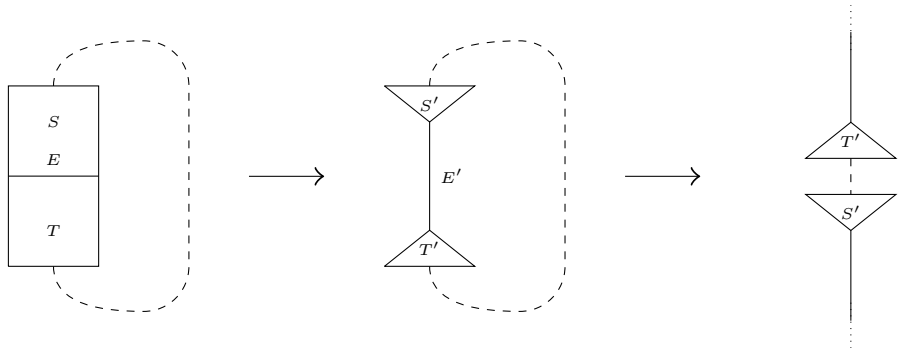

\noindent In all, we can divide the analysis of rank preserving RG flows into three cases:

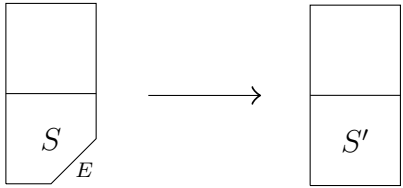
\begin{figure}
\begin{center}
\begin{tikzpicture}
	\node(1) at (0,0) {$
		\begin{tikzpicture}[scale=1.2]
		\draw (0,0) -- (0,2) -- (1,2) -- (1,.5) -- node[right,pos=.7]{\scriptsize $E$} (.5,0) -- (0,0);
		\draw (0,1) -- (1,1);
		\node at (.5,.5) { $S$};
		\end{tikzpicture}
		$};
	\node(2) at (4,0){$
		\begin{tikzpicture}[scale=1.2]
		\draw (0,0) -- (0,2) -- (1,2) -- (1,0) -- (0,0);
		\draw (0,1) -- (1,1);
		\node at (.5,.5) { $S'$};
		\end{tikzpicture}
		$};
		\draw[big arrow] ($(1)+(1.25,0)$) -- ($(2)-(1.25,0)$);
\end{tikzpicture}
\end{center}
\caption{From left to right: A $-1$ curve $E$ neither intersects the gluing curve nor is a part of it. Its flop simply leads to a surface $S'$ which is a blow-down of $S$.}
\label{blowd}
\end{figure}

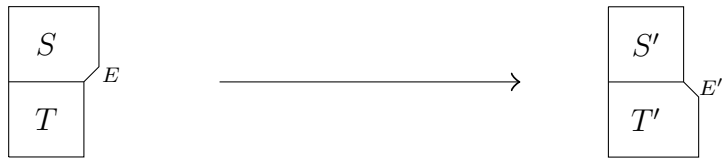
\begin{figure}
\begin{center}
\begin{tikzpicture}
	\node(1) at (0,0) {$
		\begin{tikzpicture}
		\draw (0,0) -- (1,0) -- (1,1) -- node[right,pos=.5]{\scriptsize $E$}(1.2,1.2) -- (1.2,2) -- (0,2) -- (0,0);
		\draw (0,1) -- (1,1);
		\node at (.5,1.5) {$S$};
		\node at (.5,.5){$T$};
		\end{tikzpicture}
		$};
		\node(3) at (8,0) {$
		\begin{tikzpicture}
		\draw (0,0) -- (1,0) -- (1,-1) -- node[right,pos=.3]{\scriptsize $E'$} (1.2,-1.2) -- (1.2,-2) -- (0,-2) -- (0,0);
		\draw (0,-1) -- (1,-1);
		\node at (.5,-.5) {$S'$};
		\node at (.5,-1.5) {$T'$};
		\end{tikzpicture}
		$};
		\draw[big arrow] ($(1)+(2,0)$) -- ($(3)-(2,0)$);
\end{tikzpicture}
\end{center}
\caption{A flop transition that corresponds to changing the chamber of Coulomb branch. $S'$ is a blow-down of $S$ and $T'$ is a blow-up of $T$.}
\label{chamber}
\end{figure}

\ben
\item If $E$ is part of the gluing locus between $S$ and some other compact surface $T$, then after the flop $E'$ will separate $S'$ and $T'$ (where $S'$ and $T'$ are the images of $S$ and $T$ after the flop). This can lead to a rank preserving RG flow only if $S'$ and $T'$ are linked by a chain of surfaces, which is equivalent to saying that $S$ and $T$ were part of a loop of surfaces before the flop. See Figure \ref{disjoin}. In other words, the RG flow is implemented by blowing down $E$ inside $S$ and $T$, and deleting the gluing corresponding to $E$ between $S$ and $T$.
\item If the $E$ is neither a part of gluing locus between any two compact surfaces nor intersects the gluing locus between any two compact surfaces, then it can be flopped into an adjacent non-compact surface and its volume can subsequently be sent to infinity leading to an RG flow. See Figure \ref{blowd}. In other words, the geometry after the RG flow is simply described by blowing down $E$ inside $S$, and everything else remains the same.
\item If $E$ is not a part of gluing locus between any two compact surfaces, but rather intersects the gluing locus between $S$ and some other compact surface $T$ at some point, then it can be flopped into $T$. Now we have a collection of surfaces in which $S$ and $T$ have been replaced by $S'$ and $T'$, which might carry new $-1$ curves each of which can then be subsequently flopped. In other words, we have blow down $E$ inside $S$ and blown-up the point inside $T$ corresponding to intersection with $E$. This takes us to a different chamber of the Coulomb branch with, in general, a different set of $-1$ curves that can lead to RG flows as in the above two criteria. See Figure \ref{chamber}.
\een

%
%

\subsection{Illustration:  \texorpdfstring{$\cN=2\to\cN=1$}{}}\label{(2,0)}
Consider $6d$ $A_1$ $(2,0)$ SCFT compactified on a circle. It is well known that the KK theory in this case is $5d$ $\cN=2$ pure super Yang-Mills with gauge group $SU(2)$. Viewed as an $\cN=1$ theory, it has a hyper in the adjoint representation of $SU(2)$. Turning on the mass of this hyper and sending it to infinity induces an RG flow of $\cN=2$ pure super Yang-Mills to $\cN=1$ pure super Yang-Mills based on $SU(2)$, which is known to be a $5d$ SCFT. 

In Section \ref{A}, we found two geometries for the KK theory. One of them descended from $\text{I}_0$ fibered over a $-2$ curve and the other descended from $\text{I}_1$ fibered over a $-2$ curve in the base. The geometry descending from $\text{I}_0$ has no $-1$ curves, and hence we do not see this RG flow there. We will now show that the geometry descending from $\text{I}_1$ allows one to see this RG flow. This disparity is due to the fact that since the string theory setup in the $\text{I}_0$ case preserves 16 supercharges, a supersymmetry breaking deformation can only be induced by bringing new ingredients (for example, some D-branes) from infinity in the threefold. Since our method of computing RG flows cares only about the data of the surfaces in the deep interior of the threefold, it is only natural that we do not see this flow.

Let us start by reproducing here the geometry of the Calabi-Yau corresponding to $\text{I}_1$ for our convenience

\begin{align}
\begin{array}{c}
\begin{tikzpicture}[scale=2]
\draw  (-3.5,4) ellipse (0.5 and 0.5);
\node (v1) at (-3.5,4) {$0^2_0$};
\draw  (-3.2902,4.2051) rectangle (-3.1053,4.0203);
\node at (-3.1906,4.1161) {\tiny{-2}};
\begin{scope}[shift={(-0.0137,-0.5476)}]
\draw  (-3.3913,4.3642) rectangle (-3.2064,4.1794);
\node at (-3.2917,4.2752) {\tiny{0}};
\end{scope}
\begin{scope}[shift={(-0.5545,-0.2807)}]
\draw  (-3.3913,4.3642) rectangle (-3.2064,4.1794);
\node at (-3.2917,4.2752) {\tiny{0}};
\end{scope}
\node at (-3.4709,4.3287) {\tiny{$h-\sum x_i$}};
\node at (-3.1558,3.7385) {\tiny{$h$}};
\node at (-3.8461,4.1604) {\tiny{$e$}};
\draw (-3.1935,3.5857) .. controls (-1.2581,2.9132) and (-1.2077,4.4972) .. (-3.0281,4.1613);
\end{tikzpicture}
\end{array}
\end{align}
We first flop one of the exceptional curves, say $x_1$, to obtain

\begin{align}
\begin{array}{c}
\begin{tikzpicture}[scale=2]
\draw  (-3.5,4) ellipse (0.5 and 0.5);
\node (v1) at (-3.5,4) {$0^2_0$};
\draw  (-3.2902,4.2051) rectangle (-3.1053,4.0203);
\node at (-3.1906,4.1161) {\tiny{-1}};
\begin{scope}[shift={(-0.0137,-0.5476)}]
\draw  (-3.3913,4.3642) rectangle (-3.2064,4.1794);
\node at (-3.2917,4.2752) {\tiny{-1}};
\end{scope}
\begin{scope}[shift={(-0.5545,-0.2807)}]
\draw  (-3.3913,4.3642) rectangle (-3.2064,4.1794);
\node at (-3.2917,4.2752) {\tiny{0}};
\end{scope}
\node at (-3.3892,4.3042) {\tiny{$h-x_2$}};
\node at (-3.1558,3.7385) {\tiny{$h$}};
\node at (-3.8461,4.1604) {\tiny{$e$}};
\draw (-3.2009,3.5983) .. controls (-1.2581,2.9132) and (-1.2077,4.4972) .. (-3.0281,4.1613);
\node at (-3.6061,3.753) {\tiny{$e-x_1$}};
\end{tikzpicture}
\end{array}
\end{align}

Since the gluing curve itself is a $-1$ curve now, we can flop it and, since the surface was glued to itself in a loop before the flop, we obtain a rank preserving RG flow to the geometry described by a local $\F_0$

\begin{align}
\begin{array}{c}
\begin{tikzpicture}[scale=2]
\draw  (-3.5,4) ellipse (0.5 and 0.5);
\node (v1) at (-3.5,4) {$0_0$};
\end{tikzpicture}
\end{array}
\end{align}

This geometry is indeed known to describe the Coulomb branch of $5d$ $SU(2)$ $\cN=1$ pure SYM.

\section{Future direction: Higher rank}\label{conclusion}
It seems straightforward to extend our results to higher rank $6d$ SCFTs. Say $C$ and $D$ are two curves in the base which intersect each other transversely. The intersection corresponds to gluing $S_C$ and $S_D$ with each other along the components of the degenerate elliptic fibers on $C$ and $D$. However, there is a subtlety that one might have to do some flops (corresponding to going to a different chamber of the Coulomb branch) on either $S_C$ or $S_D$ for a gluing to be permissible. We will discuss some examples of such gluings in this section\footnote{A general story for higher rank is currently a work in progress.}.

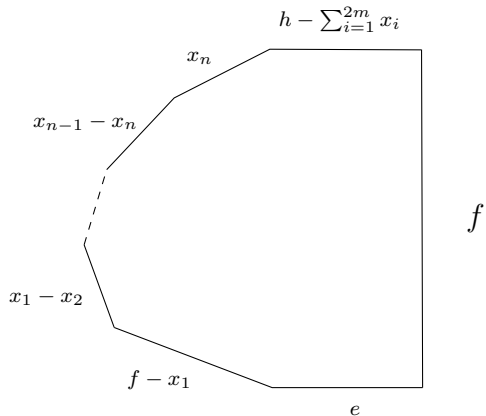
\begin{figure}
\begin{center}
\begin{tikzpicture}
\draw[dashed] (-9.4,4.5)-- (-9.1,5.5) node (v1) {};
\draw (-9,3.4) -- (-6.9,2.6) -- (-4.9,2.6);
\node at (-6,7.5) {\scriptsize{$h-\sum_{i=1}^{2m} x_i$}};
\node at (-7.8598,6.988) {\scriptsize{$x_n$}};
\node at (-9.383,6.1215) {\scriptsize{$x_{n-1}-x_n$}};
\node at (-8.4,2.7) {\scriptsize{$f-x_1$}};
\node at (-5.8,2.3) {\scriptsize{$e$}};
\draw (-9.4,4.5)--(-9,3.4);
\node at (-9.9,3.8) {\scriptsize{$x_1-x_2$}};
\draw (-9.1,5.5) -- (-8.2024,6.4543) -- (-6.9318,7.1022) -- (-4.9208,7.0938);
\draw (-4.9141,7.099)--(-4.9,2.6);
\node at (-4.1975,4.8529) {$f$};
\end{tikzpicture}
\end{center}
\caption{A non-generic blow-up of $\F_m$ as explained in the text. The edges represent curves inside the surface.}
\label{high}
\end{figure}

\bit
\item \ubf{$A_2$ $(2,0)$ SCFT:} The F-theory construction involves two $-2$ curves in the base intersecting each other at one point. The surface associated to each curve is $\P^1\times T^2$. We simply glue the two $T^2$ together, which satisfies the Calabi-Yau condition \ref{CY} because the self-intersection of each $T^2$ is zero and the genus is one.
\item \ubf{Bifundamental between $SU(m)$ and $SU(n)$:} We have two $-2$ curves $C$ and $D$ in the base intersecting each other at a point with $\text{I}_m$ fiber on $C$ and $\text{I}_n$ fiber on $D$. We have computed $S_C$ and $S_D$ in Section \ref{A}. To glue them, we first have to make the blow-ups slightly more non-generic. We require $n$ out of the $2m$ blow-ups on $h$ in $\F_{m}$ inside $S_C$ to take place such that $x_i-x_{i+1}$ for $i=1,\cdots, n$ are effective curves after blow-ups. In this way, we obtain $n-1$ number of $-2$ curves, namely $x_i-x_{i+1}$ for $i=1,\cdots, n$. We also get two $-1$ curves, namely $x_n$ and $f-x_1$. See Figure \ref{high} for a description of the resulting surface. Similarly we make $m$ out of $2n$ blowups on $\F_n$ in $S_D$ to be non-generic in exactly the same fashion leading to $m-1$ number of $-2$ curves and two $-1$ curves. The gluing can now be described as follows. The $n-1$ number of $-2$ curves in $\F^{2m}_m$ are glued to the $f$ curves inside all the surfaces in $S_D$ except for $\F_{n}^{2n}$. Notice that, since $S_D$ contains $n$ surfaces, the $f$ curves participating in the gluing are $n-1$ in number too. Similarly, the $m-1$ number of $-2$ curves in $\F_n^{2n}$ are glued to the $f$ curves inside all the surfaces in $S_C$ except for $\F_{m}^{2m}$. The two $-1$ curves in $\F_m^{2m}$ are glued to the two $-1$ curves inside $\F^{2n}_n$.
\eit

\section*{Acknowledgments}
The authors thank Hee-Cheol Kim, Kantaro Ohmori and Cumrun Vafa for useful discussions.

Part of this work was completed at String-Math 2018 and Strings 2018. The authors thank the organizers of these conferences for providing a stimulating environment for scientific collaboration.

The research of LB was partially supported by the NSF grant PHY-1719924 and partially supported by Perimeter Institute for Theoretical Physics. Research at Perimeter Institute is supported by the Government of Canada through Industry Canada and by the Province of Ontario through the Ministry of Economic Development and Innovation.

The research of PJ is supported in part by the NSF grant PHY-1067976. PJ is thankful to Cumrun Vafa for his guidance and continued support. 

\appendix
\section{Mathematical background}\label{background}

\subsection{Some useful geometric notions}
 
 Let $X$ be a smooth projective variety. The set of all curve classes in $X$ with holomorphic representatives is called the \emph{Mori cone}, where we view two representatives as being equivalent if they have identical intersection numbers with all divisors of $X$. The Mori cone can be represented as the real span of a set of generators $C_\mu$; consequently, an arbitrary curve $C$ with holomorphic representative can be expressed as a non-negative linear combination of these generators:
	\begin{align}
		C= a_\mu C_\mu, ~~a_\mu \in \mathbb Z_{\geq 0},
	\end{align}
Note that $C_\mu$ themselves are linear combinations of a basis $C_i$ of all (possibly non-holomorphic) curve classes. The dual of the Mori cone (in the sense of convex geometry) is called the nef cone, and consists of all divisors $D$ such that 
	\begin{align}
		(D \cdot C)_{X} \geq 0
	\end{align}		
for all $C$ belonging to the Mori cone of $X$. 

Let $X=S$ be a surface with canonical class $K$. Suppose $\pi : S' \rightarrow S$ is the blowup of $S$ at a single point $p$ on a curve $C$, and assume that $C$ has multiplicity $m$ at $p$. Then the proper transform $C'$ of $C$ in $S'$ is denoted by 
	\begin{align}
		C' = \pi^*(C) - m E. 
	\end{align}	
Moreover, the canonical class $K'$ of $S'$ can be expressed in terms of $K, S$ as 
	\begin{align}
		K' = \pi^* K + E. 	
	\end{align}
Finally, consider gluing a surface $S$ to itself by identifying two curves $C_1,C_2 \in S$. This induces a birational map $\pi: S \rightarrow S'$. The canonical class $K'$ of $S'$ is then given by 
	\begin{align}
		\pi^* K' = K + C_1 + C_2. 
	\end{align}
In practice we omit explicit pullback maps when the context is clear.

\subsection{Ruled surfaces} \label{ruled}
A ruled surface $\mathbb F_{n,g}$ is a projective surface which can be viewed as a $\mathbb P^1$ bundle $\mathbb F_{n,g} \rightarrow e$ over a curve $e$ with genus $g$ and degree $n = -(e^2)_{\mathbb F_{n,g}}$. When $g=0$, $\mathbb F_{n,0} \equiv \mathbb F_{n} = \mathbb P [ \mathcal O \oplus \mathcal O(-n) ] \rightarrow e$ is a Hirzebruch surface. The Mori cone of $\mathbb F_{n,g}$ is the positive real span of the curve classes $e, f$, namely all curves of the form 
	\begin{align}
		a f + b e, ~~ (a,b) \in \mathbb Z^2_{\geq 0},
	\end{align}
where 
	\begin{align}
		(e^2)_{\mathbb F_{n,g}} = -n,~~  (e\cdot f)_{\mathbb F_{n,g}} = 1, ~~ (f^2)_{\mathbb F_{n,g}} = 0. 
	\end{align}
It is also useful to define the curve class 
	\begin{align}
		h = e + nf, ~~  (h^2)_{\mathbb F_{n,g}}=n,~~  (h\cdot f)_{\mathbb F_{n,g}} = 1, ~~ ( h \cdot e)_{\mathbb F_{n,g}}=0,
	\end{align}
in terms of which the canonical class is
	\begin{align}
		K = -2h + (2g -2 + n) f.
	\end{align}
(Note that the canonical class can be derived by starting with the parametrization $K = a f + b e$ and then demanding that adjunction is satisfied for curves of known genus.) Given a curve $c= a f + be$, the genus $g(c)$ can be expressed as a function of the genus and self-intersection of $e$ using the adjunction formula:
	\begin{align}
		2g(c)-2 =  ((K+c) \cdot c)_{\mathbb F_{n,g}} = 2 a (b-1)+b (-b n+2 g+n-2).
	\end{align}
	
\subsection{Total transforms versus proper transforms}\label{transforms}
In this paper, we often distinguish between the total transform and the proper transform of a curve in a surface. In this subsection, we explain the distinction between these two notions, and what this distinction implies about self-intersections of curve classes. Let $f : S' \rightarrow S$ be the blowup of a surface $S$ at a point $p \in S$. The exceptional divisor of this blowup is a curve $E \cong \mathbb P^1$. Now, consider a smooth curve $C \subset S$ which passes through the point $p$. By construction, 
	\begin{align}
		f^*(C) = C' + E,~~ f_*(C') = C, ~~f_*(E) = 0.
	\end{align}
We call the curve $C'$ the \emph{proper transform} of $C$, while we call the curve $f^*(C)$ the \emph{total transform} of the curve $C$. Informally, the proper transform can be thought of as the inverse image of all parts of the curve $C$ away from the blowup locus. More formally, $C'$ is the closure (in the Zariski topology) of the curve $f^{-1}(C-\{p\})$. The total transform, by contrast, is simply the inverse image of the entire curve $C$. 

What is the self-intersection of the curve $C'$ in $S'$? Observe that the curve $C'$ meets $E$ transversally at one point and hence we have 
	\begin{align}
		(C' \cdot E)_{S'} =1.
	\end{align}
Using the above, it follows from a useful result called the \emph{projection formula} that 
	\begin{align}
		0 =f^* (f_*(E) \cdot C)_S = (E \cdot f^*(C) )_{S'} = (E^2)_{S'} + (E \cdot C')_{S'}
	\end{align}
and hence 
	\begin{align}
		(E \cdot f^*(C))_{S'} = 0,~~(E^2)_{S'} = - (E \cdot C')_{S'} = -1. 
	\end{align}
Therefore, we may write 
	\begin{align}
		(C'^2)_{S} = (f^*(C)-E)^2_{S'} = (f^*(C))^2_{S'} + (E^2)_{S'} - 2 (f^*(C) \cdot E)_{S'} = C^2_{S} - 1. 
	\end{align}
Thus we see that the self-intersection of the curve $C'$ is reduced by 1. In practice, we suppress the pushforward and pullback notation and simply write
	\begin{align}
		C' = C - E,
	\end{align}
keeping in mind that $(C \cdot E)_{S'} = 0$. This construction has a simple generalization to the case where several points $p_i \in S$ are blown up. Let $C$ be a curve which passes through each point $p_i$ with multiplicity $m_i$. Then the proper transform of $C$ is 
	\begin{align}
		C' = C-\sum m_i E_i,
	\end{align}	
where we have $(C \cdot E_i)_{S'} = 0, (E_i^2)_{S'} = -1$.

\section{Example computations} \label{sample}
In this appendix, we provide explicit details in some examples where we compute the geometry of a resolved Calabi-Yau threefold starting from a singular elliptically fibered one. In the following examples, we use the notation established in Section \ref{w_models}. We compute the degrees and some example triple intersection numbers associated to the fibral divisors $S_i$ following the strategies explained there. For brevity, we suppress notation indicating pullbacks with respect to the blowups and projection of the elliptic fibration whenever the context is clear.

We have chosen these examples carefully to illustrate some key points. In Appendix \ref{1}, we discuss theories with $SO(10)$ gauge group along with $4-k$ hypers in spinor representation and $6-k$ hypers in fundamental representation, for $1\le k\le4$. For $k=4$, we only have fundamental hypers and we naively expect to see surfaces intersecting in the fashion of an affine $D_5$ Dynkin graph. However, our computations reveal that the surfaces have some extra intersections which do not modify the intersection pattern of the components of the degenerate $D_5$ elliptic fiber. In Appendix \ref{2}, we discuss rank one $6d$ SCFTs having $G_2$ gauge group with $n_f=1,4,7$ hypers in fundamental representation. These examples illustrate that non-split fibers can sometimes lead to ruled surfaces over curves of non-zero genus. In Appendix \ref{3}, we discuss theories with $SU(4)$ gauge group along with $16-4k$ hypers in fundamental and $2-k$ hypers in antisymmetric for $k=1,2$. This is the simplest example where a non-fundamental matter representation first appears.

\subsection{ \texorpdfstring{$G_2$}{} theories} \label{2}

In this subsection we showcase the $G_2$-model as an illustrative example---note that many of the results presented below first appeared in \cite{Esole:2017qeh}. The non-split $\text{I}_0^{*}$ model engineering $G_2$ is defined by the following Weierstrass equation:
	\begin{align}
		W_0 = y^2 - (x^3 + e_0^2 a_{4,2} x + e_0^3 a_{6,3} )=0.
	\end{align}		
For convenience, we work in the open set $z=1$. In the following discussion we describe the resolution 
	\begin{align}
		(x,y,e_0|e_1) ,~~(y,e_1|e_2). 
	\end{align}
The $\mathbb P^2$ ambient spaces of the elliptic fibers are described by homogeneous coordinates $[x:y:z=1]$. The singular locus $W_0 = \partial_i W_0 = 0$ is given by 
	\begin{align}
		x=y=e_0 =0
	\end{align}
and hence we blow up $X_0$ along the center $(x,y,e_0)$. The homogeneous coordinates of the new ambient space $Y_1$ are $[e_1x:e_1y:z=1][x:y:e_0]$. Introducing new coordinates  $ (x,y,e_0) \to (e_1 x, e_1 y, e_1 e_0)$ and factoring out two copies of the exceptional divisor $E_1$ defined by the local equation $e_1=0$, we find that the total transform $X_1$ is given by
	\begin{align}
		W_1 = y^2 - ( e_1 x^3 +  e_1 e_0^2 a_{4,2} x + e_1e_0^3 a_{6,3}) = 0. 
	\end{align}  
The singular locus of $X_1$ is a subset of the locus
	\begin{align}
		y = e_1 =0. 
	\end{align}
We thus take $(y,e_1)$ to be the center of the second blowup and make the replacements $(y,e_1) \rightarrow (e_2 y,e_2 e_1)$. The homogeneous coordinates of the new ambient space $Y_2$ are $[e_2e_1 x:e_2^2e_1y:z=1][x:e_2 y:e_0] [y:e_1]$. Factoring out a single copy of the exceptional divisor $E_2$ with local equation $e_2=0$, the total transform $X_2$ is given by
	\begin{align}
		W_2 = e_2 y^2 - (e_1 x^3 + e_1 e_0^2 a_{4,2} x + e_1 e_0^3 a_{6,3} )=0. 
	\end{align}
One can verify by explicit computation that $X_2$ is smooth, i.e. that $W_2=\partial_i W_2 =0$ has no solutions. 

We now turn our attention to computing the geometry of these divisors, noting for reference the following divisor classes:
	\begin{align}
	\begin{split}
		[x] &= -2K_B -E_1,~~[y]= -3K_B - E_1 -E_2,~~[e_0] = E_0 - E_1\\
		[e_1]&= E_1-E_2,~~ [e_2]=E_2.
	\end{split} 
	\end{align}
We remark that by fixing $z=1$ using the scaling freedom of the original $\mathbb P^2$ coordinates $[x:y:z] \cong [\lambda_0 x : \lambda_0 y : \lambda_0 z]$ of the fibers of $Y_0$, we have eliminated the dependence of $[x],[y]$ on the divisor class $H$. The irreducible components $F_{i}$ of the resolved elliptic fiber $F = F_0 + 2 F_1 + (\sum_{j=1}^3 F_{2,j})$ are parametrized as follows:
	\begin{align}
	\begin{split}
		F_0 ~&:~ e_2 y^2 -  e_1 x^3 =0 ~~\subset ~~[\lambda_1 x: \lambda_1 e_2 y] [ \lambda_2 \lambda_1 y: (\lambda_2/\lambda_1) e_1] \\
		F_1 ~&:~ [\lambda_1 x : \lambda_1 e_0] [\lambda_2 \lambda_1 y :0]\\
		F_2~&:~x^3 +a_{4,2} e_0^2 x + a_{6,3}e_0^3 =0 ~~\subset ~~[\lambda_1 x:\lambda_1 e_0] [ \lambda_2 \lambda_1 y : (\lambda_2/\lambda_1) e_1] 
	\end{split}
	\end{align}
where $\lambda_i \in \mathbb C^{\times}$ are scale factors associated to the scaling symmetries of the above homogeneous coordinates. We ignore the dependence of $F_i$ on the coordinates of the original $\mathbb P^2$, $[e_2 e_1 x:e_2^2 e_1 y:1]$, since the only curve with a non-trivial dependence on these coordinates is $F_0$ which has a rational parametrization in terms of a single complex coordinate $x/y e_2$---see (\ref{eqn:su2fib}) below. We restrict $F_i$ to open sets and fix the projective scaling freedom by setting $\lambda_i$ equal to convenient choices of nonzero coordinates in order to determine the degrees of their associated projective bundles: 
	\begin{align}
		\begin{split}
		\label{eqn:su2fib}
			F_0 ~&:~\left(x/y e_2\right)^3 (y e_1 e_2^2 ) -1 = 0~~\subset~~ [x/y e_2:1][y e_1 e_2^2 :1]   \\
			F_1 ~&:~ [x:e_0]\\
			F_2 ~&:~ (x/e_0)^3 + a_{4,2} (x/e_0) + a_{6,3} =0 ~~ \subset ~~[x/e_0:1] [y/e_0 :e_0 e_1]
		\end{split}
	\end{align}
In the case of $F_2$, we see the defining equation above parametrizes three disjoint points on the complex line $\mathbb C$ with coordinate $x/e_0$, and hence the fibral divisor $S_2$ is a triple cover of the ruled surface over $C'$ with fiber coordinates $[y/e_0:e_0e_1]$. The projective bundles $ \mathbb P_{ C} [ \mathcal O \oplus \mathcal L_i]$ are defined by the following divisor classes $L_i$ dual to the line bundles $\mathcal L_i$:
	\begin{align}
	\begin{split}
	\label{eqn:G2fib}
		L_0& =[x/ye_2] =-2K_B -E_1 - (H-3K_B -E_1-E_2) -E_2=   K_B \\
		 L_1&= [x/e_0] =-2K_B -E_1 - (E_0-E_1) = -2K_B - E_0 \\
		 L_2&= [y/e_0^2 e_1] =  -3K_B - E_1-E_2 - 2(E_0 - E_1) - (E_1-E_2)= -3K_B-2E_0 ,
	\end{split}
	\end{align}
in terms of which the degrees $n_i =( \pi_* \circ f_* (L_i) \cdot  C')_B$  are 
	\begin{align}
	\begin{split}
		n_0 &= k-2,~~n_1  = k - 4,~~n_2  = k-6.
	\end{split}
	\end{align}
In the above equations we have used the fact that $(C')^2_B = -k$ and $g(C') = 0$. Since the non-split $\text{I}_{0}^{*} $ model contains non-split Kodaira fibers, the divisor $S_2$ is a triple cover of the projective bundle $\mathbb P_{C'} [ \mathcal O \oplus \mathcal L_2]$, and may be viewed as a ruled surface of degree $3(k-6)$ over a curve $C$ which according to (\ref{eqn:newgen}) has genus 
	\begin{align}
		g(C) = \frac{3}{2} (b- 2 +\frac{2}{3} ) = \frac{3}{2} b - 2. 
	\end{align}
In order to determine the genus of $C$ we compute the number of fibers in the ramification locus. Referring once again to the defining equation for $F_2$, we see that the ramification locus is the vanishing of the discriminant of the equation 
	\begin{align}
		(x/e_0)^3 + a_{4,2} (x /e_0) + a_{6,3} = 0, 
	\end{align}
namely
	\begin{align}
		4 a_{4,2}^3 + 27 a_{6,3}^2 =0,
	\end{align}
where two of the solutions of the above cubic equation are conjugate. Since $[4 a_{4,2}^3 + 27 a_{6,3}^2]/3 = -4K_B - 2 C$, we find that that the discriminant vanishes at
	\begin{align}
		b = 2(4-k) 
	\end{align}
points along $C$, which implies 
	\begin{align}
		g(C) = 3(4-k) - 2 = 10-3k. 
	\end{align}

Next, we compute the triple intersection numbers of this resolved non-split $\text{I}_{0}^{*}$ model. The smooth elliptic threefold $X_2$ has a natural basis of divisors $S_i$ given by 
	\begin{align}
		S_0 = E_0 -E_1,~~ S_1 = E_1 - E_2, ~~ S_2 = E_2.
	\end{align}
We also note the class of $W_2= 0$ is 
	\begin{align}
		[W_2] = 3 H-6 K_B -2 E_1 - E_2.
	\end{align}		
 Following procedure outlined in \cite{Esole:2017kyr}, we compute triple intersection numbers in terms of the following pushforwards of intersection products. As a concrete example, we compute one triple intersection number explicitly, namely $(D_0^3)_{X_2}$. First we compute the pullback of this intersection product to the ambient projective bundle $Y_2$ and express it in a basis of exceptional divisors:
 	\begin{align}
		(D_0^3)_{X_2} = (D_0^3  [W_2] )_{Y_2} = ((S-E_1)^3 (-E_2-2 E_1+3 H-6K_B))_{Y_2}.
	\end{align}
Next, we compute a sequence of pushforwards:
	\begin{align}
	\begin{split}
		&f_{2*} ((E_0-E_1)^3  (-E_2-2 E_1+3 H-6K_B))_{Y_2} \\
		=~&((E_0-E_1)^3  (-2 E_1+3 H-6 K_B))_{Y_1}	
	\end{split}\\
	\begin{split}
		& f_{1*} \circ f_{2*} ((E_0-E_1)^3  (-E_2-2 E_1+3 H-6K_B))_{Y_2}\\
		=~&(E_0 (H-2 K_B)  \left(H^2+H (-7K_B -4 E_0)+3 (E_0+2 K_B)^2\right))_{Y_0} 
	\end{split}\\
	\begin{split}
		& \pi_{*} \circ f_{1*} \circ f_{2*} ((E_0-E_1)^3  (-E_2-2 E_1+3 H-6K_B))_{Y_2}\\
		=~&-4 (C \cdot  (K_B+C))_B
	\end{split}
	\end{align}
where in the last line above we have used $\pi_* E_0 = C$. Finally, we use the fact that $C^2=-k$ to compute
	\begin{align}
		-4 (C\cdot  (K_B+C))_B &= 8. 
	\end{align}
The complete set of triple intersection numbers are summarized in (\ref{eqn:G2}). Ultimately, we find 
	\begin{align}
	\begin{split}
		S_0 = \mathbb F_{k-2} , ~~ S_1 =\mathbb F_{k-4},~~ S_2 = \mathbb F_{3(k-6),10-k}.
	\end{split}
	\end{align}

\subsection{ \texorpdfstring{$SU(4)$}{} theories}\label{3}

We turn to the $SU(4)$-model as our next example. Note that many of the results presented below first appeared in \cite{Esole:2014bka}. The split $\text{I}_4$ model engineering $SU(4)$ is described by the following Weierstrass equation:
	\begin{align}
		W_0 =y^2 + a_1 x y  - ( x^3 +  a_{2,1} e_0 x^2+a_{4,1} e_0 x  +a_{6,2}e_0^2) =0.
	\end{align}
Again for convenience, we work in the open set $z=1$. We study the resolution
	\begin{align}
		(x,y,e_0|e_1) ,~~ (y,e_1|e_2),~~(x,e_2|e_3).
	\end{align}
The ambient spaces of the elliptic fibers are described by projective coordinates $[x:y:z=1]$. The singular locus $W_0 = \partial_i W_0 =0$ is given by 
	\begin{align}
		x = y= e_0 =0.
	\end{align}
We first blow up $X_0$ along the center $(x,y,e_0)$ by making the substitution $(x,y,e_0) \to (e_1x,e_1y,e_1e_0)$, which introduces an exceptional divisor $E_1$ with local equation $e_1=0$. The homogeneous coordinates of the ambient space $Y_1$ are now $[e_0 x:e_0y:z=1][x:y:e_0]$. Factoring out two copies of $E_1$, the total transform $X_1$ is given by
	\begin{align}
		W_1 = y^2 + a_1 x y - ( e_1 x^3 + e_1 e_0 a_{2,1} x^2 + e_1 e_0^2 a_{4,2} x + e_1^2 a_{6,4} e_0^4 ).
	\end{align}
The singular locus of $X_1$ is a subset of the locus 
	\begin{align}
		y= e_1 =0,
	\end{align}
hence we select $(y,e_1)$ as the center of the second blowup. Making the substitution $(y,e_1) \to (e_2 y , e_2 e_1)$ and introducing the exceptional divisor $E_2$ with local equation $e_2 =0$, we find that the ambient space $Y_2$ is described by the homogeneous coordinates $[e_1 e_0 x: e_1^2 e_0:z=1][x:e_2 y :e_0][y:e_1]$. Factoring out a single copy of $E_2$, the proper transform $X_2$ is given by 
	\begin{align}
		W_2 =e_2 y^2 + a_1 x y- (  e_1x^3+ e_1  e_0 a_{2,1}x^2+ e_1 e_0^2a_{4,2} x+  e_2 e_1^2e_0^4a_{6,4}).
	\end{align} 
The singular locus of $X_2$ is a subset of the locus	
	\begin{align}
		x = e_2 = 0.
	\end{align}
We perform one final blowup along the center $(x,e_2)$, making the substitution $(x,e_2) \to (e_3 x,e_3 e_2)$ and introducing the exceptional divisor $E_3$ with local equation $e_3=0$. The ambient space $Y_3$ is described by the homogeneous coordinates $[e_3^2 e_2 e_1 x: e_3^2 e_2^2 e_1 y:z=1][e_3 x:e_3 e_2 y:e_0][y :e_1][x:e_2]$. Factoring out a single copy of $E_3$, the proper transform $X_3$ is given by
	\begin{align}
		W_3 =e_2 y^2 + a_1 x y- (  e_3^2e_1x^3+ e_3e_1  e_0 a_{2,1}x^2+ e_1 e_0^2a_{4,2} x+  e_2 e_1^2e_0^4a_{6,4}).
	\end{align}
One can verify by explicit computation that the $W_3= \partial_i W_3 =0$ has no solutions and hence $X_3$ is a smooth elliptic fibration.

Our next task is to determine the geometry of the divisors $S_i$. For reference, we note the following divisor classes
	\begin{align}
	\begin{split}
		[x]&= -2K_B - E_1 - E_3,~~ [y] = -3K_B - E_1 - E_2,~~ [e_0] =E_0-E_1\\
		[e_1] &=E_1-E_2,~~ [e_2] = E_2- E_3 ,~~ [e_3]=E_3.
	\end{split}
	\end{align} 
Again, we have used the scaling symmetry of the $\mathbb P^2$ fiber coordinates of $Y_0$ to set $z=1$, which removes the dependence of the classes $[x],[y]$ on the divisor class $H$ in the above expressions. 

We first study the irreducible components $F_{i=1,\dots 3} $ of the resolved elliptic fiber $F = F_0 + F_1 + F_2 + F_3 $ (we ignore $F_0$ because as in the previous example $F_0$ has a rational parametrization.) The projective coordinates and scaling symmetries introduced by the blowups are
	\begin{align}
		[\lambda_1 e_3 x: \lambda_1 e_3 e_2 y : \lambda_1 e_0][\lambda_2 \lambda_1 y: (\lambda_2/\lambda_1) e_1] [\lambda_3 \lambda_1 x: e_2 (\lambda_3/\lambda_2) ],
	\end{align}
where $\lambda_i \in \mathbb C^{\times}$. After fixing the various projective scaling symmetries, we find the irreducible components admit the following parametrizations:
	\begin{align}	
		\begin{split}
			F_1 ~&:~[x e_3 :e_0]\\
			F_2 ~&:~[x e_3 :e_0]\\
			F_3 ~&:~A (x/e_0) + B e_2 =0 ~~\subset ~~[(y/e_0) : e_0 e_1][(x/e_0) :e_2] 
		\end{split}
	\end{align}
where 
	\begin{align}
		A = a_1 (y/e_0) - a_{4,2} e_0 e_1,~~ B = (y/e_0)^2 - a_{6, 4} (e_0 e_1)^2.
	\end{align}
The above computation implies the projective bundles $\mathbb P_{C}[\mathcal O\oplus \mathcal L_i]$ are defined by the following divisor classes:
	\begin{align}
	\begin{split}
		L_0 &= -K_B\\
		 L_1 &=[x e_3/e_0]=   -2 K_B-E_0\\
		 L_2& =[x e_3/e_0]=  -2 K_B -E_0\\
		  L_3& =[y/e_0^2e_1]= -K_B.
	\end{split}
	\end{align}	
We thus find the following degrees $n_i =( \pi_* \circ f_*(L_i) \cdot C)_B$: 
	\begin{align}
		n_0 = k-2,~~n_1=k-4,~~n_2=k-4,~~n_3=k-6.
	\end{align}
We next compute the triple intersection numbers. The threefold $X_3$ has a natural basis of divisors $S_i$ given by 
	\begin{align}
		S_0 = E_0 - E_1,~~ S_1 = E_1 - E_2,~~ S_2 = E_2 - E_3 ,~~ S_3 = E_3.
	\end{align}
The divisor class $[W_3]$ of $W_3 = 0$ is 
	\begin{align}
		[W_3] = 3 H - 6 K_B - 2 E_1 - E_2 - E_3. 
	\end{align}
The triple intersection numbers and above degrees together imply
	\begin{align}
		S_0 = \mathbb F_{k-2},~~ S_1 = \mathbb F_{k-4},~~ S_2 = \mathbb F_{k-4}^{2(2-k)},~~ S_3 = \mathbb F_{k-6}^{4(4-k)}.
	\end{align}
One very interesting property of the $\text{I}_{4}^s$ model in the case $k=1$ is the existence of unusual intersection patterns, i.e. intersections which do not fit into the structure of a Dynkin diagram, due to the presence of localized matter in the model. In particular, we find that $S_1$ and $S_2$ intersect non-trivially. Notice that the irreducible components of the elliptic fiber only intersect over specific points in $B$:
	\begin{align}
		F_1 \cap F_2 ~:~ a_1 =0.
	\end{align}
Notice that $a_1$ vanishes at $(C \cdot [a_1])_B = -(K_B \cdot C)_B = 2-k$ points along $C \subset B$.

\subsection{ \texorpdfstring{$SO(10)$}{} theories}\label{1}

The split $\text{I}_{1}^{*}$ model engineering gauge group $SO(10)$ is defined by the following Weierstrass equation:
	\begin{align}
		W_0 = y^2 +e_0^2 a_3 y+ e_0 a_1x y -(x^3+e_0 a_2 x^2+e_0^3a_4 x+e_0^5a_6 ) =0.
	\end{align}
We study the resolution
	\begin{align}
		(x,y,e_0|e_1) ,~~ (y,e_1|e_2),~~(x,e_2|e_3),~~(y,e_3|e_4),~~(e_2,e_3|e_5). 
	\end{align}	
The ambient spaces of the elliptic fibers are described by the projective coordinates $[x:y:z=1]$. The singular locus of $W_0 = \partial_i W_0 = 0$ as in previous cases is given by 
	\begin{align}
		x = y = e_0.
	\end{align}
Since we have already explained the procedure for computing resolutions in previous examples, we skip the details of the resolution and simply record the final result. The smooth elliptic threefold $X_5$ is a hypersurface 
	\begin{align}
	\begin{split}
		W_5&=e_2 e_4 y^2+a_1 e_1 e_2 e_3 e_4 e_5 e_0 x y+a_3 e_1 e_2 e_0^2 y\\
		&-(e_1 e_3^2 e_4 e_5 x^3+a_2 e_1 e_3 e_0 x^2+a_4 e_1^2 e_2 e_3 e_5 e_0^3 x+a_6 e_1^3 e_2^2 e_3 e_5^2 e_0^5) =0
	\end{split}
	\end{align}
in the ambient projective bundle $Y_5$ with homogeneous coordinates (including their scaling symmetries)
	\begin{align}
	\begin{split}
		&[e_1 e_2 e_3^2 e_4^2 e_5^3 x:e_1 e_2^2 e_3^2 e_4^3 e_5^4 y:z=1][e_3 e_4 e_5 \lambda _1 x:e_2 e_3 e_4^2 e_5^2 \lambda _1 y:\lambda _1 s]\\
		&[e_4 \lambda _1 \lambda _2 y:\frac{e_1 \lambda _2}{\lambda _1}][\lambda _1 \lambda _3 x:\frac{e_2 e_5 \lambda _3}{\lambda _2}][\lambda _1 \lambda _2 \lambda _4 y:\frac{e_3 e_5 \lambda _4}{\lambda _3}][\frac{e_2 \lambda _3 \lambda _5}{\lambda _2}:\frac{e_3 \lambda _4 \lambda _5}{\lambda _3}]
	\end{split} 
	\end{align}
One can verify by explicit computation that $W_5 = \partial_i W_5 =0$ and hence $X_5$ is smooth. 

To determine the geometry of the divisors, as in previous examples, we must compute the degrees of the projective bundles $S_i$. Our starting point is again the irreducible components $F_i$, whose explicit parametrizations we will use to retrieve the degrees of their projective bundles. We note the following divisor classes:
	\begin{align}
	\begin{split}
		[x] &=  2L - E_1-E_3 ,~~ [y] = 3L - E_1 -E_2 - E_4,~~ [e_0] = E_0 - E_1\\
		[e_1] &= E_1 - E_2,~~ [e_2] = E_2-E_3-E_5,~~ [e_3] = E_3 - E_4-E_5\\
		[e_4]&=E_4,~~ [e_5]=E_5. 	
	\end{split}
	\end{align}
After fixing the various projective scaling symmetries, we find the irreducible components of the elliptic fiber admit the following parametrizations in $X_5$:	
	\begin{align}
	\begin{split}
		F_1 ~&:~[e_5 : e_0/(x e_3 e_4)]\\
		F_2 ~&:~ [(y e_4/e_0) : e_0 e_1]\\
		F_3 ~&:~[(x/e_0): e_0 e_1 e_2 e_5 ] \\
		F_4 ~ &:~[(x/e_0^2 e_1) :e_5] \\
		F_5 ~&:~[y e_4:s^2 e_1] 
	\end{split}
	\end{align}
The projective bundles $\mathbb P_{C} [ \mathcal O \oplus \mathcal L_i]$ are defined in terms of the following line bundles $L_i$:
	\begin{align}
	\begin{split}
		L_0& = K_B\\
		 L_1 &=[xe_3e_4e_5/e_0] = E_0 + 2K_B\\
		  L_2 &=[ye_4/e_0^2e_1] =2 E_0 + 3 K_B\\
		 L_3& =[x/e_0^2e_1e_2e_5]=  2 K_B + 2 E_0\\
		 L_4 &= [x/e_0^2e_1e_5]= 2 K_B + 2 E_0\\
		 L_5 &=[y e_4/s^2 es_1] = 2 E_0 + 3K_B.
	\end{split}
	\end{align}
It follows that the degrees $n_i = (\pi_* \circ f_*(L_i) \cdot C)_B$ are 
	\begin{align}
		n_0 = k-2,~~ n_1=k-4,~~ n_2 =k-6,~~ n_3 =4,~~n_4 =4,~~ n_5=k-6,
	\end{align}
Next, we compute the triple intersection numbers. The threefold $X_5$ has a natural basis of fibral divisors given by 
	\begin{align}
	\begin{split}
		S_0 &= E_0 - E_1,~~ S_1 = E_1 - E_2,~~ S_2 =-E_1 + 2 E_2 - E_3 - E_5\\
		S_3 &= E_3 - E_4 -E_5,~~ S_4=E_4,~~ S_5 = E_5.
	\end{split}
	\end{align}		
The divisor class of $W_5=0$ is given by
	\begin{align}
		[W_5]= 3H -6K_B -2E_1 - E_2 -E_3 - E_4 -E_5. 
	\end{align}
The triple intersection numbers are summarized in (\ref{eqn:SO10samp}). The degrees and triple intersections together imply that the divisors have the following geometry:
	\begin{align}
		S_0 = \mathbb F_{k-2} ,~~ S_1 = \mathbb F_{k-4},~~S_2= \mathbb F_{k-6} ,~~ S_3=\mathbb F_4,~~ S_4 = \mathbb F_4^{16-3k},~~ S_5 = \mathbb F_{6-k}^{4-k}.
	\end{align}

\bibliographystyle{ytphys}
\let\bbb\bibitem\def\bibitem{\itemsep4pt\bbb}
\bibliography{ref}

\providecommand{\href}[2]{#2}\begingroup\raggedright\begin{thebibliography}{10}

\bibitem{Cordova:2016xhm}
C.~Cordova, T.~T. Dumitrescu, and K.~Intriligator, ``{Deformations of
  Superconformal Theories},''
  \href{http://dx.doi.org/10.1007/JHEP11(2016)135}{{\em JHEP} {\bfseries 11}
  (2016) 135},
\href{http://arxiv.org/abs/1602.01217}{{\ttfamily arXiv:1602.01217 [hep-th]}}.

\bibitem{Cordova:2016emh}
C.~Cordova, T.~T. Dumitrescu, and K.~Intriligator, ``{Multiplets of
  Superconformal Symmetry in Diverse Dimensions},''
\href{http://arxiv.org/abs/1612.00809}{{\ttfamily arXiv:1612.00809 [hep-th]}}.

\bibitem{Jefferson:2018irk}
P.~Jefferson, S.~Katz, H.-C. Kim, and C.~Vafa, ``{On Geometric Classification
  of 5d SCFTs},'' \href{http://dx.doi.org/10.1007/JHEP04(2018)103}{{\em JHEP}
  {\bfseries 04} (2018) 103},
\href{http://arxiv.org/abs/1801.04036}{{\ttfamily arXiv:1801.04036 [hep-th]}}.

\bibitem{Douglas:1996xp}
M.~R. Douglas, S.~H. Katz, and C.~Vafa, ``{Small instantons, Del Pezzo surfaces
  and type I-prime theory},''
  \href{http://dx.doi.org/10.1016/S0550-3213(97)00281-2}{{\em Nucl. Phys.}
  {\bfseries B497} (1997) 155--172},
\href{http://arxiv.org/abs/hep-th/9609071}{{\ttfamily arXiv:hep-th/9609071
  [hep-th]}}.

\bibitem{Morrison:1996xf}
D.~R. Morrison and N.~Seiberg, ``{Extremal transitions and five-dimensional
  supersymmetric field theories},''
  \href{http://dx.doi.org/10.1016/S0550-3213(96)00592-5}{{\em Nucl. Phys.}
  {\bfseries B483} (1997) 229--247},
\href{http://arxiv.org/abs/hep-th/9609070}{{\ttfamily arXiv:hep-th/9609070
  [hep-th]}}.

\bibitem{Bhardwaj:2019hhd}
L.~Bhardwaj, ``{Revisiting the classifications of $6d$ SCFTs and LSTs},''
\href{http://arxiv.org/abs/1903.10503}{{\ttfamily arXiv:1903.10503 [hep-th]}}.

\bibitem{Bhardwaj:2018jgp}
L.~Bhardwaj, D.~R. Morrison, Y.~Tachikawa, and A.~Tomasiello, ``{The frozen
  phase of F-theory},'' \href{http://dx.doi.org/10.1007/JHEP08(2018)138}{{\em
  JHEP} {\bfseries 08} (2018) 138},
\href{http://arxiv.org/abs/1805.09070}{{\ttfamily arXiv:1805.09070 [hep-th]}}.

\bibitem{Tachikawa:2015wka}
Y.~Tachikawa, ``{Frozen singularities in M and F theory},''
  \href{http://dx.doi.org/10.1007/JHEP06(2016)128}{{\em JHEP} {\bfseries 06}
  (2016) 128},
\href{http://arxiv.org/abs/1508.06679}{{\ttfamily arXiv:1508.06679 [hep-th]}}.

\bibitem{Heckman:2015bfa}
J.~J. Heckman, D.~R. Morrison, T.~Rudelius, and C.~Vafa, ``{Atomic
  Classification of 6D SCFTs},''
  \href{http://dx.doi.org/10.1002/prop.201500024}{{\em Fortsch. Phys.}
  {\bfseries 63} (2015) 468--530},
\href{http://arxiv.org/abs/1502.05405}{{\ttfamily arXiv:1502.05405 [hep-th]}}.

\bibitem{Heckman:2013pva}
J.~J. Heckman, D.~R. Morrison, and C.~Vafa, ``{On the Classification of 6D
  SCFTs and Generalized ADE Orbifolds},''
  \href{http://dx.doi.org/10.1007/JHEP06(2015)017,
  10.1007/JHEP05(2014)028}{{\em JHEP} {\bfseries 05} (2014) 028},
  \href{http://arxiv.org/abs/1312.5746}{{\ttfamily arXiv:1312.5746 [hep-th]}}.
[Erratum: JHEP06,017(2015)].

\bibitem{Bhardwaj:2015xxa}
L.~Bhardwaj, ``{Classification of 6d $ \mathcal{N}=\left(1,0\right) $ gauge
  theories},'' \href{http://dx.doi.org/10.1007/JHEP11(2015)002}{{\em JHEP}
  {\bfseries 11} (2015) 002},
\href{http://arxiv.org/abs/1502.06594}{{\ttfamily arXiv:1502.06594 [hep-th]}}.

\bibitem{DelZotto:2017pti}
M.~Del~Zotto, J.~J. Heckman, and D.~R. Morrison, ``{6D SCFTs and Phases of 5D
  Theories},'' \href{http://dx.doi.org/10.1007/JHEP09(2017)147}{{\em JHEP}
  {\bfseries 09} (2017) 147},
\href{http://arxiv.org/abs/1703.02981}{{\ttfamily arXiv:1703.02981 [hep-th]}}.

\bibitem{Jefferson:2017ahm}
P.~Jefferson, H.-C. Kim, C.~Vafa, and G.~Zafrir, ``{Towards Classification of
  5d SCFTs: Single Gauge Node},''
\href{http://arxiv.org/abs/1705.05836}{{\ttfamily arXiv:1705.05836 [hep-th]}}.

\bibitem{Seiberg:1996bd}
N.~Seiberg, ``{Five-dimensional SUSY field theories, nontrivial fixed points
  and string dynamics},''
  \href{http://dx.doi.org/10.1016/S0370-2693(96)01215-4}{{\em Phys. Lett.}
  {\bfseries B388} (1996) 753--760},
\href{http://arxiv.org/abs/hep-th/9608111}{{\ttfamily arXiv:hep-th/9608111
  [hep-th]}}.

\bibitem{Intriligator:1997pq}
K.~A. Intriligator, D.~R. Morrison, and N.~Seiberg, ``{Five-dimensional
  supersymmetric gauge theories and degenerations of Calabi-Yau spaces},''
  \href{http://dx.doi.org/10.1016/S0550-3213(97)00279-4}{{\em Nucl. Phys.}
  {\bfseries B497} (1997) 56--100},
\href{http://arxiv.org/abs/hep-th/9702198}{{\ttfamily arXiv:hep-th/9702198
  [hep-th]}}.

\bibitem{Witten:1996qb}
E.~Witten, ``{Phase transitions in M theory and F theory},''
  \href{http://dx.doi.org/10.1016/0550-3213(96)00212-X}{{\em Nucl. Phys.}
  {\bfseries B471} (1996) 195--216},
\href{http://arxiv.org/abs/hep-th/9603150}{{\ttfamily arXiv:hep-th/9603150
  [hep-th]}}.

\bibitem{Grauert}
H.~Grauert, ``{{\"U}ber Modifikationen und exzeptionelle analytische Mengen},''
  {\em Math. Ann.} (1962) 331--368.

\bibitem{Vafa:1996xn}
C.~Vafa, ``{Evidence for F theory},''
  \href{http://dx.doi.org/10.1016/0550-3213(96)00172-1}{{\em Nucl. Phys.}
  {\bfseries B469} (1996) 403--418},
\href{http://arxiv.org/abs/hep-th/9602022}{{\ttfamily arXiv:hep-th/9602022
  [hep-th]}}.

\bibitem{Morrison:1996na}
D.~R. Morrison and C.~Vafa, ``{Compactifications of F theory on Calabi-Yau
  threefolds. 1},'' \href{http://dx.doi.org/10.1016/0550-3213(96)00242-8}{{\em
  Nucl. Phys.} {\bfseries B473} (1996) 74--92},
\href{http://arxiv.org/abs/hep-th/9602114}{{\ttfamily arXiv:hep-th/9602114
  [hep-th]}}.

\bibitem{Morrison:1996pp}
D.~R. Morrison and C.~Vafa, ``{Compactifications of F theory on Calabi-Yau
  threefolds. 2.},'' \href{http://dx.doi.org/10.1016/0550-3213(96)00369-0}{{\em
  Nucl. Phys.} {\bfseries B476} (1996) 437--469},
\href{http://arxiv.org/abs/hep-th/9603161}{{\ttfamily arXiv:hep-th/9603161
  [hep-th]}}.

\bibitem{Katz:2011qp}
S.~Katz, D.~R. Morrison, S.~Schafer-Nameki, and J.~Sully, ``{Tate's algorithm
  and F-theory},'' \href{http://dx.doi.org/10.1007/JHEP08(2011)094}{{\em JHEP}
  {\bfseries 08} (2011) 094},
\href{http://arxiv.org/abs/1106.3854}{{\ttfamily arXiv:1106.3854 [hep-th]}}.

\bibitem{Esole:2017rgz}
M.~Esole, P.~Jefferson, and M.~J. Kang, ``{The Geometry of F$_4$-Models},''
\href{http://arxiv.org/abs/1704.08251}{{\ttfamily arXiv:1704.08251 [hep-th]}}.

\bibitem{Esole:2017kyr}
M.~Esole, P.~Jefferson, and M.~J. Kang, ``{Euler Characteristics of Crepant
  Resolutions of Weierstrass Models},''
\href{http://arxiv.org/abs/1703.00905}{{\ttfamily arXiv:1703.00905 [math.AG]}}.

\bibitem{Grassi:2011hq}
A.~Grassi and D.~R. Morrison, ``{Anomalies and the Euler characteristic of
  elliptic Calabi-Yau threefolds},''
  \href{http://dx.doi.org/10.4310/CNTP.2012.v6.n1.a2}{{\em Commun. Num. Theor.
  Phys.} {\bfseries 6} (2012) 51--127},
\href{http://arxiv.org/abs/1109.0042}{{\ttfamily arXiv:1109.0042 [hep-th]}}.

\bibitem{Esole:2014bka}
M.~Esole, S.-H. Shao, and S.-T. Yau, ``{Singularities and Gauge Theory
  Phases},'' \href{http://dx.doi.org/10.4310/ATMP.2015.v19.n6.a2}{{\em Adv.
  Theor. Math. Phys.} {\bfseries 19} (2015) 1183--1247},
\href{http://arxiv.org/abs/1402.6331}{{\ttfamily arXiv:1402.6331 [hep-th]}}.

\bibitem{Esole:2014hya}
M.~Esole, S.-H. Shao, and S.-T. Yau, ``{Singularities and Gauge Theory Phases
  II},'' \href{http://dx.doi.org/10.4310/ATMP.2016.v20.n4.a2}{{\em Adv. Theor.
  Math. Phys.} {\bfseries 20} (2016) 683--749},
\href{http://arxiv.org/abs/1407.1867}{{\ttfamily arXiv:1407.1867 [hep-th]}}.

\bibitem{Esole:2011sm}
M.~Esole and S.-T. Yau, ``{Small resolutions of SU(5)-models in F-theory},''
  \href{http://dx.doi.org/10.4310/ATMP.2013.v17.n6.a1}{{\em Adv. Theor. Math.
  Phys.} {\bfseries 17} no.~6, (2013) 1195--1253},
\href{http://arxiv.org/abs/1107.0733}{{\ttfamily arXiv:1107.0733 [hep-th]}}.

\bibitem{Esole:2015xfa}
M.~Esole and S.-H. Shao, ``{M-theory on Elliptic Calabi-Yau Threefolds and 6d
  Anomalies},''
\href{http://arxiv.org/abs/1504.01387}{{\ttfamily arXiv:1504.01387 [hep-th]}}.

\bibitem{Esole:2017qeh}
M.~Esole, R.~Jagadeesan, and M.~J. Kang, ``{The Geometry of G$_2$, Spin(7), and
  Spin(8)-models},''
\href{http://arxiv.org/abs/1709.04913}{{\ttfamily arXiv:1709.04913 [hep-th]}}.

\bibitem{Huang:2018gpl}
Y.-C. Huang and W.~Taylor, ``{Comparing elliptic and toric hypersurface
  Calabi-Yau threefolds at large Hodge numbers},''
\href{http://arxiv.org/abs/1805.05907}{{\ttfamily arXiv:1805.05907 [hep-th]}}.

\end{thebibliography}\endgroup


\end{document}